\documentclass{aa}

\usepackage[utf8]{inputenc}
\usepackage[T1]{fontenc}

\usepackage{graphicx}	
\usepackage{amsmath}	
\usepackage{amssymb}	
\usepackage{upgreek}    
\usepackage{multirow}
\usepackage{verbatim}
\usepackage[varg]{txfonts}
\usepackage{hyperref}
\usepackage[all]{hypcap}
\usepackage{fixmath}
\usepackage[usenames, dvipsnames]{color}

\newcommand{\homsun}{\,h_{70}^{-1} {\rm M_\odot}}
\newcommand{\hmsun}{\,h_{70}^{-2} {\rm M_\odot}}
\newcommand{\msun}{\,{\rm M_\odot}}
\newcommand{\hkpc}{\,h_{70}^{-1}{\rm{kpc}} }
\newcommand{\hMpc}{\,h_{70}^{-1}{\rm{Mpc}} }
\newcommand{\magn}{\,{\rm mag} }
\newcommand{\mpss}{\,{\rm m}\,{\rm s}^{-2} }
\newcommand{\am}{\,{\rm arcmin}}
\newcommand{\as}{\,{\rm arcsec}}
\newcommand{\dex}{\,{\rm dex}}

\renewcommand{\Upomega}{{\rm \Omega}}
\renewcommand{\Upsigma}{{\rm \Sigma}}
\renewcommand{\Updelta}{{\rm \Delta}}
\renewcommand{\Uplambda}{{\rm \Lambda}}

\newcommand{\lan}{\langle}
\newcommand{\ran}{\rangle}
\newcommand{\lcdm}{{\rm \Lambda CDM}}
\newcommand*{\meanl}[1]{\overline{#1}}
\newcommand*{\meanb}[1]{\langle{#1}\rangle}
\newcommand*{\E}[1]{\times 10^{#1}}
\newcommand{\un}[1]{_{\rm #1}}

\begin{document}

\title{The weak lensing radial acceleration relation: Constraining modified gravity and cold dark matter theories with KiDS-1000}

\author{
Margot M. Brouwer\inst{1,2}
\and Kyle A. Oman\inst{1,3}
\and Edwin A. Valentijn\inst{1}
\and Maciej Bilicki\inst{4}
\and Catherine Heymans\inst{5,6}
\and Henk Hoekstra\inst{7}
\and Nicola R. Napolitano\inst{8}
\and Nivya Roy\inst{8}
\and Crescenzo Tortora\inst{9}
\and Angus H. Wright\inst{6}
\and Marika Asgari\inst{5}
\and Jan Luca van den Busch\inst{6}
\and Andrej Dvornik\inst{6}
\and Thomas Erben\inst{10}
\and Benjamin Giblin\inst{5}
\and Alister W. Graham\inst{11}
\and Hendrik Hildebrandt\inst{6}
\and Andrew M. Hopkins\inst{12}
\and Arun Kannawadi\inst{13}
\and Konrad Kuijken\inst{7}
\and Jochen Liske\inst{14}
\and HuanYuan Shan\inst{15,16}
\and Tilman Tr\"{o}ster\inst{5}
\and Erik Verlinde\inst{2}
\and Manus Visser\inst{17}.
}
\institute{Kapteyn Astronomical Institute, University of Groningen, PO Box 800, NL-9700 AV Groningen, the Netherlands.
\and Institute for Theoretical Physics, University of Amsterdam, Science Park 904, 1098 XH Amsterdam, The Netherlands.
\and Institute for Computational Cosmology, Department of Physics, Durham University, South Road, Durham DH1 3LE, UK.
\and Center for Theoretical Physics, Polish Academy of Sciences, al. Lotnik\'{o}w 32/46, 02-668 Warsaw, Poland.
\and Institute for Astronomy, University of Edinburgh, Royal Observatory, Blackford Hill, Edinburgh, EH9 3HJ, UK.
\and Ruhr University Bochum, Faculty of Physics and Astronomy, Astronomical Institute (AIRUB), German Centre for Cosmological Lensing, 44780 Bochum, Germany.
\and Leiden Observatory, Leiden University, P.O.Box 9513, 2300RA Leiden, The Netherlands.
\and School of Physics and Astronomy, Sun Yat-sen University, Guangzhou 519082, Zhuhai Campus, P.R. China.
\and INAF -- Osservatorio Astronomico di Capodimonte, Salita Moiariello 16, Napoli 80131, Italy.
\and Argelander-Institut für Astronomie, Auf dem Hügel 71, 53121 Bonn, Germany.
\and Centre for Astrophysics and Supercomputing, Swinburne University of Technology, Hawthorn, VIC 3122, Australia.
\and Australian Astronomical Optics, Macquarie University, 105 Delhi Road, North Ryde, NSW 2113, Australia.
\and Department of Astrophysical Sciences, Princeton University, 4 Ivy Lane, Princeton NJ 08544, USA.
\and Hamburger Sternwarte, University of Hamburg, Gojenbergsweg 112, 21029 Hamburg, Germany.
\and Shanghai Astronomical Observatory (SHAO), Nandan Road 80, Shanghai 200030, China.
\and University of the Chinese Academy of Sciences, Yuquanlu 19A, Beijing 100049, China.
\and Department of Theoretical Physics, University of Geneva, 24 quai Ernest-Ansermet, 1211 Gen\`{e}ve 4, Switzerland.
\\
\\
\email{margot.brouwer@gmail.com}}

\date{Received ...; Accepted ...}

\keywords{Gravitational lensing: weak -- Methods: statistical -- Surveys -- Galaxies: haloes -- Cosmology: dark matter, theory -- Gravitation.}

\titlerunning{The lensing RAR: testing MG and CDM with KiDS-1000}
\authorrunning{M. M. Brouwer et al.}

\abstract{We present measurements of the radial gravitational acceleration around isolated galaxies, comparing the expected gravitational acceleration given the baryonic matter ($g\un{bar}$) with the observed gravitational acceleration ($g\un{obs}$), using weak lensing measurements from the fourth data release of the Kilo-Degree Survey (KiDS-1000). These measurements extend the radial acceleration relation (RAR), traditionally measured using galaxy rotation curves, by 2 decades in $g\un{obs}$ into the low-acceleration regime beyond the outskirts of the observable galaxy. We compare our RAR measurements to the predictions of two modified gravity (MG) theories: modified Newtonian dynamics (MOND) and Verlinde’s emergent gravity (EG). We find that the measured relation between $g\un{obs}$ and $g\un{bar}$ agrees well with the MG predictions. In addition, we find a difference of at least $6\sigma$ between the RARs of early- and late-type galaxies (split by S\'{e}rsic index and $u-r$ colour) with the same stellar mass. Current MG theories involve a gravity modification that is independent of other galaxy properties, which would be unable to explain this behaviour, although the EG theory is still limited to spherically symmetric static mass models. The difference might be explained if only the early-type galaxies have significant ($M\un{gas} \approx M_\star$) circumgalactic gaseous haloes. The observed behaviour is also expected in ${\rm \Lambda}$-cold dark matter ($\lcdm$) models where the galaxy-to-halo mass relation depends on the galaxy formation history. We find that MICE, a $\lcdm$ simulation with hybrid halo occupation distribution modelling and abundance matching, reproduces the observed RAR but significantly differs from BAHAMAS, a hydrodynamical cosmological galaxy formation simulation. Our results are sensitive to the amount of circumgalactic gas; current observational constraints indicate that the resulting corrections are likely moderate. Measurements of the lensing RAR with future cosmological surveys (such as Euclid) will be able to further distinguish between MG and $\lcdm$ models if systematic uncertainties in the baryonic mass distribution around galaxies are reduced.}

\maketitle

\clearpage



\section{Introduction}
\label{sec:Introduction}

It has been known for almost a century that the outer regions of galaxies rotate faster than would be expected from Newtonian dynamics based on their luminous, or `baryonic', mass \cite[]{kapteyn1922,oort1932,oort1940,babcock1939}. This was also demonstrated by \cite{gottesman1966} and \cite{bosma1981} through measurements of hydrogen profiles at radii beyond the optical discs of galaxies, and by \cite{rubin1983} through measurements of galactic rotation curves within the optical discs. The excess gravity implied by these measurements has generally been attributed to an unknown and invisible substance named dark matter (DM), a term coined more than 40 years prior by \cite{zwicky1933} when he discovered the so-called missing mass problem through the dynamics of galaxies in clusters. More recently, new methods such as weak gravitational lensing \cite[]{hoekstra2004,mandelbaum2006,clowe2006,heymans2013,linden2014}, baryon acoustic oscillations \cite[]{eisenstein2005,blake2011}, and the cosmic microwave background \cite[CMB;][]{debernardis2000,spergel2003,planck2014} have contributed unique evidence to the missing mass problem.

Among many others, these observations have contributed to the fact that cold dark matter\footnote{DM particles that moved at non-relativistic speeds at the time of recombination, as favoured by measurements of the CMB \cite[]{planck2014} and the Lyman-$\alpha$ forest \cite[]{viel2013}.} (CDM) has become a key ingredient of the current standard model of cosmology: the $\lcdm$ model. In this paradigm, CDM accounts for a fraction $\Upomega\un{CDM}=0.266$ of the critical density $\rho_\textrm{crit}=3H_0^2/8\pi G$ in the Universe, while baryonic matter only accounts for $\Upomega\un{bar}=0.049$ \cite[]{planck2020}. The cosmological constant $\Uplambda$, which is necessary to explain the accelerated expansion of the Universe \cite[]{riess1998,perlmutter1999} and is a special case of dark energy (DE), accounts for the remaining $\Upomega\un{\Lambda}=0.685$ in our flat space-time \cite[]{debernardis2000}.

Although the $\lcdm$ model successfully describes the observations on a wide range of scales, no conclusive direct evidence for the existence of DM particles has been found so far \cite[despite years of enormous effort; for an overview, see][]{bertone2005,bertone2018}. Combined with other current open questions in physics, such as the elusive unification of general relativity (GR) with quantum mechanics and the mysterious nature of DE, this leaves room for alternative theories of gravity. Two modified gravity (MG) theories that do not require the existence of particle DM are modified Newtonian dynamics \cite[MOND;][]{milgrom1983} and the more recent theory of emergent gravity \cite[EG;][]{verlinde2017}. In these theories all gravity is due to the baryonic matter (or, in the case of EG, the interaction between baryons and the entropy associated with DE). Hence, one of the main properties of these theories is that the mass discrepancy in galaxies correlates strongly with their baryonic mass distribution.

Such a correlation has indeed been observed, such as via the Tully-Fisher relation \cite[]{tully1977} between the luminosity of a spiral galaxy and its asymptotic rotation velocity \cite[]{pierce1988,bernstein1994}. This relation was later generalised as the baryonic Tully-Fisher relation \cite[]{mcgaugh2000,mcgaugh2012} to include non-stellar forms of baryonic matter. Even earlier, astronomers had found a strong correlation between the observed rotation velocity as a function of galaxy radius $v\un{obs}(r)$ and the enclosed luminous mass $M\un{bar}(<r)$ \cite[]{sanders1986,sanders1996,mcgaugh2004,sanders2007,wu2015}. Since $M\un{bar}(<r)$ corresponds to the expected gravitational acceleration $g\un{bar}(r)$ from baryonic matter, and the observed gravitational acceleration can be calculated through $g\un{obs}(r)=v\un{obs}^2(r)/r$, this relation has also been named the radial acceleration relation (RAR)\footnote{Another closely related (though slightly different) relation is the mass-discrepancy acceleration relation, which shows the expected baryonic acceleration against the discrepancy between the baryonic and the observed mass: $M\un{obs}-M\un{bar}$ \cite[see][]{mcgaugh2004}. Although measuring this relation requires the same data, we prefer the RAR because the two observables ($g\un{bar}$ and $g\un{obs}$) are uncorrelated.}.

\citet[][hereafter M16]{mcgaugh2016} in particular measured the RAR with unprecedented accuracy, using the Spitzer Photometry and Accurate Rotation Curves \cite[SPARC;][]{lelli2016b} data of 153 late-type galaxies. Their results again showed a tight correlation between $g\un{obs}$ and $g\un{bar}$, which they could describe using a simple double power law (eq.~4 in M16) that depends only on $g\un{bar}$ and one free parameter: the acceleration scale $g\un{\dagger}$ where Newtonian gravity appears to break down. This rekindled the interest of scientists working on alternative theories of gravity \cite[]{lelli2017a,lelli2017b,burrage2017,li2018,obrien2019}, but also of those seeking an explanation of the RAR within the $\lcdm$ framework, employing correlations between the masses, sizes, and DM content of galaxies \cite[]{dicintio2016,keller2017,desmond2017,ludlow2017, navarro2017,tenneti2018}.

\citet[][hereafter N17]{navarro2017} used a range of simplifying assumptions based on galaxy observations and DM simulations in order to create an analytical galaxy model including the baryonic and halo components. With this model they reconstruct the RAR inside galaxy discs, in particular the value of $a\un{0}$, the acceleration scale where the relation transitions from the baryon-dominated to the DM-dominated regime (which is equivalent to $g\un{\dagger}$), and $a\un{min}$, the minimum acceleration probed by galaxy discs. Based on their results, they claim that the RAR can be explained within the $\lcdm$ framework at the accelerations probed by galaxy rotation curves (within the galaxy disc, i.e. $g\un{obs}>a\un{min}$). However, since their model relies on the fact that luminous kinematic tracers in galaxies only probe a limited radial range, N17 predicted that extending observations to radii beyond the disc (which correspond to lower gravitational accelerations) would lead to systematic deviations from the simple double power law proposed by M16. Although some progress has been made using globular clusters \cite[]{bilek2019a, bilek2019b, muller2020}, using kinematic tracers to measure the RAR beyond the outskirts of visible galaxies remains difficult.

The goal of this work is to extend observations of the RAR to extremely low accelerations that cannot currently be detected through galaxy rotation curves or any other kinematic measurement. To this end, we use gravitational lensing: the perturbation of light inside a gravitational potential as described by relativistic theories such as GR. Both weak and strong gravitational lensing were used by \cite{tian2020} to measure the RAR from observations of 20 galaxy clusters targeted by the CLASH survey. However, due to the high cluster masses, the accelerations probed by these measurements were of the same order as those measurable with galaxy rotation curves. In this work, we use the method of galaxy-galaxy lensing (GGL): the statistical measurement of the coherent image distortion (shear) of a field of background galaxies (sources) by the gravitational potential of a sample of individual foreground galaxies \cite[lenses; for examples, see e.g.][]{brainerd1996,fischer2000,hoekstra2004,mandelbaum2006,uitert2016}. Using GGL we can measure the average (apparent) density distribution of isolated galaxies up to a radius of $3\,{\rm Mpc}$, roughly $100$ times larger than the radius of the luminous disc ($\sim 30\,{\rm kpc}$). At our stellar mass scale of interest -- $\log(M_\star/\hmsun) \approx 10.5$ -- this radius corresponds to $g\un{bar}\approx10^{-15} \mpss$, which is three orders of magnitude lower than the baryonic accelerations of the M16 rotation curves\footnote{We note that this value of $g\un{bar}$ only takes into account the stellar and cold gas mass of the galaxy. In Section~\ref{sec:Missing_Baryons} we show that the contributions of additional hot gas, dust and `missing baryons' could increase this value to $g\un{bar}\approx10^{-14}\mpss$, which is still two orders of magnitude lower than the accelerations measurable with galaxy rotation curves.}.

Our main goal is to use the lensing RAR of isolated galaxies at lower accelerations (beyond the observable galaxy disc) to distinguish which of the aforementioned MG and $\lcdm$ models best describe this result. To achieve this, we first measure the total and baryonic density profiles of our galaxies through their GGL profiles and luminosities. These measurements will be performed using $1006 \deg^2$ of weak lensing data from the Kilo-Degree Survey \cite[KiDS-1000;][]{dejong2013,kuijken2019}, and nine-band photometric data from KiDS and the VISTA Kilo-Degree Infrared Galaxy Survey \cite[VIKING]{edge2013}. We then translate these measurements into the observed and baryonic radial accelerations, $g\un{obs}$ and $g\un{bar}$. Finally, we compare the resulting RAR to predictions from different MG theories (MOND and EG) and $\lcdm$. To test the MG theories, we need to make the assumption that the deflection of light by gravitational potentials (as described in GR) holds in these modified theories, which we motivate in the relevant sections. This work can be seen as an extension of \cite{brouwer2017}, where we tested the predictions of EG using KiDS GGL on foreground galaxies from $180 \deg^2$ of the Galaxy and Mass Assembly (GAMA) survey. Instead of GAMA, we now use a selection of $\sim1$ million foreground galaxies from KiDS-1000 to achieve a fivefold increase in survey area.

The $\lcdm$ predictions will not only be provided by the N17 analytical model, but also by mock galaxy catalogues based on two different DM simulations. One is the Marenostrum Institut de Ci{\`e}ncies de l'Espai (MICE) Galaxy and Halo Light-cone catalogue \cite[]{carretero2015,hoffmann2015}, which is based on the MICE Grand Challenge lightcone simulation \cite[]{fosalba2015b,fosalba2015a,crocce2015}. The other mock galaxy catalogue is based on a suite of large-volume cosmological hydrodynamical simulations, called the BAryons and HAloes of MAssive Systems (BAHAMAS) project \cite[]{mccarthy2017}.

Having $\sim1$ million foreground galaxies at our disposal allows us to select specific galaxy samples, designed to optimally test the predictions from the aforementioned MG and $\lcdm$ models. Particularly, we note that the analytical models (MOND, EG and N17) mostly focus on the description of individual, isolated galaxies. In order to test them, we select a sample of galaxies whose GGL profiles are minimally affected by neighbouring galaxies (e.g. satellites) within the radius of our measurement. In contrast, the predictions from simulations can be tested with both isolated and non-isolated galaxy samples.

In addition, our sample of $\sim350\,000$ isolated lens galaxies allows us to analyse the RAR as a function of colour, S\'ersic index and stellar mass. Because MG and $\lcdm$ give different predictions regarding the dependence of the RAR on these observables, this allows us to better distinguish between the different models. Specifically: according to the MOND and EG theories the relation between $g\un{bar}$ and $g\un{obs}$ should remain fixed in the regime beyond the baryon-dominated galaxy disc, and hence be independent of galaxy observables. Within the $\lcdm$ paradigm, the relation between $g\un{bar}$ and $g\un{obs}$ is related to the stellar-to-halo-mass relation (SHMR) that is not necessarily constant as a function of galaxy stellar mass or other observables.

Our paper is structured as follows: In Section~\ref{sec:Theory} we describe the methodology behind the GGL measurements and their conversion into the RAR, in addition to the theoretical predictions to which we compare our observations: MOND, EG and the N17 analytical DM model. In Section~\ref{sec:Data} we introduce the KiDS-1000 and GAMA galaxy surveys used to perform both the GGL and stellar mass measurements. Section~\ref{sec:Simulations} describes the MICE and BAHAMAS simulations and mock galaxy catalogues to which we compare our results. In Section~\ref{sec:Results} we present our lensing RAR measurements and compare them to the different models, first using all isolated galaxies and then separating the galaxies by different observables. Section~\ref{sec:Discussion_Conclusion} contains the discussion and conclusion. In Appendix~\ref{app:Isolation} we validate our isolated galaxy selection, and Appendix~\ref{app:PPL_method} contains a description of the piecewise-power-law method of translating the lensing measurement into $g\un{obs}$. Finally, Appendix~\ref{app:Results_N17} shows the comparison of the N17 analytical DM model with our lensing RAR.

Throughout this work we adopt the WMAP 9-year \cite[]{hinshaw2013} cosmological parameters: $\Upomega\un{m}=0.2793$, $\Upomega\un{b}=0.0463$, $\Upomega\un{\Lambda}=0.7207$, $\sigma_8=0.821$ and $H_0 = 70 \, {\rm km \, s^{-1} Mpc^{-1}}$, which were used as the basis of the BAHAMAS simulation. When analysing the MICE simulations we use the cosmological parameters used in creating MICE, which are: $\Upomega\un{m}=0.25$, $\sigma_8=0.8$, $\Upomega\un{\Lambda}=0.75$, and $H_0 = 70 \, {\rm km \, s^{-1} Mpc^{-1}}$. Throughout the paper we use the reduced Hubble constant $h_{70} = \ H_0 / (70 \, {\rm km \, s^{-1} Mpc^{-1}})$. Due to the relatively low redshift of our lens galaxies ($z\sim0.2$) the effect of differences in the cosmological parameters on our results is small.

\section{Theory}
\label{sec:Theory}

\subsection{Mass measurements with weak gravitational lensing}
\label{sec:Lensing}

To estimate the gravitational acceleration around galaxies we used GGL: the measurement of the coherent image distortion of a field of background galaxies (sources) by the gravitational potential of a sample of foreground galaxies (lenses). Because the individual image distortions are very small (only $\sim1\%$ compared to the galaxy's unknown original shape), this method can only be performed statistically for a large sample of sources. We averaged their projected ellipticity component tangential to the direction of the lens galaxy, $\epsilon\un{t}$, which is the sum of the intrinsic tangential ellipticity component $\epsilon\un{t}^{\rm int}$ and the tangential shear $\gamma\un{t}$ caused by weak lensing. Assuming no preferential alignment in the intrinsic galaxy shapes ($\lan\epsilon\un{t}^{\rm int}\ran=0$), the average $\lan\epsilon\un{t}\ran$ is an estimator for $\gamma\un{t}$. By measuring this averaged quantity in circular annuli around the lens centre, we obtained the tangential shear profile $\gamma\un{t}(R)$ as a function of projected radius $R$. Because our final goal is to compute the observed gravitational acceleration $g\un{obs}$ as a function of that expected from baryonic matter $g\un{bar}$, we chose our $R$-bins such that they corresponded to $15$ logarithmic bins between $1\E{-15} < g\un{bar} < 5\E{-12} \mpss$. For each individual lens the calculation of these $g\un{bar}$-bins was based on the baryonic mass of the galaxy $M\un{gal}$ (see Section~\ref{sec:KiDS-bright}). In real space this binning approximately corresponds to the distance range used in \cite{brouwer2017}: $0.03 < R < 3 \hMpc$.

The lensing shear profile can be related to the physical excess surface density (ESD, denoted $\Updelta\Upsigma$) profile through the critical surface density $\Upsigma\un{crit}$:
\begin{equation}
\Updelta \Upsigma (R) = \Upsigma\un{crit} \gamma\un{t}(R) = \meanb{\Upsigma}(<R) - \Upsigma(R) \, ,
\label{eq:deltasigma}
\end{equation}
which is the surface density $\Upsigma(R)$ at projected radius $R$, subtracted from the average surface density $\meanb{\Upsigma}(<R)$ within $R$. See Section~\ref{sec:KiDS} for more information on how this is computed.

The error values on the ESD profile were estimated by the square-root of the diagonal of the analytical covariance matrix, which is described in section~3.4 of \cite{viola2015}. The full covariance matrix was calculated based on the contribution of each individual source to the ESD profile, and incorporates the correlation between sources that contribute to the ESD in multiple bins, both in projected distance $R$ and in galaxy observable.

\subsection{The radial acceleration relation (RAR)}
\label{sec:Conversion}

After measuring the lensing profile around a galaxy sample, the next step is to convert it into the corresponding RAR. We started from the ESD as a function of projected radius $\Updelta\Upsigma(R)$ and the measured stellar masses of the lens galaxies $M_\star$, aiming to arrive at their observed radial acceleration $g\un{obs}$ as a function of their expected baryonic radial acceleration $g\un{bar}$. The latter can be calculated using Newton's law of universal gravitation:
\begin{equation}\label{eq:grav}
g(r) = \frac{G \, M(<r)}{r^2} \, ,
\end{equation}
which defines the radial acceleration $g$ in terms of the gravitational constant $G$ and the enclosed mass $M(<r)$ within spherical radius $r$. Assuming spherical symmetry here is reasonable, given that for lensing measurements thousands of galaxies are stacked under many different angles to create one average halo profile. 

The calculation of $g\un{bar}$ requires the enclosed baryonic mass $M\un{bar}(<r)$ of all galaxies. We discuss our construction of $M\un{bar}(<r)$ in Section~\ref{sec:KiDS-bright}. The calculation of $g\un{obs}$ requires the enclosed observed mass $M\un{obs}(<r)$ of the galaxy sample, which we obtained through the conversion of our observed ESD profile \mbox{$\Updelta\Upsigma(R)$}.

When calculating $g\un{obs}$ we started from our ESD profile measurement, which consists of the value $\Updelta\Upsigma(R)$ measured in a set of radial bins. At our measurement radii ($R>30\hkpc$) the ESD is dominated by the excess gravity, which means the contribution from baryonic matter can be neglected. We adopted the simple assumption that our observed density profile $\rho\un{obs}(r)$ is roughly described by a Singular Isothermal Sphere (SIS) model:
\begin{equation}\label{eq:rho_SIS}
\rho\un{SIS}(r) = \frac{\sigma^2}{2 G \pi r^2} \, . 
\end{equation}
The SIS is generally considered to be the simplest parametrisation of the spatial distribution of matter in an astronomical system (such as galaxies, clusters, etc.). If interpreted in a $\lcdm$ context, the SIS implies the assumption that the DM particles have a Gaussian velocity distribution analogous to an ideal gas that is confined by their combined spherically symmetric gravitational potential, where $\sigma$ is the total velocity dispersion of the particles. In a MG context, however, the SIS profile can be considered to represent a simple $r^{-2}$ density profile as predicted by MOND and EG in the low-acceleration regime outside a baryonic mass distribution, with $\sigma$ as a normalisation constant. The ESD derived from the SIS profile is:
\begin{equation}\label{eq:ESD_SIS}
\Updelta\Upsigma\un{SIS}(R) = \frac{\sigma^2}{2 G R} \, .
\end{equation}
From \cite{brouwer2017} we know that, despite its simple form, it provides a good approximation of the GGL measurements around isolated galaxies. The SIS profile is therefore well-suited to analytically model the total enclosed mass distribution of our lenses, which can then be derived as follows:
\begin{equation}\label{eq:mass_SIS}
M\un{SIS}(<r) = 4 \pi \int_{0}^{r}\rho\un{SIS}(r') r'^2 {\rm d} r' = \frac{2\sigma^2 r}{G} \, .
\end{equation}

Now, for each individual observed ESD value $\Updelta\Upsigma_{{\rm obs},m}$ at certain projected radius $R_{m}$, we assumed that the density distribution within $R_{m}$ is described by an SIS profile with $\sigma$ normalised such that $\Updelta\Upsigma\un{SIS}(R_{m}) = \Updelta\Upsigma_{{\rm obs},m}$. Under this approximation, we combined equations \ref{eq:ESD_SIS} and \ref{eq:mass_SIS} to give a relation between the lensing measurement $\Delta\Sigma$ and the deprojected, spherically enclosed mass $M_{\rm obs}$:
\begin{equation}\label{eq:Mobs}
M\un{obs}(<r) = 4 \Updelta\Upsigma\un{obs}(r) \, r^2 \, .
\end{equation}
Through Eq.~\ref{eq:grav}, this results in a very simple expression for the observed gravitational acceleration:
\begin{equation}
\label{eq:gobs_from_ESD}
g\un{obs}(r) = \frac{G \, [4 \Updelta\Upsigma\un{obs}(r) \, r^2]}{r^2} = 4 G \Updelta\Upsigma\un{obs}(r) \, .
\end{equation}
Throughout this work, we have used the SIS approximation to convert the ESD into $g\un{obs}$. In Section~\ref{sec:Conversion_test} we validate this approach by comparing it to a more elaborate method and testing both on the BAHAMAS simulation.

\subsection{The RAR with modified Newtonian dynamics}
\label{sec:MOND}

With his theory, MOND, \cite{milgrom1983} postulated that the missing mass problem in galaxies is not caused by an undiscovered fundamental particle, but that instead our current gravitational theory should be revised. Since MOND is a non-relativistic theory, performing GGL measurements to test it requires the assumption that light is curved by a MONDian gravitational potential in the same way as in GR. This assumption is justified since \citet[][while testing the MOND paradigm using GGL data from the Canada-France-Hawaii Telescope Lensing survey]{milgrom2013}, states that non-relativistic MOND is a limit of relativistic versions that predict that gravitational potentials determine lensing in the same way as Newtonian potentials in GR. For this reason GGL surveys can be used as valuable tools to test MOND and similar MG theories, as was done for instance by \cite{tian2009} using Sloan Digital Sky Survey (SDSS) and Red-sequence Cluster Survey data.

MOND's basic premise is that one can adjust Newton's second law of motion ($F=ma$) by inserting a general function $\mu(a/a_0)$, which only comes into play when the acceleration $a$ of a test mass $m$ is much smaller than a critical acceleration scale $a_0$. This function predicts the observed flat rotation curves in the outskirts of galaxies, while still reproducing the Newtonian behaviour of the inner disc. In short, the force $F$ becomes:
\begin{align}\label{eq:mond_f}
& F(a) = m \, \mu \left( \frac{a}{a_0} \right) \, a \, ,
& \mu(x \gg 1) \approx 1 \, , \, \mu(x \ll 1) \approx x \, .
\end{align}
This implies that $a\gg a_0$ represents the Newtonian regime where $F\un{N} = m \, a\un{N}$ as expected, while $a \ll a_0$ represents the `deep-MOND' regime where $F\un{MOND} = m \, a\un{MOND}^2/a_0$. In a circular orbit, this is reflected in the deep-MOND gravitational acceleration $g\un{MOND} \equiv a\un{MOND}$ as follows:
\begin{align}\label{eq:mond_g}
& F\un{MOND} = m \frac{a\un{MOND}^2}{a_0} = \frac{G \, M m}{r^2}
& \rightarrow \,  g\un{MOND} = \sqrt{a_0 \frac{G M}{r^2}} \, .
\end{align}
This can be written in terms of the expected baryonic acceleration $g\un{bar}=GM/r^2$ as follows:
\begin{equation}
g\un{MOND}(g\un{bar}) = \sqrt{a_0 \, g\un{bar}} \, .
\end{equation}

This demonstrates that MOND predicts a very simple relation for the RAR: $g\un{obs}=g\un{bar}$ in the Newtonian regime ($g\un{obs} \gg a_0$) and Eq. \ref{eq:mond_g} in the deep-MOND regime ($g\un{obs} \ll a_0$). However, since $\mu(a/a_0)$, also known as the interpolating function, is not specified by \cite{milgrom1983}, there is no specific constraint on the behaviour of this relation in between the two regimes. In the work of \cite{milgrom2008}, several families of interpolation functions are discussed. Selecting the third family (given by their eq.~13) with constant parameter $\alpha=1/2$, provides the function that M16 later used to fit to their measurement of the RAR using rotation curves of $153$ galaxies. This relation can be written as:
\begin{equation}\label{eq:rar_mcgaugh}
g\un{obs}(g\un{bar}) = \frac{g\un{bar}}{1 - e^{-\sqrt{g\un{bar}/a_0}}} \, ,
\end{equation}
where $a_0 \equiv g\un{\dagger}$ corresponds to the fitting parameter constrained by M16 to be $g\un{\dagger}=1.20\pm0.26\E{-10} \mpss$. Since Eq. \ref{eq:rar_mcgaugh} (equal to eq. 4 in M16) is also considered a viable version of the MOND interpolation function by \cite{milgrom2008}, we will consider it the baseline prediction of MOND in this work. As the baseline value of $a_0$, we will likewise use the value of $g\un{\dagger}$ measured by M16 since it exactly corresponds to the value of $a_0=1.2\E{-10} \mpss$ considered canonical in MOND since its first measurement by \cite{begeman1991}, using the rotation curves of 10 galaxies.

One of the main characteristics of the MOND paradigm, is that it gives a direct and fixed prediction for the total acceleration based only on the system's baryonic mass, given by Eq. \ref{eq:rar_mcgaugh}. The main exception to this rule is the possible influence by neighbouring mass distributions through the external field effect (EFE), predicted by \cite{milgrom1983} and studied analytically, observationally and in simulations by \cite{banik2015,banik2020,chae2020}. Since we explicitly selected isolated galaxies in this work (see Appendix~\ref{app:Isolation}), this effect is minimised as much as possible. However, since total isolation cannot be guaranteed, a small EFE might remain. In order to describe this effect, we used eq. 6 from \cite{chae2020}:
\begin{equation}
g\un{MOND}(g\un{bar}) = \nu\un{e}(z) \, g\un{bar} \, ,
\end{equation}
with:
\begin{equation} \label{eq:mond_efe}
\nu\un{e}(z) = \frac{1}{2} - \frac{A\un{e}}{z} + \sqrt{\left( \frac{1}{2} - \frac{A\un{e}}{z} \right)^2 + \frac{B\un{e}}{z}} \, .
\end{equation}
Here $z \equiv g\un{bar}/g_\dagger$, $A\un{e} \equiv e(1+e/2)/(1+e)$, and $B\un{e} \equiv (1+e)$. The strength of the EFE is parametrised through: $e = g\un{ext}/g_\dagger$, determined by the external gravitational acceleration $g\un{ext}$. Although the interpolation functions differ, the result of Eq. \ref{eq:mond_efe} corresponds almost exactly to the M16 fitting function given in Eq. \ref{eq:rar_mcgaugh} in the limit $e=0$ (no EFE). Positive values of $e$ result in reduced values of the predicted $g\un{obs}$ at very low accelerations (see Fig.~\ref{fig:RAR_kids_gama_verlinde} in Section~\ref{sec:Results-MG_theories}, and fig. 1 of \citealp{chae2020}). It should be noted that this fitting function represents an idealised model and could be subject to deviations in real, complex, 3D galaxies.

\subsection{The RAR with emergent gravity}
\label{sec:EG}

The work of \citet[][V17 hereafter]{verlinde2017}, which is embedded in the framework of string theory and holography, shares the view that the missing mass problem is to be solved through a revision of our current gravitational theory. Building on the ideas from \cite{jacobson1995,jacobson2016,padmanabhan2010,verlinde2011,faulkner2015}, V17 abandons the notion of gravity as a fundamental force. Instead, it emerges from an underlying microscopic description of space-time, in which the notion of gravity has no \emph{a priori} meaning.

V17 shows that constructing an EG theory in a universe with a negative cosmological constant (`anti-de Sitter') allows for the re-derivation of Einstein's laws of GR. A distinguishing feature of V17 is that it attempts to describe a universe with a positive cosmological constant (`de Sitter'), that is, one that is filled with a DE component. This results in a new volume law for gravitational entropy caused by DE, in addition to the area law normally used to retrieve Einsteinian gravity. According to V17, energy that is concentrated in the form of a baryonic mass distribution causes an elastic response in the entropy of the surrounding DE. This results in an additional gravitational component at scales set by the Hubble acceleration scale $a\un{0} = c H_0/6$. Here $c$ is the speed of light, and $H_0$ is the current Hubble constant that measures the Universe's expansion velocity.

Because this extra gravitational component aims to explain the effects usually attributed to DM, it is conveniently expressed as an apparent dark matter (ADM) distribution:
	\begin{equation}
	M\un{ADM}^2 (r) = \frac{cH_0 r^2}{6G} \frac{{\rm d}\left[ M\un{bar}(r) r \right]}{{\rm d}r} \, .
	\label{eq:eg_mdm}
	\end{equation}
Thus the ADM distribution is completely defined by the baryonic mass distribution $M\un{bar}(r)$ as a function of the spherical radius $r$, and a set of known physical constants.

Since we measured the ESD profiles of galaxies at projected radial distances $R>30 \hkpc$, we can follow \cite{brouwer2017} in assuming that their baryonic component is equal to the stars+cold gas mass enclosed within the minimal measurement radius (for further justification of this assumption, see Section~\ref{sec:Missing_Baryons}). This is equivalent to describing the galaxy as a point mass $M\un{bar}$, which allows us to simplify Eq. \ref{eq:eg_mdm} to:
\begin{equation}
M\un{ADM}(r)=\sqrt{\frac{cH_{0} \, M\un{bar}}{6 \, G}} \, r \, .
\end{equation}
Now the total enclosed mass $M\un{EG}(r) = M\un{bar} + M\un{ADM}(r)$ can be used to calculate the gravitational acceleration $g\un{EG}(r)$ predicted by EG, as follows:
\begin{equation}
g\un{EG}(r) = \frac{G M\un{EG}(r)}{r^2} = \frac{G M\un{bar}}{r^2} + \sqrt{\frac{cH_{0}}{6}} \, \frac{\sqrt{G M\un{bar}}}{r} \, .
\end{equation}
In terms of the expected baryonic acceleration $g\un{bar}(r) = G M\un{bar}/r^2$, this simplifies even further to:
\begin{equation}
\label{eq:rar_verlinde}
g\un{EG}(g\un{bar}) = g\un{bar} + \sqrt{\frac{cH_{0}}{6}} \, \sqrt{g\un{bar}} \, .
\end{equation}

We emphasise that Eq.~\ref{eq:eg_mdm} is only a macroscopic approximation of the underlying microscopic phenomena described in V17, and is thus only valid for static, spherically symmetric and isolated baryonic mass distributions. For this reason, we selected only the most isolated galaxies from our sample (see Appendix~\ref{app:Isolation}), such that our GGL measurements are not unduly influenced by neighbouring galaxies. Furthermore, the current EG theory is only valid in the acceleration range $g\un{bar}<a_0$, often called the deep-MOND regime. Therefore, the prediction of Eq. \ref{eq:rar_verlinde} should be taken with a grain of salt for accelerations $g\un{bar}>1.2\E{-10} \mpss$. This will not affect our analysis since weak lensing takes place in the weak gravity regime. In addition, cosmological evolution of the $H_0$ parameter is not yet implemented in the theory, restricting its validity to galaxies with relatively low redshifts. However, we calculated that at our mean lens redshift, $\meanb{z}\sim0.2$, using an evolving $H(z)$ would result in only a $\sim5\%$ difference in our ESD measurements, based on the background cosmology used in this work.

In order to test EG using the standard GGL methodology, we needed to assume that the deflection of photons by a gravitational potential in this alternative theory corresponds to that in GR. This assumption is justified because, in EG's original (anti-de Sitter) form, Einstein's laws emerge from its underlying description of space-time. The additional gravitational force described by ADM does not affect this underlying theory, which is an effective description of GR. Therefore, we assumed that the gravitational potential of an ADM distribution produces the same lensing shear as an equivalent distribution of actual matter.

\subsection{The RAR in $\Lambda$CDM}
\label{sec:N17}

To help guide an intuitive interpretation of the lensing RAR within the framework of the $\lcdm$ theory, we made use of the simple model of N17, which combines a basic model of galactic structure and scaling relations to predict the RAR. We refer to N17 for a full description, but give a summary here. A galaxy of a given stellar (or baryonic -- there is no distinction in this model) mass occupies a DM halo of a mass fixed by the abundance matching relation of \citet{behroozi2013}. The dark halo concentration is fixed to the cosmological mean for haloes of that mass \citep{ludlow2014}. The baryonic disc follows an exponential surface density profile with a half-mass size fixed to $0.2\times$ the scale radius of the dark halo. This model is sufficient to specify the cumulative mass profile of both the baryonic and dark components of the model galaxy; calculating $g\un{obs}$ and $g\un{bar}$ is then straightforward. However, since the N17 model is merely a simple analytical description, our main $\lcdm$ test utilised more elaborate numerical simulations (see Section~\ref{sec:Simulations}).

\section{Data}
\label{sec:Data}

\subsection{The Kilo-Degree Survey (KiDS)}
\label{sec:KiDS}

We measured the gravitational potential around a sample of foreground galaxies (lenses), by measuring the image distortion (shear) of a field of background galaxies (sources). These sources were observed using OmegaCAM \cite[]{kuijken2011}: a 268-million pixel CCD mosaic camera mounted on the Very Large Telescope (VLT) Survey Telescope \cite[]{capaccioli2011}. Over the past ten years these instruments have performed KiDS, a photometric survey in the $ugri$ bands, which was especially designed to perform weak lensing measurements \cite[]{dejong2013}.

GGL studies with KiDS have hitherto been performed in combination with the spectroscopic GAMA survey (see Section~\ref{sec:GAMA}), with the KiDS survey covering $180 \deg^2$ of the GAMA area. Although the final KiDS survey will span $1350 \deg^2$ on the sky, the current state-of-the-art is the $4^{\rm th}$ Data Release \cite[KiDS-1000;][]{kuijken2019} containing observations from $1006$ square-degree survey tiles. We therefore used a photometrically selected `KiDS-bright' sample of lens galaxies from the full KiDS-1000 release, as described in Section~\ref{sec:KiDS-bright}. The measurement and calibration of the source shapes and photometric redshifts are described in \cite{kuijken2019,giblin2020} and \cite{hildebrandt2020}.

The measurements of the galaxy shapes are based on the $r$-band data since this filter was used during the darkest time (moon distance $> 90 \deg$) and with the best atmospheric seeing conditions ($<0.8 \as$). The $r$-band observations were co-added using the {\scshape Theli} pipeline \cite[]{erben2013}. From these images the galaxy positions were detected through the \textsc{SExtractor} algorithm \cite[]{bertin1996}. After detection, the shapes of the galaxies were measured using the \emph{lens}fit pipeline \cite[]{miller2007,miller2013}, which includes a self-calibration algorithm based on \cite{fenechconti2017} that was validated in \cite{kannawadi2019}. Each shape is accompanied by a \emph{lens}fit weight $w\un{s}$, which was used as an estimate of the precision of the ellipticity measurement.

For the purpose of creating the photometric redshift and stellar mass estimates, 9 bands were observed in total. The $ugri$ bands were observed by KiDS, while the VIKING survey \cite[]{edge2013} performed on the VISTA telescope adds the $ZYJHK\un{s}$ bands. All KiDS bands were reduced and co-added using the Astro-WISE pipeline \cite[AW;][]{mcfarland2013}. The galaxy colours, which form the basis of the photometric redshift measurements, were measured from these images using the Gaussian Aperture and PSF pipeline \cite[\textsc{GAaP};][]{kuijken2008,kuijken2015}.

The addition of the lower frequency VISTA data allowed us to extend the redshift estimates out to $0.1<z\un{B}<1.2$, where $z\un{B}$ is the best-fit photometric redshift of the sources \cite[]{benitez2000,hildebrandt2012}. However, when performing our lensing measurements (see Section~\ref{sec:Lensing}) we used the total redshift probability distribution function $n(z\un{s})$ of the full source population. This $n(z\un{s})$ was calculated using a direct calibration method (see \citealp{hildebrandt2017} for details), and circumvents the inherent bias related to photometric redshift estimates of individual sources.

We note that this is a different redshift calibration method than that used by the KiDS-1000 cosmology analyses \citep{asgari2020,heymans2020,troster2020}, who used a self-organising map to remove (primarily high-redshift) sources whose redshifts could not be accurately calibrated due to incompleteness in the spectroscopic sample \citep{wright2020,hildebrandt2020}. Following Robertson et al. (in prep.) we prioritised precision by analysing the full KiDS-1000 source sample (calibrated using the direct calibration method) since percent-level biases in the mean source redshifts do not significantly impact our analysis.

For the lens redshifts $z\un{l}$, we used the \textsc{ANNz2} (Artificial Neural Network) machine-learning redshifts of the KiDS foreground galaxy sample (KiDS-bright; see Section~\ref{sec:KiDS-bright}). We implemented the contribution of $z\un{l}$ by integrating over the individual redshift probability distributions $p(z\un{l})$ of each lens. This $p(z\un{l})$ is defined by a normal distribution centred at the lens' $z\un{ANN}$ redshift, with a standard deviation: $\sigma\un{z}/(1+z) = 0.02$ (which is equal to the standard deviation of the KiDS-bright redshifts compared to their matched spectroscopic GAMA redshifts). For the source redshifts $z\un{s}$ we followed the method used in \cite{dvornik2018}, integrating over the part of the redshift probability distribution $n(z\un{s})$ where $z\un{s}>z\un{l}$. In addition, sources only contribute their shear to the lensing signal when $z\un{B}+\Updelta z > z\un{l}$ -- when the sum of their best-fit photometric redshift $z\un{B}$ and the redshift buffer $\Updelta z=0.2$ is greater than the lens redshift. Hence, when performing the lensing measurement in Section~\ref{sec:Lensing} the critical surface density\footnote{As derived in Appendix~C of \cite{dvornik2018}, there are two possible definitions of $\Upsigma\un{crit}$: proper and comoving. In this work we used the proper $\Upsigma\un{crit}$, and we compute $\Updelta\Upsigma(R)$ as a function of proper transverse separation $R$. This choice is reasonable because, within a $3 \hMpc$ range, the measured ESD profiles are expected to be approximately stationary in proper coordinates.} (the conversion factor between $\gamma\un{t}$ and $\Updelta\Upsigma$, whose inverse is also called the lensing efficiency) was calculated as follows:
\begin{equation}
\Upsigma\un{crit}^{-1} = \frac{4\pi G}{c^2} \int_{0}^{\infty} D(z\un{l}) \left( \int_{z\un{l}}^{\infty}  \frac{D(z\un{l}, z\un{s})}{D(z\un{s})} n(z\un{s}) \, {\rm d} z\un{s} \right) p(z\un{l}) \, {\rm d} z\un{l} \, .
\label{eq:sigmacrit}
\end{equation}
Here $D(z\un{l})$ and $D(z\un{s})$ are the angular diameter distances to the lens and the source respectively, and $D(z\un{l}, z\un{s})$ the distance between them. The constant multiplication factor is defined by Newton's gravitational constant $G$ and the speed of light $c$.

The ESD profile was averaged (or `stacked') for large samples of lenses to increase the signal-to-noise ($S/N$) ratio of the lensing signal. We defined a lensing weight $W\un{ls}$ that depends on both the \emph{lens}fit weight $w\un{s}$ and the lensing efficiency $\Upsigma\un{crit}^{-1}$:
\begin{equation}
W_{ls} = w_s \left( \Upsigma_{{\rm crit},ls}^{-1} \right)^2 \, ,
\label{eq:weights}
\end{equation}
and used it to optimally sum the measurements from all lens-source pairs into the average ESD:
\begin{equation}
\Updelta\Upsigma = \frac{1}{1+\mu} \frac{\sum_{ls} W_{ls} \, \epsilon_{{\rm t},ls} \, \Upsigma_{{\rm crit},ls} }{ \sum_{ls}{W_{ls}} }  \, .
\label{eq:ESDmeasured}
\end{equation}
Here the factor (1+$\mu$) calibrates the shear estimates \cite[]{fenechconti2017,kannawadi2019}. Extending the method of \cite{dvornik2017} to the higher KiDS-1000 redshifts, $\mu$ denotes the mean multiplicative calibration correction calculated in $11$ linear redshift bins between \mbox{$0.1<z\un{B}<1.2$} from the individual source calibration values $m$:
\begin{equation}
\mu=\frac{\sum_{s} w_{s} m_{s}}{\sum_{s} w_{s}} \, ,
\label{eq:biascorr}
\end{equation}
The value of this correction is $\mu\approx0.014$, independent of the projected distance from the lens.

We also corrected our lensing signal for sample variance on large scales by subtracting the ESD profile measured around $\sim5$ million uniform random coordinates, $50$ times the size of our total KiDS-bright sample. These random coordinates mimic the exact footprint of KiDS, excluding the areas masked by the `nine-band no AW-$r$-band' mask that we applied to the KiDS-bright lenses (see Section~\ref{sec:KiDS-bright}). In order to create random redshift values that mimic the true distribution, we created a histogram of the KiDS-bright redshifts divided into $80$ linear bins between $0.1<z\un{ANN}<0.5$. In each bin, we created random redshift values equal to the number of real lenses in that bin. Because of the large contiguous area of KiDS-1000, we found that the random ESD profile is very small at all projected radii $R$, with a mean absolute value of only $1.85\pm0.75\%$ of the lensing signal of the full sample of isolated KiDS-bright galaxies.

\subsection{The Galaxy and Mass Assembly (GAMA) survey}
\label{sec:GAMA}

Although the most contraining RAR measurements below were performed using exclusively KiDS-1000 data, the smaller set of foreground galaxies observed by the spectroscopic GAMA survey \cite[]{driver2011} functions both as a model and validation sample for the KiDS foreground galaxies. The survey was performed by the Anglo-Australian Telescope with the AAOmega spectrograph, and targeted more than $238\,000$ galaxies selected from the Sloan Digital Sky Survey \cite[SDSS;]{abazajian2009}. For this study we used GAMA~II observations \cite[]{liske2015} from three equatorial regions (G09, G12, and G15) containing more than $180\,000$ galaxies. These regions span a total area of $\sim180 \deg^2$ on the sky, completely overlapping with KiDS.

GAMA has a redshift range of $0<z<0.5$, with a mean redshift of $\meanb{z}=0.22$. The survey has a redshift completeness of $98.5\%$ down to Petrosian $r$-band magnitude $m_{r, {\rm Petro}} = 19.8 \magn$. We limited our GAMA foreground sample to galaxies with the recommended redshift quality: $n\un{Q}\geq3$. Despite being a smaller survey, GAMA's accurate spectroscopic redshifts were highly advantageous when measuring the lensing profiles of galaxies (see Section~\ref{sec:Lensing}). The GAMA redshifts were used to train the photometric machine-learning (ML) redshifts of our larger sample of KiDS foreground galaxies (see Section~\ref{sec:KiDS-bright}). Also, in combination with its high redshift completeness, GAMA allows for a more accurate selection of isolated galaxies. We therefore checked that the results from the KiDS-only measurements are consistent with those from KiDS-GAMA.

To measure the RAR with KiDS-GAMA, we need individual stellar masses $M_\star$ for each GAMA galaxy. We used the \cite{taylor2011} stellar masses, which are calculated from $ugrizZY$ spectral energy distributions\footnote{The spectral energy distributions were constrained to the rest frame wavelength range $3\,000-11\,000$ \AA.} measured by SDSS and VIKING by fitting them with \cite{bruzual2003} Stellar Population Synthesis (SPS) models, using the Initial Mass Function (IMF) of \cite{chabrier2003}. Following the procedure described by \cite{taylor2011}, we accounted for flux falling outside the automatically selected aperture using the `flux-scale' correction.

\subsection{Selecting isolated lens galaxies with accurate redshifts and stellar masses}
\label{sec:KiDS-bright}

Because of its accurate spectroscopic redshifts, the GAMA lenses would be an ideal sample for the selection of isolated galaxies and the measurement of accurate stellar masses \cite[as was done in][]{brouwer2017}. However, since the current KiDS survey area is $>5$ times larger than that of GAMA, we selected a KiDS-bright sample of foreground galaxies from KiDS-1000 that resembles the GAMA survey. We then used the GAMA redshifts as a training sample to compute neural-net redshifts for the KiDS-bright lenses \cite[see e.g.][]{bilicki2018}, from which accurate stellar masses could subsequently be derived. The details of the specific sample used in this work are provided in \cite{bilicki2021}. Here we give an overview relevant for this paper.

To mimic the magnitude limit of GAMA ($m_{r, {\rm Petro}}<19.8 \magn$), we applied a similar cut to the (much deeper) KiDS survey. Because the KiDS catalogue does not contain Petrosian magnitudes we used the Kron-like elliptical aperture $r$-band magnitudes from \textsc{SExtractor}, calibrated for $r$-band extinction and zero-point offset\footnote{${\rm MAG\_AUTO\_CALIB = MAG\_AUTO \, + \, DMAG \, - \,}$ ${\rm EXTINCTION\_R}$}, which have a very similar magnitude distribution. Through matching the KiDS and GAMA galaxies and seeking the best trade-off between completeness and purity, we decided to limit our KiDS-bright sample to $m\un{r,auto}<20.0$. In addition we removed KiDS galaxies with a photometric redshift $z>0.5$, where GAMA becomes very incomplete.

To remove stars from our galaxy sample, we applied a cut based on galaxy morphology, nine-band photometry and the \textsc{SExtractor} star-galaxy classifier\footnote{Our star-galaxy separation corresponds to applying the following flags: SG2DPHOT=0, SG\_FLAG=1, CLASS\_STAR<0.5.}. Through applying the IMAFLAGS\_ISO=0 flag, we also removed galaxies that are affected by readout and diffraction spikes, saturation cores, bad pixels, or by primary, secondary or tertiary haloes of bright stars\footnote{The IMAFLAGS\_ISO cut corresponds to applying all MASK values (1, 2, 4, 8, 16, 32 and 64) described in App. A.1.1 of \cite{kuijken2019}.}. We applied the recommended mask that was also used to create the KiDS-1000 shear catalogues\footnote{This mask corresponds to the nine-band KiDS MASK bit values 2 to 11, 13 and 14, described in App. A.2 of \cite{kuijken2019}.}. In addition, objects that are not detected in all 9 bands were removed from the sample. Our final sample of KiDS-bright lenses consists of $\sim1$ million galaxies, more than fivefold the number of GAMA galaxies. This increased lens sample allowed us to verify the results from \cite{brouwer2017} with increased statistics, and to study possible dependencies of the RAR on galaxy observables.

To use the KiDS-bright sample as lenses to measure $g\un{obs}$, we needed accurate individual redshifts for all galaxies in our sample. These photometric redshifts $z\un{ANN}$ were derived from the full nine-band KiDS+VIKING photometry by training on the spectroscopic GAMA redshifts (see Section~\ref{sec:GAMA}) using the \textsc{ANNz2} (Artificial Neural Network) machine learning method \cite[]{sadeh2016}. When comparing this $z\un{ANN}$ to the spectroscopic GAMA redshifts $z\un{G}$ measured for the same galaxies, we found that their mean offset $\meanb{(z\un{ANN} - z\un{G})/(1+z\un{G})} = 9.3\E{-4}$. However, this offset is mainly caused by the low-redshift galaxies: $z\un{ANN}<0.1$. Removing these reduces the mean offset to $\meanb{\delta z/(1+z\un{G})} = -6\E{-5}$, with a standard deviation $\sigma\un{z} = \sigma(\delta z) = 0.026$. This corresponds to a redshift-dependent deviation of $\sigma\un{z}/(1+\meanb{z\un{ANN}})=0.02$ based on the mean redshift $\meanb{z\un{ANN}} = 0.25$ of KiDS-bright between $0.1<z<0.5$, which is the lens redshift range used throughout this work for all lens samples.

In order to measure the expected baryonic acceleration $g\un{bar}$, we computed the KiDS-bright stellar masses $M_\star$ based on these \textsc{ANNz2} redshifts and the nine-band \textsc{GAaP} photometry. Because the \textsc{GAaP} photometry only measures the galaxy magnitude within a specific aperture size, the stellar mass was corrected using the `fluxscale' parameter\footnote{This fluxscale correction of the stellar mass $M_\star$ was applied to \textsc{Lephare}'s best-fit mass value as follows: $M_\star = {\rm MASS\_BEST} + ({\rm MAG\_GAAP\_r} - {\rm MAG\_AUTO\_CALIB}) / 2.5$, where the latter are the \textsc{GAaP} and calibrated elliptical $r$-band magnitudes.} The stellar masses were computed using the \textsc{LePhare} algorithm \cite[]{arnouts1999, ilbert2006}, which performs SPS model fits on the stellar component of the galaxy spectral energy distribution. We used the \cite{bruzual2003} SPS model, with the IMF from \citet[][equal to those used for the GAMA stellar masses]{chabrier2003}. \textsc{LePhare} provides both the best-fit logarithmic stellar mass value `MASS\_BEST' of the galaxy template's probability distribution function, and the $68\%$ confidence level upper and lower limits. We used the latter to estimate the statistical uncertainty on $M_\star$. For both the upper and lower limit, the mean difference with the best-fit mass is approximately: $|\log_{10}\meanb{M\un{lim}/M\un{best}}| \approx 0.06 \dex$.

Another way of estimating the statistical uncertainty in the stellar mass is to combine the estimated uncertainties from the input: the redshifts and magnitudes. The redshift uncertainty $\sigma\un{z}/\meanb{z\un{G}} = 0.11$ corresponds to an uncertainty in the luminosity distance of: $\sigma(\delta D\un{L})/\meanb{D\un{L}} = 0.12$. We took0 the flux $F$ to remain constant between measurements, such that: $4 \pi D\un{L}^2 F \propto D\un{L}^2 \propto L$. Assuming that approximately $L \propto M_\star$ leads to an estimate: 
\begin{equation}
\frac{M_\star + \delta M_\star}{M_\star} = \frac{D\un{L}(\meanl{z}) + D\un{L}(\meanl{z}+\delta z)^2}{D\un{L}(\meanl{z})^2} \, ,
\end{equation}
which finally gives our adopted stellar mass uncertainty resulting from the KiDS-bright redshifts: $\log_{10}(1+\delta M_\star/M_\star) = 0.11 \dex$. The uncertainty resulting from the KiDS-bright magnitudes is best estimated by comparing two different KiDS apparent magnitude measurements: the elliptical aperture magnitudes `MAG\_AUTO\_CALIB' from \textsc{SExtractor} and the S\'ersic magnitudes `MAG\_2dphot' from \textsc{2DPHOT} \cite[]{barbera2008}. The standard deviation of their difference, $\delta m = m\un{2dphot} - m\un{calib}$, is $\sigma(\delta m) = 0.69$, which corresponds to a flux ratio of $F\un{2dphot}/F\un{calib} = 1.88$ (or $0.27 \dex$). Using the same assumption, now taking $D\un{L}$ to remain constant, results in: $4 \pi D\un{L}^2 F \propto F \propto L \propto M_\star$. This means our flux ratio uncertainty directly corresponds to our estimate of the $M_\star$ uncertainty. Quadratically combining the $0.11 \dex$ uncertainty from the redshifts and the $0.27 \dex$ uncertainty from the magnitudes gives an estimate of the total statistical uncertainty on the stellar mass of $\sim0.29\dex$. This is much larger than that from the \textsc{LePhare} code. Taking a middle ground between these two, we have assumed twice the \textsc{LePhare} estimate: $\sigma\un{M_\star} = 0.12 \dex$. However, we have confirmed that using the maximal estimate $\sigma\un{M_\star} = 0.29\dex$ throughout our analysis does not change the conclusions of this work, in particular those of Section~\ref{sec:Results-Types}.

When comparing $M\un{\star,ANN}$ with the GAMA stellar masses $M\un{\star,G}$ of matched galaxies, we found that its distribution is very similar, with a standard deviation of $0.21 \dex$ around the mean. Nevertheless there exists a systematic offset of $\log(M\un{\star,ANN}) - \log(M\un{\star,G}) = -0.056 \dex$, which is caused by the differences in the adopted stellar mass estimation methods. In general, it has been found impossible to constrain stellar masses to within better than a systematic uncertainty of $\Updelta M_\star \approx 0.2 \dex$ when applying different methods, even when the same SPS, IMF and data are used \cite[]{taylor2011,wright2017}. We therefore normalised the $M\un{\star,ANN}$ values of our KiDS-bright sample to the mean $M\un{\star,G}$ of GAMA, while indicating throughout our results the range of possible bias due to a $\Updelta M_\star = 0.2 \dex$ systematic shift in $M_\star$. We estimated the effect of this bias by computing the RAR with $\log_{10}(M_\star)\pm\Updelta M_\star$ as upper and lower limits.

In order to compare our observations to the MG theories, the measured lensing profiles of our galaxies should not be significantly affected by neighbouring galaxies, which we call `satellites'. We defined our isolated lenses (Appendix~\ref{app:Isolation}) such that they do not have any satellites with more than a fraction $f\un{M_\star} \equiv M\un{\star,sat}/M\un{\star,lens}$ of their stellar mass within a spherical radius $r\un{sat}$ (where $r\un{sat}$ was calculated from the  projected and redshift distances between the galaxies). We chose $f\un{M_\star}=0.1$, which corresponds to $10\%$ of the lens stellar mass, and $r\un{sat}=3\hMpc$, which is equal to the maximum projected radius of our measurement. In short: $r\un{sat}(f\un{M_\star}>0.1)>3\hMpc$. We also restricted our lens stellar masses to $M_\star < 10^{11} \hmsun$ since galaxies with higher masses have significantly more satellites (see Section~2.2.3 of \citealp{brouwer2017}). This provided us with an isolated lens sample of $259\,383$ galaxies. We provide full details of our choice of isolation criterion and an extensive validation of the isolated galaxy sample in Appendix~\ref{app:Isolation}. Based on tests with KiDS, GAMA and MICE data we found that this is the optimal isolation criterion for our data. The ESD profile of our isolated sample is not significantly affected by satellite galaxies and that our sample is accurate to $\sim80\%$, in spite of it being flux-limited. Using the MICE simulation we also estimated that the effect of the photometric redshift error is limited.

\section{Simulations}
\label{sec:Simulations}

In order to compare our observations to $\lcdm$-based predictions, we used two different sets of simulations: MICE and BAHAMAS. Here MICE is an $N$-body simulation, which means that galaxies are added to the DM haloes afterwards, while BAHAMAS is a hydrodynamical simulation that incorporates both stars and gas through sub-grid physics. MICE, however, has a simulation volume at least two orders of magnitude larger than BAHAMAS. Below we explain the details of each simulation, and how we utilised their unique qualities for our analysis.

\subsection{MICE mock catalogues}
\label{sec:MICE}

The MICE $N$-body simulation contains $\sim 7\E{10}$ DM particles in a $(3072 \hMpc)^3$ comoving volume \cite[]{fosalba2015b}. From this simulation the MICE collaboration constructed a $\sim5000\deg^2$ lightcone with a maximum redshift of $z=1.4$. The DM haloes in this lightcone were identified using a Friend-of-Friend algorithm on the particles. These DM haloes were populated with galaxies using a hybrid halo occupation distribution (HOD) and halo abundance matching (HAM) prescription \cite[]{carretero2015,crocce2015}. The galaxy luminosity function and colour distribution of these galaxies were constructed to reproduce local observational constraints from SDSS \cite[]{blanton2003a, blanton2003b, blanton2005}.

In the MICECATv2.0 catalogue\footnote{The MICECATv2.0 catalogue is available through CosmoHub (\url{https://cosmohub.pic.es}).}, every galaxy had sky coordinates, redshifts, comoving distances, apparent magnitudes and absolute magnitudes assigned to them. Of the total MICE lightcone we used $1024\deg^2$, an area similar to the KiDS-1000 survey. We used the SDSS apparent $r$-band magnitudes $m\un{r}$ as these most closely match those from KiDS \cite[see][]{brouwer2018}. We could therefore limit the MICE galaxies to the same apparent magnitude as the KiDS-bright sample: $m\un{r}<20 \, {\rm mag}$, in order to create a MICE foreground galaxy (lens) sample. We used the same redshift limit: $0.1<z<0.5$, resulting in a mean MICE lens redshift $\meanb{z}=0.23$, almost equal to that of GAMA and KiDS-bright within this range. The absolute magnitudes of the mock galaxies go down to $M\un{r} - 5\log_{10}(h_{100}) < -14 \magn$, which corresponds to the faintest GAMA and KiDS-bright galaxies. Each galaxy was also assigned a stellar mass $M_\star$, which is needed to compute the RAR (see Section~\ref{sec:Conversion}). These stellar masses were determined from the galaxy luminosities $L$ using \cite{bell2001} $M_\star/L$ ratios.

In addition, each galaxy had a pair of lensing shear values associated with it ($\gamma_1$ and $\gamma_2$, with respect to the Cartesian coordinate system). These shear values were calculated from {\sc healpix} weak lensing maps that were constructed using the `onion shell method' \cite[]{fosalba2008, fosalba2015a}. The lensing map of MICECATv2.0 has a pixel size of $0.43 \am$. We did not use MICE results within a radius $R\un{res}$ corresponding to 3 times this resolution. We calculated $R\un{res}$ and the corresponding $g\un{bar}$ using the mean angular diameter distance and baryonic mass of the MICE lens sample. For the full sample of isolated MICE galaxies these values are: $R\un{res}=0.25 \hMpc$ and $g\un{bar}=6.60\E{-14} \mpss$.

At scales larger than this resolution limit, the MICE shears allowed us to emulate the GGL analysis and conversion to the RAR that we performed on our KiDS-1000 data (as described in Section~\ref{sec:Theory}) using the MICE simulation. To create a sample of MICE background galaxies (sources) for the lensing analysis, we applied limits on the MICE mock galaxies' redshifts and apparent magnitudes, which are analogous to those applied to the KiDS source sample: $0.1 < z < 1.2$, $m\un{r}>20$ (see \citealp{hildebrandt2017} and Section~\ref{sec:KiDS}; uncertainties in the KiDS $z\un{B}$ are not accounted for in this selection). We also applied an absolute magnitude cut of $M\un{r}>-18.5 \, {\rm mag}$, in order to reproduce the KiDS source redshift distribution more closely.

The MICE mock catalogue also features very accurate clustering. At lower redshifts ($z<0.25$) the clustering of the mock galaxies as a function of luminosity was constructed to reproduce the \cite{zehavi2011} clustering observations, while at higher redshifts ($0.45<z<1.1$) the MICE clustering was validated against the Cosmic Evolution Survey \cite[COSMOS;][]{ilbert2009}. The accurate MICE galaxy clustering allowed us to analyse the RAR at larger scales ($>0.3\hMpc$) where clustered neighbouring galaxies start to affect the lensing signal. MICE also allowed us to test our criteria defining galaxy isolation (see Appendix.~\ref{app:Isolation}).

\subsection{BAHAMAS mock catalogue}
\label{sec:Bahamas}

The second set of simulations that we utilised is BAHAMAS \cite[]{mccarthy2017}. The BAHAMAS suite are smoothed-particle hydrodynamical realisations of $(400 \, h_{100}^{-1} {\rm Mpc})^3$ volumes and include prescriptions for radiative cooling and heating, ionising background radiation, star formation, stellar evolution and chemical enrichment, (kinetic wind) supernova feedback, supermassive black hole accretion, and merging and thermal feedback from active galactic nuclei (AGN). The simulations were calibrated to reproduce the stellar and hot gas content of massive haloes, which makes them particularly well suited for our study of the matter content around haloes out to distances of $1$--$3\hMpc$. The masses of DM and baryonic resolution elements are $3.85\times 10^9 \, h_{100}^{-1} {\rm M_\odot}$ and $7.66\times 10^8 \, h_{100}^{-1} {\rm M_\odot}$ respectively, and the gravitational softening is fixed at $\epsilon = 4 \, h_{100}^{-1} {\rm kpc} = 5.71 \hkpc$.

Haloes and galaxies were identified in the simulations using the friends-of-friends \citep{davis1985} and \textsc{Subfind} \citep{springel2001a,dolag2009} algorithms. We labeled the most massive sub-halo in each Friend-of-Friend group as the `central' and other sub-haloes as `satellites'. We constructed an `isolated' galaxy sample by restricting the selection to central sub-haloes that have no other sub-haloes (satellites or centrals) more massive than $10\%$ of their mass within $3 \hMpc$. We randomly selected $100$ galaxies per $0.25$ dex bin in $M_{200}$ between $10^{12}$ and $10^{13.5} \hmsun$. In the last two bins there were fewer than $100$ candidates, so we selected them all. All galaxies have a redshift $z=0.25$. For each selected galaxy we constructed an integrated surface density map, integrated along the line-of-sight for $\pm 15 \, {\rm comoving} \, h_{100}^{-1} {\rm Mpc}$ around the target halo. We also extracted the cumulative spherically averaged mass profile of each target sub-halo, decomposed into DM, stars, and gas. For both the maps and profiles, we included mass contributions from all surrounding (sub)structures: we did not isolate the haloes from their surrounding environment.

We used the integrated surface density map of each galaxy to calculate its mock ESD profile as a function of the projected distance $R$ from the lens centre, in order to mimic the effect of GGL and the conversion to the RAR on the BAHAMAS results. Each pixel on these maps corresponds to $15 \, {\rm comoving} \, h_{100}^{-1} {\rm kpc}$, which in our physical units is: $15 / (1+z) \, 0.7^{-1} h_{70}^{-1} {\rm kpc} = 17.14 \hkpc$. The density maps each have a dimensionality of $400 \times 400$ pixels. Hence the total area of each map is $(6.86 \hMpc)^2$. In calculating the lensing profiles and RAR with BAHAMAS we followed, as closely as possible, the GGL procedure and conversion to the RAR as described in Section~\ref{sec:Theory}. We truncated our lensing profiles at $10$ times the gravitational softening length: $10 \, \epsilon = 0.057 \hMpc$, to avoid the numerically poorly converged central region \cite[]{power2003}. For a typical galaxy in our sample of isolated BAHAMAS galaxies, this corresponds to $g\un{bar}\sim2.38\E{-12} \mpss$.

\subsection{The BAHAMAS RAR: Quantifying the missing baryon effect}
\label{sec:Missing_Baryons}

The calculation of the expected baryonic radial acceleration $g\un{bar}$ requires the enclosed baryonic mass $M\un{bar}(<r)$ within a spherical radius $r$ around the galaxy centre. Since we are dealing with measurements around isolated galaxies at $R>30\hkpc$, we can approximate $M\un{bar}(<r)$ as a point mass $M\un{gal}$ mainly composed of the mass of the lens galaxy itself. $M\un{gal}$ can be subdivided into stars and gas, and the latter further decomposed into cold and hot gas.

\begin{figure}
	\resizebox{\hsize}{!}{\includegraphics{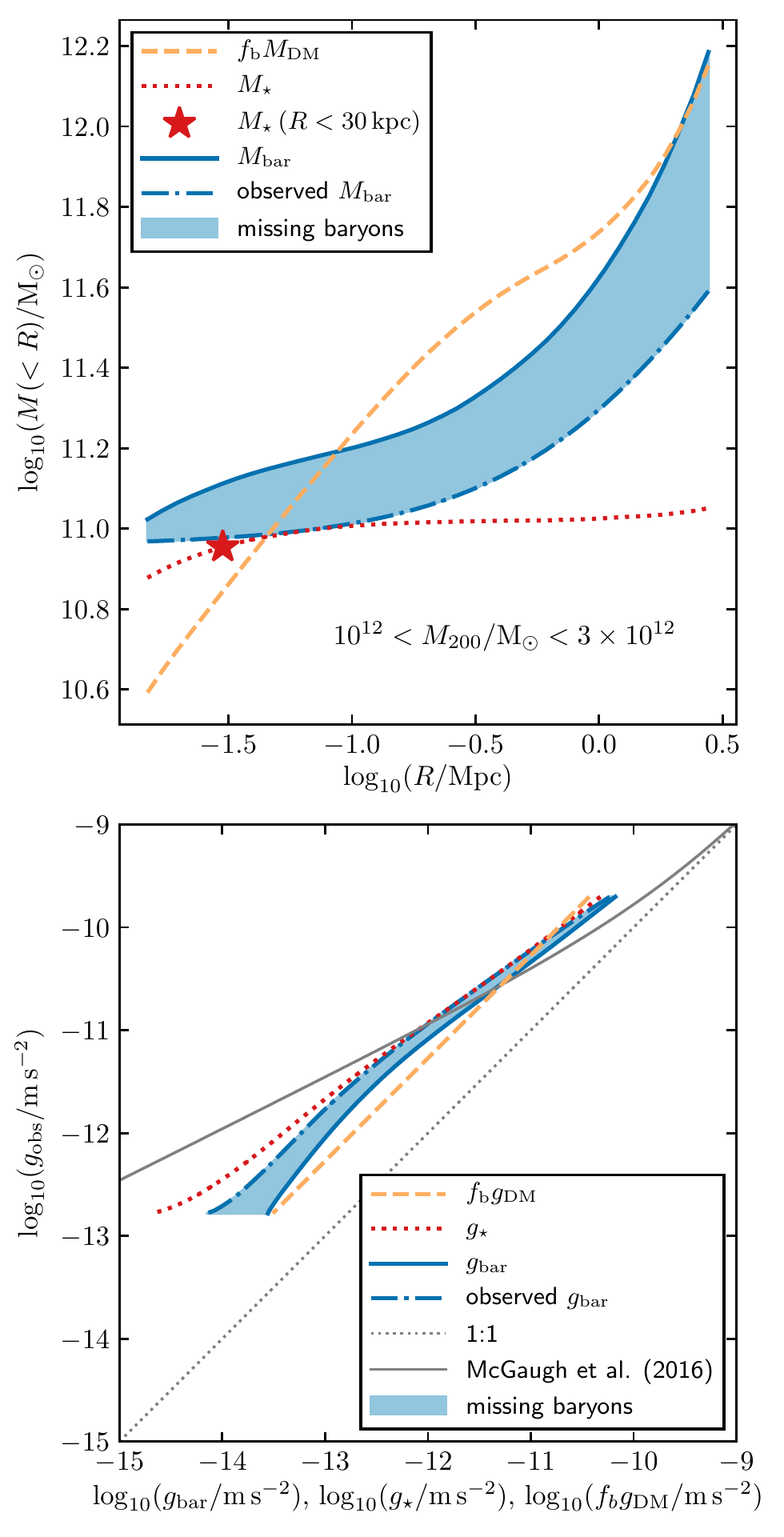}}
	\caption{Mass profiles and RAR of BAHAMAS galaxies. \emph{Upper panel:} Cumulative mass profiles of stars (red dotted line) and total baryons (blue solid line) for BAHAMAS galaxies with $1<M_{200}/(10^{12}\hmsun)<3$. The star marker indicates the stellar mass within a $30 \hkpc$ aperture, indicative of what is typically regarded as the stellar mass of a galaxy. The blue dash-dotted line shows the typical baryonic mass profile of observed galaxies of similar mass, estimated based on an extrapolation of the compilation in fig.~7 of \citet{tumlinson2017}. In the inner galaxy the discrepancy (light blue shaded region) between the observed and simulated $M\un{bar}$ is relatively small, but in the outer galaxy the majority of the baryons predicted to be present in BAHAMAS consist of currently unobserved, missing baryons. The orange dashed line shows the expected baryonic mass profile if the baryon density is everywhere equal to a fixed fraction $f\un{b}=\Upomega\un{b}/\Upomega\un{m}$ of the local DM density. At large enough radii ($\gtrsim 2 \hMpc$), the baryon-to-DM ratio converges to the cosmic average. \emph{Lower panel: } As in upper panel, but in acceleration space. The cosmic baryon fraction provides a strong theoretical upper limit on $g\un{bar}$ at low accelerations in the context of the $\lcdm$ cosmology.}
	\label{fig:missing-baryons}
\end{figure}

How we obtained the stellar masses of our GAMA, KiDS-bright, MICE and BAHAMAS galaxies is described in Sections~\ref{sec:Data} and \ref{sec:Simulations}. From these $M_\star$ values, the fraction of cold gas $f\un{cold} = M\un{cold}/M_\star$ can be estimated using scaling relations based on H\,{\sc i} and CO observations. Following \cite{brouwer2017} we used the best-fit scaling relation found by \cite{boselli2014}, based on the Herschel Reference Survey \cite[]{boselli2010}:
\begin{equation}\label{eq:fcold}
\log(f\un{cold}) = -0.69 \, \log(M_\star/\hmsun) + 6.63 \, .
\end{equation}
We applied this equation to all observed and simulated values of $M_\star$ in order to arrive at the total galaxy mass: $M\un{gal} = M_\star + M\un{cold} = M_\star (1 + f\un{cold})$. The spatial distribution of the stellar and cold gas mass are similar \cite[]{pohlen2010,crocker2011,cooper2012,davis2013} and can therefore be considered a single mass distribution, especially for the purposes of GGL, which only measures the ESD profile at scales larger than the galaxy disc ($R>30\hkpc$). We illustrate this in Fig.~\ref{fig:missing-baryons}, which shows the enclosed mass profiles (upper panel) and RAR (lower panel) for different baryonic components in the BAHAMAS simulation. For these mock galaxies, the stellar mass within $30 \hkpc$ (red star) gives a good approximation of the $M_\star$ distribution across all radii that we consider. We therefore modeled the baryonic mass of our galaxies as a point mass $M\un{gal}$, containing both the stellar and cold gas mass.

We recognise that the total baryonic mass distribution $M\un{bar}$ of galaxies may include a significant amount of additional mass at larger distances, notably in the hot gas phase. This is illustrated in Fig.~\ref{fig:missing-baryons}. In the upper panel, we show the average baryonic mass profile for BAHAMAS galaxies with $1<M_{200}/(10^{12}\hmsun)<3$. In addition, we show an estimate of the typical baryonic mass profile for galaxies in the same mass range, based on an extrapolation to larger radii of the compilation of observations in \citet{tumlinson2017}; including stars, cold gas ($<10^4\,{\rm K}$, traced by absorption lines such as H\,{\sc i}, Na\,{\sc i} and Ca\,{\sc ii}), cool gas ($10^4$-$10^5\,{\rm K}$, traced by many UV absorption lines, e.g. Mg\,{\sc ii}, C\,{\sc ii}, C\,{\sc iii}, Si\,{\sc ii}, Si\,{\sc iii}, N\,{\sc ii}, N\,{\sc iii}), warm gas ($10^5$-$10^6\,{\rm K}$, traced by C\,{\sc iv}, N\,{\sc v}, O\,{\sc vi} and Ne\,{\sc vii} absorption lines), hot gas ($>10^6\,{\rm K}$, traced by its X-ray emission) and dust (estimated from the reddening of background QSOs, and Ca\,{\sc ii} absorption). The light blue shaded region therefore illustrates a component of missing baryons predicted by these simulations but not (yet) observed, possibly related to the cosmological missing baryons \citep[e.g.][]{fukugita1998,fukugita2004,shull2012}. There are several possibilities: (i) there may be additional gas present in a difficult-to-observe phase \citep[e.g. hot, low-density gas, see for instance][]{nicastro2018}; (ii) the simulations do not accurately reflect reality, for example: galaxies may eject substantially more gas from their surroundings than is predicted by these simulations; (iii) there may be less baryonic matter in the Universe than expected in the standard cosmology based on big bang nucleosynthesis \cite[BBN;][]{kirkman2003} calculations and CMB measurements \cite[]{spergel2003,planck2014}.

The lower panel of Fig.~\ref{fig:missing-baryons} illustrates the magnitude of the resulting systematic uncertainties in $g_{\rm bar}$. In the $\lcdm$ cosmology, the expectation at sufficiently large radii is given by $g\un{obs}=f^{-1}\un{b} g\un{bar}$ where $f\un{b}$ is the cosmic baryon fraction $f\un{b} = \Upomega\un{b}/\Upomega\un{m} = 0.17$ \cite[]{hinshaw2013}. BAHAMAS, and generically any $\lcdm$ galaxy formation simulation, converges to this density at low enough accelerations (large enough radii). The most optimistic extrapolation of currently observed baryons falls a factor of $\sim 3$ short of this expectation, while the stellar mass alone is a further factor of $\sim 3$ lower. The unresolved uncertainty around these missing baryons is the single most severe limitation of our analysis. Given that we are interested in both $\lcdm$ and alternative cosmologies, we will use the stellar+cold gas mass $M\un{gal}$ as our fiducial estimate of the total baryonic mass $M\un{bar}$, which is translated into the baryonic acceleration $g\un{bar}$, throughout this work. This serves as a secure lower limit on $g\un{bar}$. We note that the eventual detection, or robust non-detection, of the missing baryons has direct implications for the interpretation of the results presented in Section~\ref{sec:Results}. In Section~\ref{sec:Results-MG_theories} we address the possible effect of extended hot gas haloes on $g\un{bar}$. We discuss this issue further in Section~\ref{sec:Discussion_Conclusion}.

Concerning $g\un{obs}$, omitting the contribution of hot gas will not have a large effect on the prediction within the $\lcdm$ framework (e.g. from simulations) since the total mass distribution at the considered scales is heavily dominated by DM. Within MG frameworks such as EG and MOND, where the excess gravity is sourced by the baryonic matter, it is slightly more complicated. \cite[][see section~2.2]{brouwer2017} carefully modelled the distribution of all baryonic components, based on observations from both GAMA and the literature, including their effect on the excess gravity in the EG framework. They found that, for galaxies with $M_\star<10^{11}\hmsun$, the contribution to the ESD profile (and hence to $g\un{obs}$) from hot gas and satellites was small compared to that of the stars and cold gas. Although this analysis was done for the EG theory, the effect of these extended mass distributions within MOND are similar or even less. This allows us to use a point mass $M\un{gal}$ as a reasonable approximation for the baryonic mass distribution $M\un{bar}(<r)$ within our measurement range when computing $g\un{obs}$ as predicted by MOND and EG (see Section~\ref{sec:MOND} and \ref{sec:EG}).

\subsection{The BAHAMAS RAR: Testing the ESD to RAR conversion}
\label{sec:Conversion_test}

\begin{figure*}
	\centering
    \includegraphics[width=17cm]{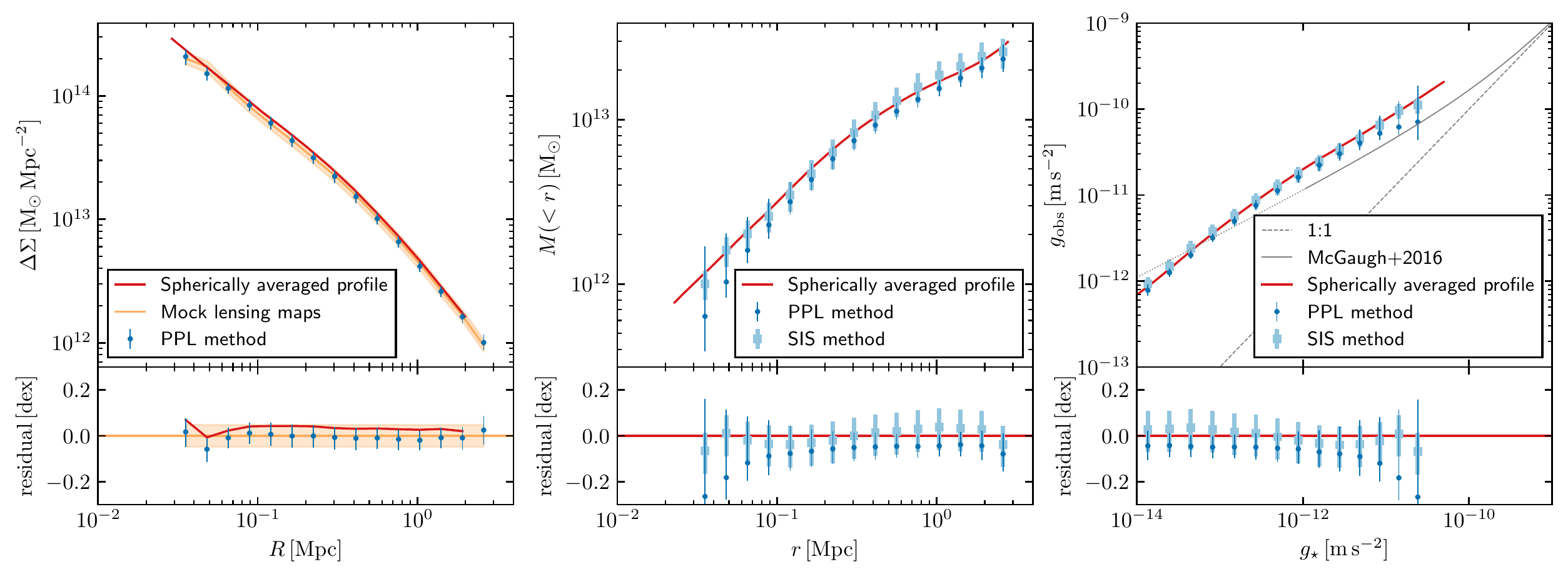}
	\caption{Illustration of the recovery of the acceleration profile from simulated weak lensing observations. \emph{Left}: Average ESD profile of a subset of our sample of BAHAMAS galaxies with $10^{13}<M_{200}/(\hmsun)<10^{13.1}$, derived from the spherically averaged mass profile (red line) and the mock lensing maps (yellow line, with an assumed $0.1 \dex$ Gaussian uncertainty). The PPL method recovery of the ESD profile is shown with the blue points; error bars represent $68\%$ confidence intervals. \emph{Centre}: The SIS (light blue squares) and PPL (dark blue points) method recover the spherically averaged enclosed mass profile. The uncertainties on the SIS points are derived by sampling the uncertainties on the mock lensing ESD profile. \emph{Right}: The resulting dynamical acceleration profile $g_{\rm obs}$ and uncertainties, plotted as a function of the acceleration due to stars $g_\star=GM_\star(<r)/r^2$.}
	\label{fig:compare_method}
\end{figure*}

We used BAHAMAS to test the accuracy of our SIS method (outlined in Section~\ref{sec:Conversion}) in estimating $g\un{obs}$ from our GGL measurement of $\Updelta\Upsigma\un{obs}$, by comparing it against the more sophisticated piece-wise power law (PPL) method outlined in Appendix~\ref{app:PPL_method}. As a test system, we used the $28$ galaxies from our BAHAMAS sample with $10^{13}<M_{200}/(\hmsun)<10^{13.1}$. We combined these into a stacked object by averaging the individual ESD profiles as derived from their mock lensing maps. The stacked ESD as measured from the lensing mocks is shown in the left panel of Fig.~\ref{fig:compare_method}. Since the mock ESD profiles are derived from convergence maps (rather than the shapes of background galaxies), they have no associated measurement uncertainty -- for simplicity, we assumed a constant $0.1 \dex$ uncertainty, which is similar to that for the KiDS measurements. We also combined the spherically averaged enclosed mass profiles of the galaxies out to $3 \hMpc$ by averaging them. From this average mass profile we analytically calculated the ESD profile shown in the left panel of Fig.~\ref{fig:compare_method}. We found that the $\Updelta\Upsigma$ calculated from the spherically averaged mass profile is $\sim 0.05 \dex$ higher than the direct measurement of the stacked lensing mocks. This primarily results from the fact that the spherically averaged mass profile does not take into account the additional matter outside the $3\hMpc$ spherical aperture, whereas the mock surface density maps are integrated along the line-of-sight for $\pm 15 \, {\rm comoving} \, h_{100}^{-1} {\rm Mpc}$ around the lens.

The PPL method described in Appendix~\ref{app:PPL_method} attempts to reproduce the ESD profile by converging to an appropriate volume density profile. The resulting recovered ESD profile and its $68\%$ confidence interval is shown with blue points and error bars in the left panel of Fig.~\ref{fig:compare_method} -- the fit to the mock data is excellent. In the centre panel we show the enclosed mass profile as recovered by both the PPL and SIS methods, in addition to the true enclosed mass profile. Both estimators recover the profile within their stated errors. The PPL method systematically underestimates it by $\sim 0.1 \dex$ across most of the radial range. This is directly caused by the difference between the spherically averaged and mock lensing ESD profiles (left panel). The somewhat wider confidence intervals at small radii are caused by the lack of information in the mock data as to the behaviour of the profile at $r < 30 \hkpc$; the PPL model marginalises over all possibilities. Once the enclosed mass is dominated by the contribution at radii covered by the measurement, the uncertainties shrink. To account for the added uncertainty resulting from the conversion to the RAR, we added $0.1 \dex$ to the error bars of our RAR measurements throughout this work.

The SIS method instead slightly underestimates the enclosed mass at small radii, and overestimates it at large radii. The apparent improved performance relative to the PPL method is actually due to a fortuitous partial cancellation of two errors. First, the SIS calculation suffers from the same underestimation of the spherically averaged enclosed mass profile as the PPL method, due to the difference between the mock lensing and spherically averaged ESD profiles. However, in addition to this, the SIS method assumes a density profile $\rho(r)\propto r^{-2}$ at all radii. At small radii, the power-law slope is in reality about $-2.1$. This results in a slight overestimate of the enclosed mass, which partially compensates the underestimate described above, resulting in a net underestimate. At larger radii, the slope of the density profile becomes progressively steeper, such that the assumption of an $r^{-2}$ profile increasingly overestimates the enclosed mass, eventually resulting in a net overestimate.

The right panel of Fig.~\ref{fig:compare_method} illustrates the resulting uncertainty in the measurement of the RAR. To focus on the influence of the method used to recover $g\un{obs}$, we simply used the exact spherically averaged stellar mass profile to calculate $g\un{\star}$, plotted on the x-axis\footnote{We do not include the additional gas, which is predominantly in the hot phase, for consistency with the presentation of the results in Section~\ref{sec:Results}.}. We found that, for mock lenses within the BAHAMAS simulation, both the SIS and the PPL method yield acceptable and consistent estimates of $g\un{obs}$. We note that the BAHAMAS $g\un{obs}(g\un{\star})$ is significantly offset from the RAR as measured by M16; we will return to this point when we compare BAHAMAS to our observations in Section~\ref{sec:Results-Simulations}.

\section{Results}
\label{sec:Results}

Tables containing the ESD profile data used to create all results figures (i.e. Figures \ref{fig:Vrot_kids}, \ref{fig:RAR_kids_gama_verlinde}, \ref{fig:RAR_kids_mice_bahamas}, \ref{fig:RAR_kids_galtypebins}, \ref{fig:RAR_kids_mice_mstarbins}, \ref{fig:RAR_kids_dwarfs},  \ref{fig:RAR_kids_mice_mstarbins_all} and \ref{fig:RAR_kids_gama_Navarro}) can be found at: \url{http://kids.strw.leidenuniv.nl/sciencedata.php}.

\subsection{Lensing rotation curves}
\label{sec:Results-Rotation_curves}

\begin{figure*}
	\centering
	\includegraphics[width=17cm]{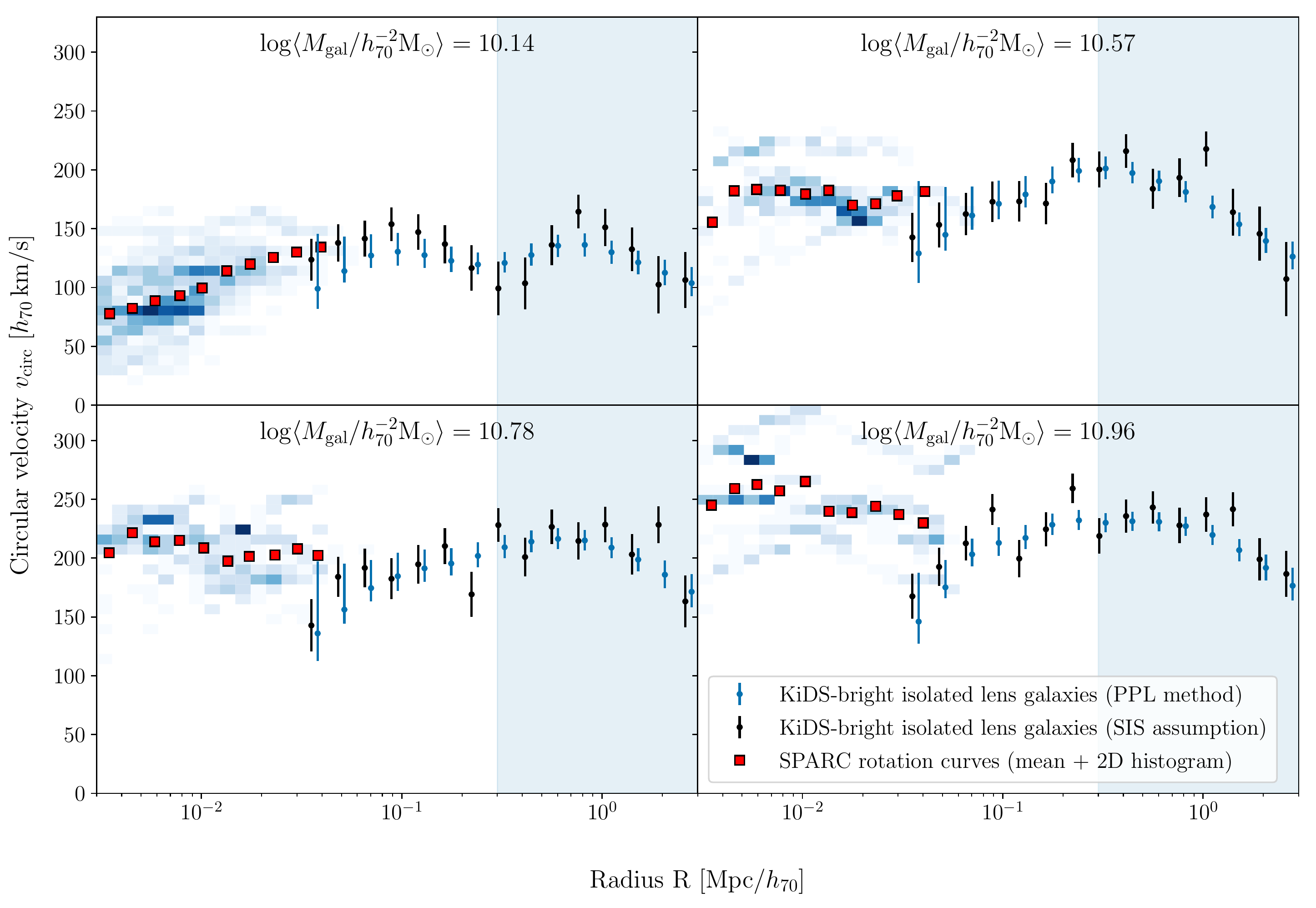}
	\caption{Measured rotation curves -- the circular velocity as a function of radius $v\un{circ}(R)$ -- of the KiDS-bright isolated lens sample, divided into four stellar mass bins. The mean galaxy mass (stars+cold gas) of the lenses is shown at the top of each panel. The light blue shaded region indicates the radii corresponding to $R > 0.3 \hMpc$, where the uncertainty in the photometric KiDS redshifts can affect the isolated lens selection (see Appendix~\ref{app:Isolation}). The black points (with $1\sigma$ error bars) show the result calculated using the SIS assumption, while the blue points (with error bars representing the 16th and 84th percentile of the fits) show the result from the more sophisticated PPL method. Our measurements are consistent between the two methods, and also with the rotation curves from SPARC (all data as the blue 2D histogram, the mean as red squares).}
	\label{fig:Vrot_kids}
\end{figure*}

As a final consistency check between the SIS assumption and the PPL method, we applied both methods to the true KiDS-1000 data. Since these methods are only used to convert $\Updelta\Upsigma(R)$ into $g\un{obs}(r)$, we can leave $g\un{bar}$ out of the comparison and plot our results as a function of $R$. An observable closely related to the RAR that is usually plotted as a function of radius, is the traditional circular velocity curve:
\begin{equation}\label{eq:Vrot}
v\un{circ}(r) = \sqrt{\frac{G M\un{obs}(<r)}{r}} \, ,
\end{equation}
an observable that indeed served as input to the M16 RAR measurement. We applied the SIS method described in Section~\ref{sec:Conversion} to convert our ESD profiles $\Updelta\Upsigma(R)$ into $v\un{circ}(R)$ since substituting Eq.~\ref{eq:Mobs} into Eq.~\ref{eq:Vrot} gives:
\begin{equation}
v\un{circ}(r) = \sqrt{\frac{G \, (4 \Updelta\Upsigma(r) \, r^2)}{r}} \, = \sqrt{4 G \, \Updelta\Upsigma(r) \, r} \, .
\end{equation}
We also applied Eq.~\ref{eq:Vrot} to compute $v\un{circ}(R)$ from the $M(<R)$ calculated through the PPL method described in Appendix~\ref{app:PPL_method}. We note that both the SIS and PPL method assume spherical symmetry, while in simulations DM haloes are found to deviate from sphericity, which could lead to deviations in the lensing rotation curves \cite[]{cuddeford1993}. However, the mean ellipticity of haloes is observed to be small ($\meanb{|\epsilon|} = 0.174\pm0.046$, \citealp{schrabback2020}). The stacking of thousands of lenses with approximately random orientations further reduces the impact on the lensing signal, which means the halo ellipticity will not significantly change our results.

Fig.~\ref{fig:Vrot_kids} shows the lensing rotation curves for isolated KiDS-bright galaxies, divided into four stellar mass bins using the following limits: $\log_{10}(M_\star/\hmsun) = [8.5,10.3,10.6,10.8,11.0]$. For each bin the mean galaxy mass (stars+cold gas) of the lenses, $\log_{10}\lan M\un{gal}/\hmsun \ran = [10.14, 10.57, 10.78, 10.96]$, is shown at the top of the panel. Showing the data in this way allows us to observe for the first time in this intuitive manner how the circular velocity curves of isolated galaxies continue beyond the observable disc ($r>30\hkpc$). In addition, it provides a consistency check against the SPARC rotation curves \cite[]{lelli2016b} that form the basis for the M16 RAR measurement. It is remarkable how well the mean of the SPARC rotation curves and our lensing results correspond at their intersection ($r\sim 30 \hkpc$). But most importantly, we find that the `lensing rotation curves' from the SIS assumption are consistent with the ones from the PPL method. Although the SIS assumption results in slightly more scatter, there is very little systematic bias between the results from the two methods, which have a fractional difference of $\meanb{\log(v\un{circ, SIS} / v\un{circ, PPL})} = 0.017 \dex$. Since this measurement is merely a different way of presenting the observed acceleration, which equals $g\un{obs}(r) = v\un{circ}^2 / r$, we can easily compute that the expected difference in $g\un{obs}$ would be $\meanb{\log(g\un{obs, SIS} / g\un{obs, PPL})} = 0.038 \dex$.

The consistency between the two conversion methods allows us to use the SIS assumption throughout this work. The great advantage of this method is that it allows us to convert GGL profiles binned by baryonic acceleration $\Updelta\Upsigma(g\un{bar})$, into the RAR: $g\un{obs}(g\un{bar})$. This is not the case for the PPL method, which only works on $\Updelta\Upsigma(R)$ binned by radius. The former can therefore be applied to any lens sample; the latter only to lenses within a narrow mass range (in order to convert $R$ into $g\un{bar}$ using the mean $\meanb{M\un{gal}}$). As explained in Section~\ref{sec:Conversion_test} we added $0.1 \dex$ to the error bars of all RAR measurements in this work, to account for the added uncertainty from the conversion of the ESD to the RAR. After showing that both methods yield acceptable and consistent estimates of $g\un{obs}$, we will show only the SIS measurement when presenting our results in this section to reduce clutter in the figures.

\subsection{The RAR of KiDS compared to MG theories}
\label{sec:Results-MG_theories}

\begin{figure*}
	\centering
	\includegraphics[width=17cm]{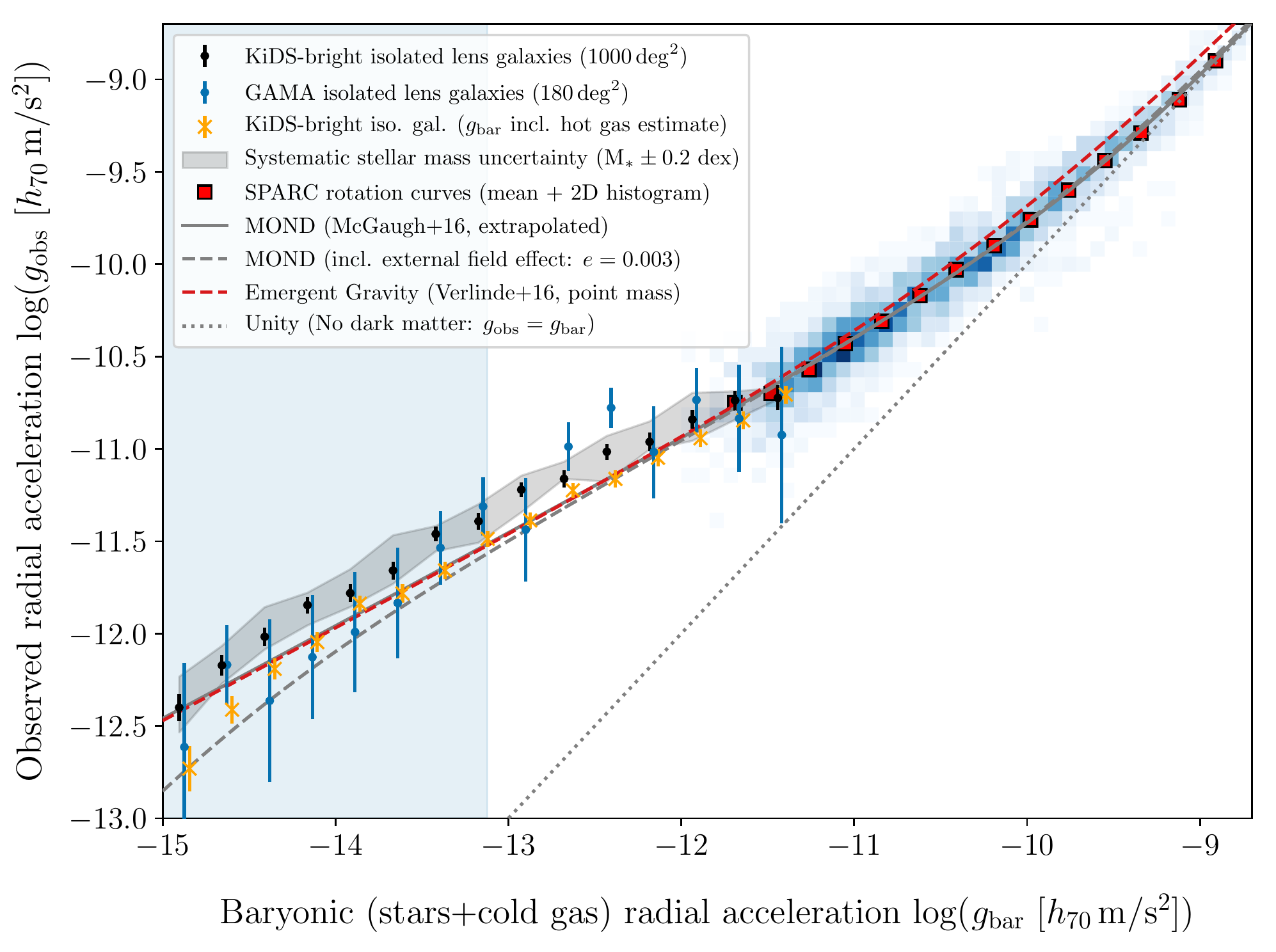}
	\caption{Measured RAR, which compares the total gravitational acceleration $g\un{obs}$ with the expected baryonic acceleration $g\un{bar}$ of galaxies. At high accelerations we show the M16 RAR measurements from galaxy rotation curves (all data as the blue 2D histogram, the mean as red squares). Using weak gravitational lensing we were able to extend this measurement to lower accelerations, using both the spectroscopic GAMA and the photometric KiDS-bright isolated lens samples (blue and black points with $1\sigma$ error bars). Comparing our lensing observations to two MG models: MOND (the M16 fitting function; grey solid line) and EG (assuming a point mass; red dashed line) we find that GAMA results are in agreement with the two models, while those from KiDS-bright are systematically higher. At very low accelerations (corresponding to $R > 0.3 \hMpc$, light blue shaded region) the uncertainty in the photometric KiDS redshifts affects the isolated lens selection, resulting in systematically higher values of $g\un{obs}$ due to the possible contribution of satellites. The results from the spectroscopic GAMA survey, however, are still reliable within this region. The impact of stellar mass uncertainty ($\Updelta M_\star=0.2 \dex$) on the measurement is shown as the grey band. We show the MOND prediction including the EFE (with $e=0.003$, see Eq. \ref{eq:mond_efe}) as the grey dashed line. In addition, we show the effect on the RAR of KiDS-bright galaxies if $g\un{bar}$ contained an additional isothermal hot gas contribution within a $100\hkpc$ radius, with a nominal gas mass equal to the stellar mass (orange crosses with $1\sigma$ error bars). We emphasise that this is only a rough order of magnitude estimate of the possible effect of gaseous haloes, which are extremely difficult to observe.}
	\label{fig:RAR_kids_gama_verlinde}
\end{figure*}

In Fig.~\ref{fig:RAR_kids_gama_verlinde} we show the RAR, with the observed radial acceleration computed from our lensing measurements through Eq. \ref{eq:gobs_from_ESD}) on the y-axis. The x-axis shows the expected baryonic (star+cold gas) radial acceleration, where the label serves as a reminder throughout this work that $g\un{bar}$ is only computed from the measured stellar masses of the galaxies and an estimate of their cold gas component.

The lensing $g\un{obs}$ was measured using the GAMA and KiDS-bright isolated galaxy samples, respectively. Due to its smaller survey area ($180$ vs. $1006 \deg^2$), the error bars using GAMA lenses are larger than those using KiDS-bright lenses. However, as explained in Appendix~\ref{app:Isolation}, the spectroscopic redshifts of the GAMA survey allow for a more reliable selection of the isolated lenses compared to KiDS (which measures photometric redshifts with a $\sigma\un{z}=0.02$ uncertainty). The effect of this uncertainty on the measured lensing profiles is modelled in Fig.~\ref{fig:isolation_test_offset}, which shows that the ESD profile of the `offset' MICE sample diverges from the truly isolated MICE galaxies at radius $R > 0.3 \hMpc$. At these large scales, the effect of satellite galaxies on the lensing signal result in a $\sim30\%$ increase in $\Updelta\Upsigma$ due to the contribution of satellite galaxies. We translated this radius into a gravitational acceleration value using Eq.~\ref{eq:grav}, based on the average $M\un{gal}$ of the lens sample. In this way we estimate that, for the full sample of isolated KiDS-bright galaxies, the isolation criterion is no longer reliable when $g\un{bar} \lessapprox 10^{-13} \mpss$, as indicated by the light blue shaded region in Fig.~\ref{fig:RAR_kids_gama_verlinde}. We note that the GAMA results, which are based on accurate spectroscopic redshift measurements, are still reliable within this region.

The grey band shows the range of possible bias due to a $\Updelta M_\star=\pm0.2 \dex$ systematic shift in stellar mass. We estimated this range by performing our analysis assuming stellar masses that are $0.2 \dex$ higher than, and then $0.2 \dex$ lower than, their best-fitting $M_\star$ values (see Section~\ref{sec:KiDS-bright}). We only show this band once, for the KiDS-bright result, but note that this uncertainty equally affects the GAMA stellar masses (and, indeed, any stellar mass measurement; see \citealt{wright2017}).

We compare our results to the M16 RAR measurements (both the full dataset: blue 2D histogram, and the mean: red squares), from SPARC galaxy rotation curves, which cover higher accelerations than our lensing measurements (corresponding to smaller scales: $R < 30 \hkpc$). At the highest-acceleration end (smallest scales), where $g\un{obs}$ is dominated by $g\un{bar}$, they follow a one-to-one relation. At lower accelerations (larger scales) their results quickly diverge from unity, signifying the start of the DM dominated regime. We find that these two fully independent RAR observations, respectively from rotation curves and lensing, are in strong agreement\footnote{Because the blinding intended to avoid observer bias in the KiDS-1000 cosmological constraints \cite[]{asgari2020,heymans2020,troster2020} only has a small effect on GGL observations, this agreement has been present since the start of our analysis (before the data were un-blinded).}.

Fig.~\ref{fig:RAR_kids_gama_verlinde} also compares the two MG models, EG and MOND, to our lensing results (for a comparison of these two models with the RAR from SPARC, see \citealt{lelli2017a}). As explained in Sections~\ref{sec:MOND} and \ref{sec:EG}, we took the MOND prediction to be equal to the extrapolated M16 fitting function (Eq. \ref{eq:rar_mcgaugh}), and that of EG as the prediction from \cite{verlinde2017} for a point mass (Eq. \ref{eq:rar_verlinde}). At high accelerations, the prediction from EG appears to lie above that of MOND and the SPARC data. However, as explained in Section~\ref{sec:EG}, the prediction of Eq. \ref{eq:rar_verlinde} should be taken with a grain of salt for accelerations $g\un{bar}>1.2\E{-10} \mpss$. Within our measurement range, the two predictions are almost indistinguishable. Both models are compatible with the GAMA data. The KiDS-bright data points, however, lie systematically above the MG predictions.

To quantify the level of agreement between the acceleration predicted by the different models $g\un{mod}$ and the observed $g\un{obs}$, we calculated the $\chi^2$ value:
\begin{equation}
\chi^2 = (g\un{obs} - g\un{mod})^\intercal \cdot C^{-1}(g\un{obs} - g\un{mod}) \, ,
\label{eq:chi2}
\end{equation}
where $C^{-1}$ is the inverse of the analytical covariance matrix (see Section~\ref{sec:Lensing}). We divided this quantity by the number of degrees of freedom $N\un{DOF}$ of the model, which gives the reduced $\chi^2$ statistic:
\begin{equation}
\chi\un{red}^2 = \frac{\chi^2}{N\un{DOF}} = \frac{\chi^2}{N\un{data} - N\un{param}} \, .
\end{equation}
Here $N\un{data}$ is the number of data points in the measurement and $N\un{param}$ is the number of free parameters in the model. Since none of the models have free parameters, $N\un{DOF}$ is simply the total number of $g\un{bar}$-bins (in this case $N\un{data}= 15$).

Comparing the GAMA data to the two MG models results in $\chi\un{red}^2$-values of $0.8$ for both MOND and EG, corresponding to a standard deviation of $0.4\sigma$. This confirms that both models agree well with the GAMA data. When using the KiDS-bright results, neither model provides a good description of the data with: $\chi\un{red}^2=4.6$ and $5.0$ for MOND and EG respectively, corresponding to $\sim6$ standard deviations ($\sim6\sigma$). Taking into account the effect of the photometric redshift uncertainty of KiDS-bright by only using the seven data points within the isolation criterion limit ($R < 3 \hMpc$) we find: $\chi\un{red}^2 = 4.0$ for MOND and $\chi\un{red}^2 = 4.4$ for EG, $\sim3.8 \sigma$ away from a good fit. Considering the $\Updelta M_\star=\pm0.2\dex$ uncertainty shown by the grey band (with the data points beyond the isolation criterion limit still removed) leads to $\chi\un{red}^2=1.5$ for $\Updelta M_\star=+0.2\dex$ and $\chi\un{red}^2=14$ for $\Updelta M_\star=-0.2\dex$ with respect to MOND, with similar results for EG. Thus, the MOND and EG predictions are able to describe our measurements within the statistical and systematic uncertainties. Whether these models are confirmed or excluded relies heavily on the systematic bias in the stellar mass measurements. This highlights the general point that GGL measurements are now so accurate in determining the total observed mass distribution that improving the RAR measurement primarily depends on obtaining better constraints on the baryonic mass distribution.

This point is highlighted further by the fact that we cannot incorporate measurements of the total baryonic mass distribution into our comparison, in particular those components that have not been detected, such as hot gaseous haloes and missing baryons. This remains a fundamental limitation of all work testing DM or MG theories at large scales (see Section~\ref{sec:Missing_Baryons}). Although there have been very recent fruitful attempts at a first detection of this barely visible baryonic component \cite[]{macquart2020,tanimura2020}, there exist no accurate measurements of its distribution around isolated galaxies. However, we can safely continue as long as all estimates of $g\un{bar}$ (in the measurements, models and simulations) are based on the same components (in our case: stars+cold gas). This way our RAR results remain purely observational, based on actual measurements along both axes.
	
However, a qualitative idea of the possible effect of an additional extended ionised gas component on $g\un{bar}$ is depicted in Fig.~\ref{fig:RAR_kids_gama_verlinde}. In addition to our standard stars-and-cold-gas point mass used to calculate $g\un{bar}$, we modeled the hot gas as a simple isothermal density profile ($\rho(r) \propto r^{-2}$), truncated at the accretion radius $R\un{acc}$. Based on \cite{valentijn1988}, we derived that $R\un{acc} \approx 100 \hkpc$ for hot gas haloes around galaxies with $M_\star \approx 10^{11} \hmsun$. Finding an accurate estimate of the additional gas mass $M\un{gas}$ within this radius is no easy matter. \cite{brouwer2017} assumed a total hot gas mass $M\un{gas} = 3 M_\star$, based on results from the OWLS hydrodynamical simulations by \cite{fedeli2014b}. They found that, in simulations with AGN feedback, OWLS galaxies with a total mass $M_{200} = 10^{12} \homsun$ (corresponding to $M_\star \approx 10^{10} \hmsun$, a lower limit on the typical stellar masses in our sample) have a gas-to-stellar-mass fraction of $M\un{gas} / M_\star \approx 3$. One of the few observational scaling relations for hot gas is derived by \cite{babyk2018}, using Chandra X-ray observations of 94 early-type galaxies. In their Fig.~7, which shows the X-ray gas mass versus the total galaxy mass, galaxies with $M\un{tot} = 10^{12} \msun$ have gas fractions ranging from $0.1 - 1$. However, \cite{babyk2018} measured both $M\un{tot}$ and $M\un{gas}$ within $5$ effective radii of their galaxies, which means that the hot gas fraction on larger scales could be as high as $3$ in extreme cases. These relatively high hot gas masses motivated by the \cite{babyk2018} observations are possibly biased towards a high X-ray surface brightness and are an order of magnitude higher than the hot gas masses presented in Fig.~7 of \cite{tumlinson2017}. As this gas mass outweighs the possible contribution of various cooler gas and dust components, this case provides a good guide for our evaluation. Based on all these considerations, we assumed a nominal gas-to-stellar-mass fraction of $M_\star/M\un{gas} = 1$, emphasising that this is only an order of magnitude estimate due to the challenging nature of observing circumgalactic gas.

In Fig.~\ref{fig:RAR_kids_gama_verlinde} we include the RAR of KiDS-bright galaxies with our nominal estimate of the hot gas distribution added to $g\un{bar}$ on the x-axis. At the highest accelerations measurable by lensing, we find that these results are almost indistinguishable from the original KiDS-bright measurements. As the acceleration decreases, the $g\un{bar}$ values including hot gas shift further to the right (higher values) due to the increased enclosed hot gas mass. This causes a steepening downward slope of the RAR, such that it finally diverges from the $g\un{obs} \propto \sqrt{g\un{bar}}$ relation at very low accelerations ($g\un{bar}<10^{-14} \mpss$). The same effect is as also seen in the BAHAMAS results in Fig.~\ref{fig:missing-baryons}. As expected, we find that this steepening of the RAR increases for higher assumed gaseous halo masses $M\un{gas}$, and decreases for lower values. This implies that, if gaseous haloes more massive than in our example ($M\un{gas} \gtrsim M_\star$) were detected directly and incorporated into the measurement, the observed RAR would diverge from the current MOND and EG predictions at low accelerations.

In the case of MOND a steep downward slope at low accelerations is not expected unless, despite our best efforts, our isolated galaxy sample is not truly isolated. In that case undetected satellites might cause an external field effect (EFE). To evaluate this effect we use the results of \cite{chae2020} for the isolated SPARC galaxies. Based on their results, we have assumed $e = g\un{ext}/g_\dagger = 0.003$ as a reasonable estimate of the external gravitational acceleration $g\un{ext}$ compared to the critical acceleration scale $g_\dagger$ (see Section~\ref{sec:MOND}) for our isolated lenses. We use the fitting function in Eq.~\ref{eq:mond_efe}, which represents the EFE for an idealised model of galaxies within their environment, to depict the EFE on the predicted MOND RAR in Fig.~\ref{fig:RAR_kids_gama_verlinde}. The extrapolated M16 fitting function represents the MOND prediction without any EFE ($e=0$). As expected the MOND prediction including the EFE diverges from the one without, tending towards a steeper downward slope at low accelerations ($g\un{bar} < 10^{-12} \mpss$). Hence the EFE moves the MOND prediction away from our main observational result: the lensing RAR from the KiDS-bright sample without an estimate for the additional hot gas, which we explore throughout the rest of this work. We will therefore maintain the use of the M16 fitting function as our main MOND prediction since this represents the optimal case considering our observations. Regarding the KiDS-bright result including an estimate for the hot gas, it turns out that the steeper downward slope resulting from the MOND EFE is not steep enough to be consistent with our measured RAR including an estimate of the additional hot gas. This is illustrated by the fact that, for our chosen value $e=0.003$, the MOND prediction including EFE and our RAR observation including hot gas reach the same value of $g\un{obs}$ at $g\un{bar}\approx10^{-15} \mpss$. However, the observation reaches this depth within a much smaller span in $g\un{bar}$ ($-15 < \log_{10}(g\un{bar}/ \mpss) < -14$). Choosing a different value for the EFE strength $e$ does not solve this problem, and the effect becomes stronger for higher assumed values of $M\un{gas}$. It is therefore unlikely that the MOND EFE can explain the effect of massive ($M\un{gas} \gtrsim M_\star$) hot gaseous haloes, if such haloes are detected. In the case of EG it is not yet known whether and, if so, how external gravitational fields affect its prediction (Verlinde, priv. comm.).

\subsection{The RAR of KiDS compared to $\Lambda$CDM simulations}
\label{sec:Results-Simulations}

\begin{figure*}
	\centering
	\includegraphics[width=17cm]{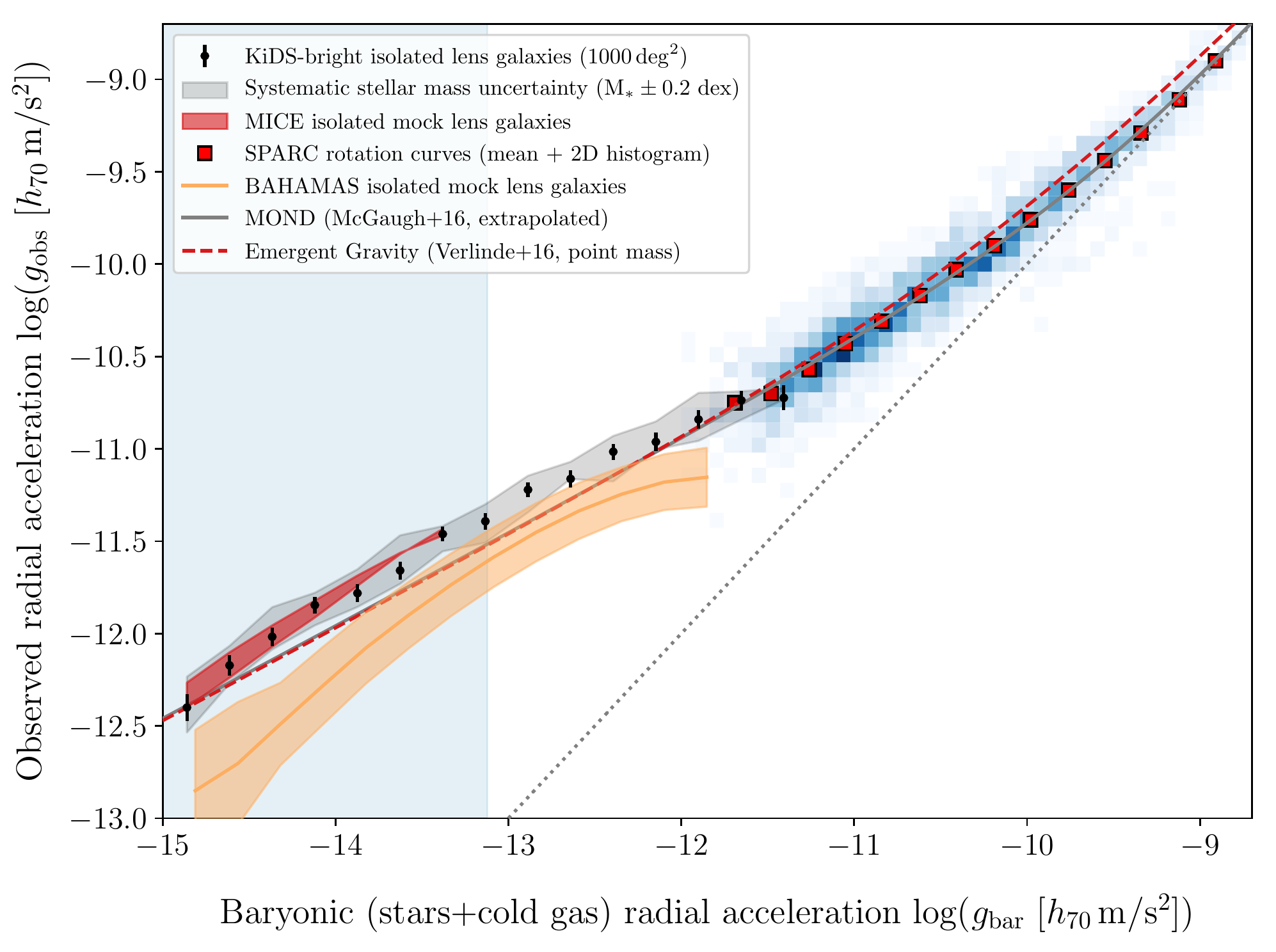}
	\caption{Measured RAR of the KiDS-bright isolated lens sample (black points with $1\sigma$ error bars) compared to two $\lcdm$ simulations: MICE and BAHAMAS. The accelerations where uncertainty in the photometric KiDS redshifts affects the KiDS-bright isolated lens selection is indicated by the light blue shaded region. The MICE results (red band) emulate the effect of the redshift uncertainty in KiDS, while the BAHAMAS results (orange band) reflect the median and $16^{\rm th}$ and $84^{\rm th}$ percentiles of the simulated lens galaxies. The MICE simulation, though limited to low accelerations by its resolution, succeeds in reproducing the lensing data. The result from the BAHAMAS simulation runs approximately parallel to the MICE curve, but underestimates our measurement by $0.5 \dex$ due to the biased SHMR of the BAHAMAS isolated galaxies (see Section~\ref{sec:Results-Simulations}).}
	\label{fig:RAR_kids_mice_bahamas}
\end{figure*}

In this section we compare the KiDS-1000 RAR with numerical $\lcdm$ simulations\footnote{The first $\lcdm$ model we test is that of N17, but find that this simple analytical model is not sufficient to describe our data (see Appendix~\ref{app:Results_N17}).}. In order to obtain the predictions from these simulations, we applied the same isolation criterion, GGL procedures and RAR conversion to mock galaxy samples from the MICE and BAHAMAS simulations (see Section~\ref{sec:Simulations}). In Fig.~\ref{fig:RAR_kids_mice_bahamas}, BAHAMAS (orange band) is shown as the median result of all lens galaxies, with the upper and lower limit of the band representing the $16^{\rm th}$ and $84^{\rm th}$ percentiles. For MICE (red band) we show the result for isolated lenses selected using the true redshifts (lower limit) and using redshifts with a normally distributed random offset of $\sigma\un{z}/(1+z) = 0.02$ (upper limit), in order to emulate the effect of the redshift uncertainty in KiDS on the isolated galaxy selection (see Appendix~\ref{app:Isolation}). This means that the upper limit of the MICE prediction is considered reliable even at high accelerations (blue shaded region), where uncertainties in the galaxy isolation could affect the RAR measurement. The RAR observations are the same KiDS-bright lensing and M16 rotation curve results as shown in Fig.~\ref{fig:RAR_kids_gama_verlinde}, this time compared to the predictions from the two simulations.

We find a good agreement between the MICE simulation and our measurements. The MICE measurements are limited to the low $g\un{bar}$ regime, owing to the resolution of the MICE simulations. The MICE scale limit of $R > 0.25 \hMpc$ is within the angular scale where satellites missed by the isolation criterion might impact the lensing signal ($R > 0.3 \hMpc$, light blue shaded region). The effect of the KiDS-bright redshift uncertainty $\sigma\un{z}$ on the isolation criterion is however mimicked in the MICE simulation (upper limit of the red band), which means we can safely compare MICE with our low-acceleration measurements. The limited width of red band shows that this effect is relatively small ($\sim 30\%$). The MICE prediction (with the $\sigma\un{z}$ offset) results in a reduced $\chi^2$ value of $\chi\un{red}^2 = 2.3$, corresponding to $2.3 \sigma$.

Figure \ref{fig:RAR_kids_mice_bahamas} shows poor agreement between the lensing RAR for isolated BAHAMAS galaxies and the KiDS measurement. The reason for this is straightforward to understand: the BAHAMAS measurement in Fig.~\ref{fig:RAR_kids_mice_bahamas} runs approximately parallel to both the KiDS and MICE curves, as a result of a constant offset in the stellar-to-halo-mass relation (SHMR) between BAHAMAS and MICE. Both simulations reproduce the observed SHMR in an overall sense, as shown in fig.~6 of \citet{mccarthy2017} and \citet{jakobs2018} for BAHAMAS, and guaranteed by construction as described in \citet{carretero2015} for MICE. However, while in MICE our isolated galaxy sample follows essentially the same SHMR as the parent sample, in BAHAMAS isolated galaxies have, on average, triple the stellar mass at fixed halo mass compared to the global BAHAMAS galaxy population. This difference fully accounts for the $0.5 \dex$ horizontal offset between the MICE and BAHAMAS curves in Fig.~\ref{fig:RAR_kids_mice_bahamas}. The failure of BAHAMAS to reproduce the observed lensing RAR could therefore be regarded as a possible shortcoming of the galaxy formation model used in those simulations, rather than a general failure of their cosmological paradigm. However, we note that the offset in the SHMR as a function of local galaxy density is theoretically expected, and (indirectly) observed \cite[e.g.][]{dutton2010,correa2020}. It is therefore curious that MICE, which does not reproduce this observed bias, turns out to be in reasonable agreement with our measurements. The discrepancy between KiDS-bright and BAHAMAS must therefore arise due to some more subtle underlying reason that we have yet to identify; we hope to follow this up in future work. We initially selected BAHAMAS for our analysis due to its large volume -- required to produce enough of the rare isolated, relatively massive galaxies of interest -- and readily available mock lensing data. It will be interesting to revisit the lensing RAR as cosmological hydrodynamical galaxy formation simulations continue to improve in terms of realism, simulated volume, and resolution.

\subsection{The RAR for early- and late-type KiDS galaxies}
\label{sec:Results-Types}

\begin{figure}
	\resizebox{\hsize}{!}{\includegraphics{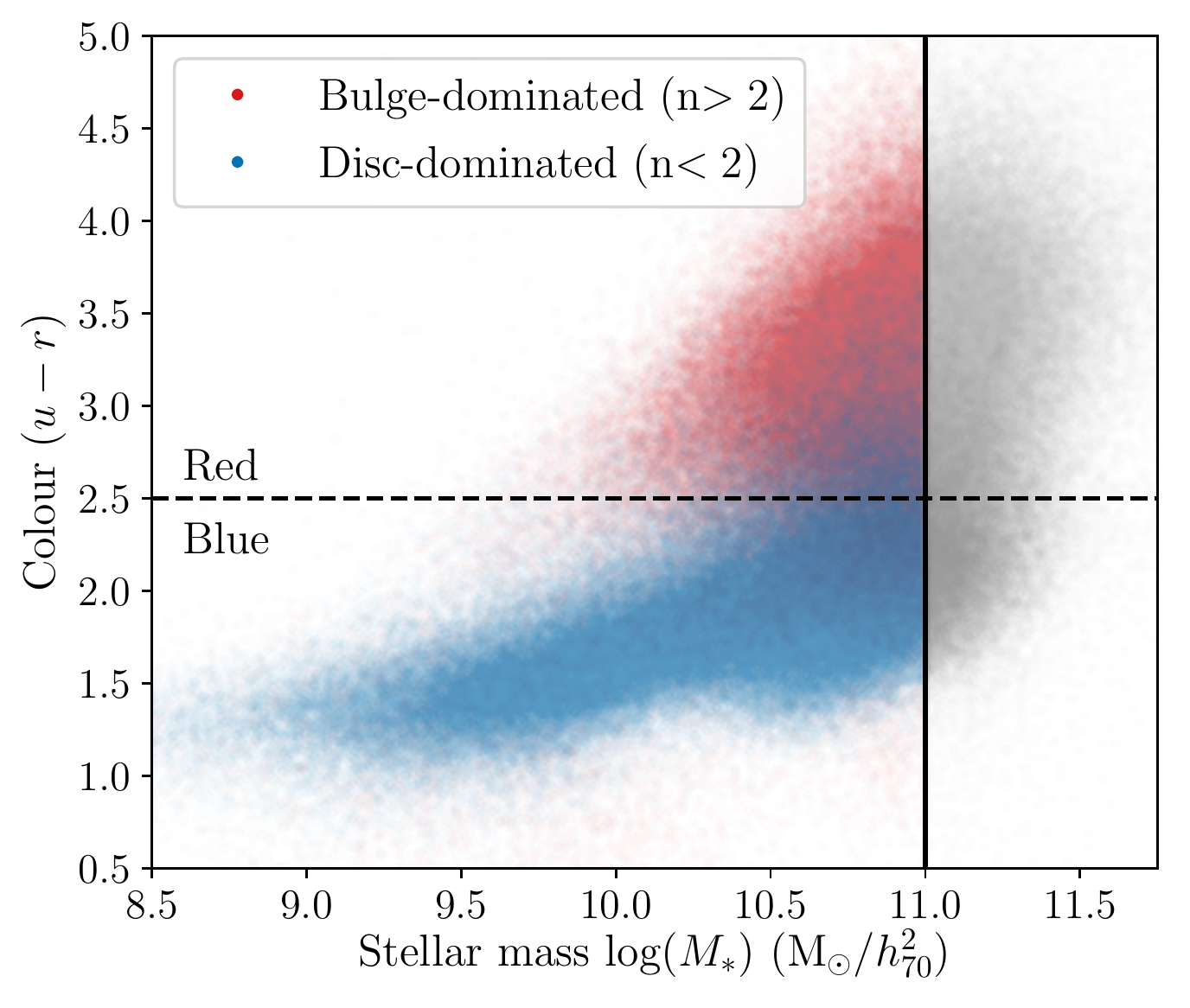}}
	\caption{2D histogram of the $u-r$ colour and stellar mass of isolated KiDS-bright galaxies. We divide our galaxies into canonically early- and late-type galaxies, based on either S\'ersic index $n$ or $u-r$ magnitude. When dividing by S\'ersic index, we define bulge-dominated (early-type) galaxies as those with $n>2$ and disc-dominated (late-type) galaxies as those with $n<2$ (red and blue points). When dividing colour we define red (early-type) galaxies as those with $m\un{u} - m\un{r} > 2.5$ and blue (late-type) galaxies as those with $m\un{u} - m\un{r} < 2.5$ (above and below the dashed horizontal line).}
	\label{fig:galtypes_scatterplot}
\end{figure}

\begin{figure}
	\resizebox{\hsize}{!}{\includegraphics{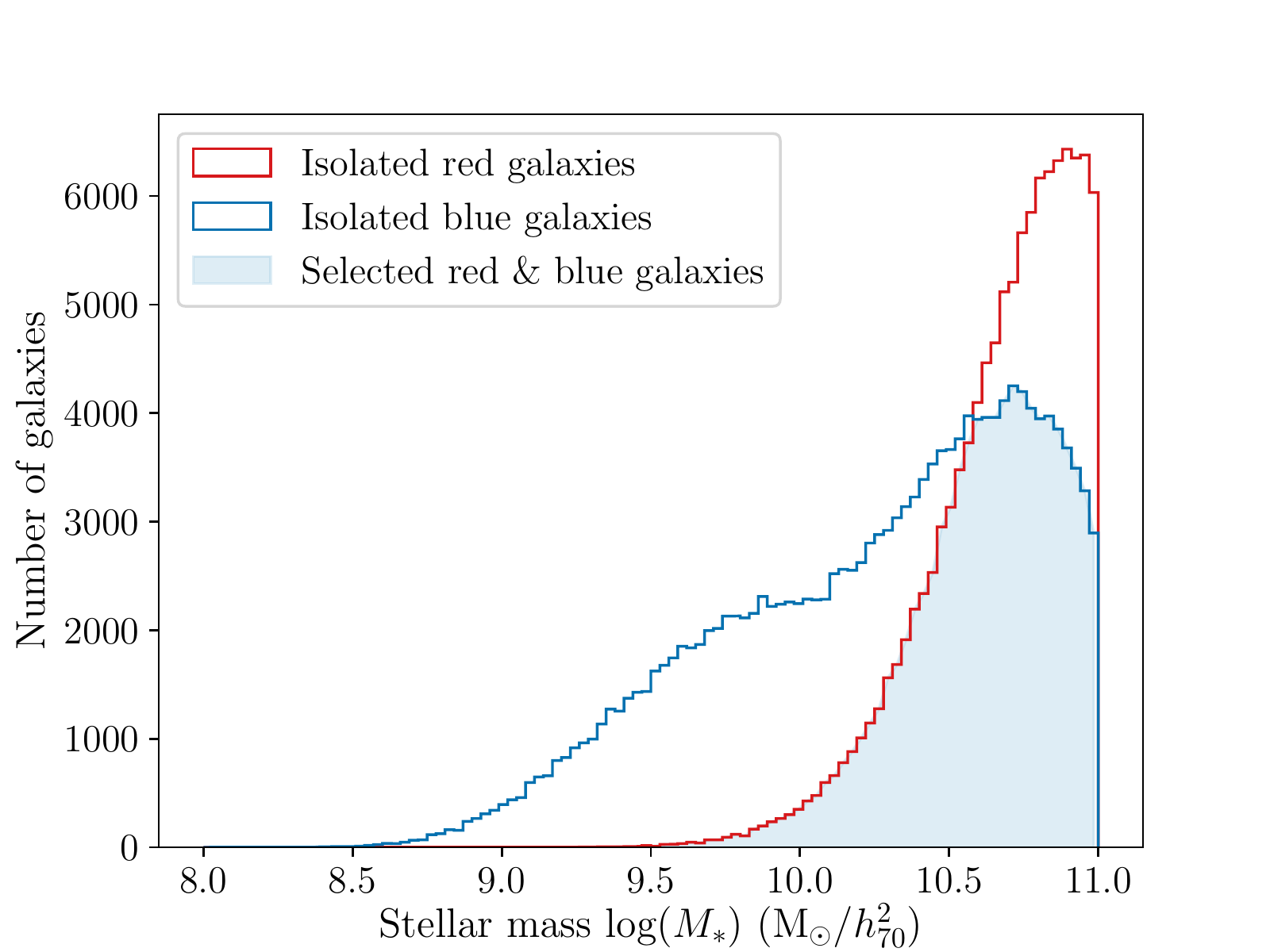}}
	\caption{Stellar mass histogram of the red (early-type) and blue (late-type) isolated KiDS-bright galaxies (red and blue lines), divided by $u-r$ colour ($m\un{u} - m\un{r} \lessgtr 2.5 \magn$). To isolate the effect of galaxy type on the RAR from that of $M_\star$, we select two samples with the same stellar mass distribution by randomly removing galaxies from both samples until only the overlapping region (light blue shaded region) remains.}
	\label{fig:galtypes_masshist}
\end{figure}

\begin{figure*}
	\centering
	\includegraphics[width=17cm]{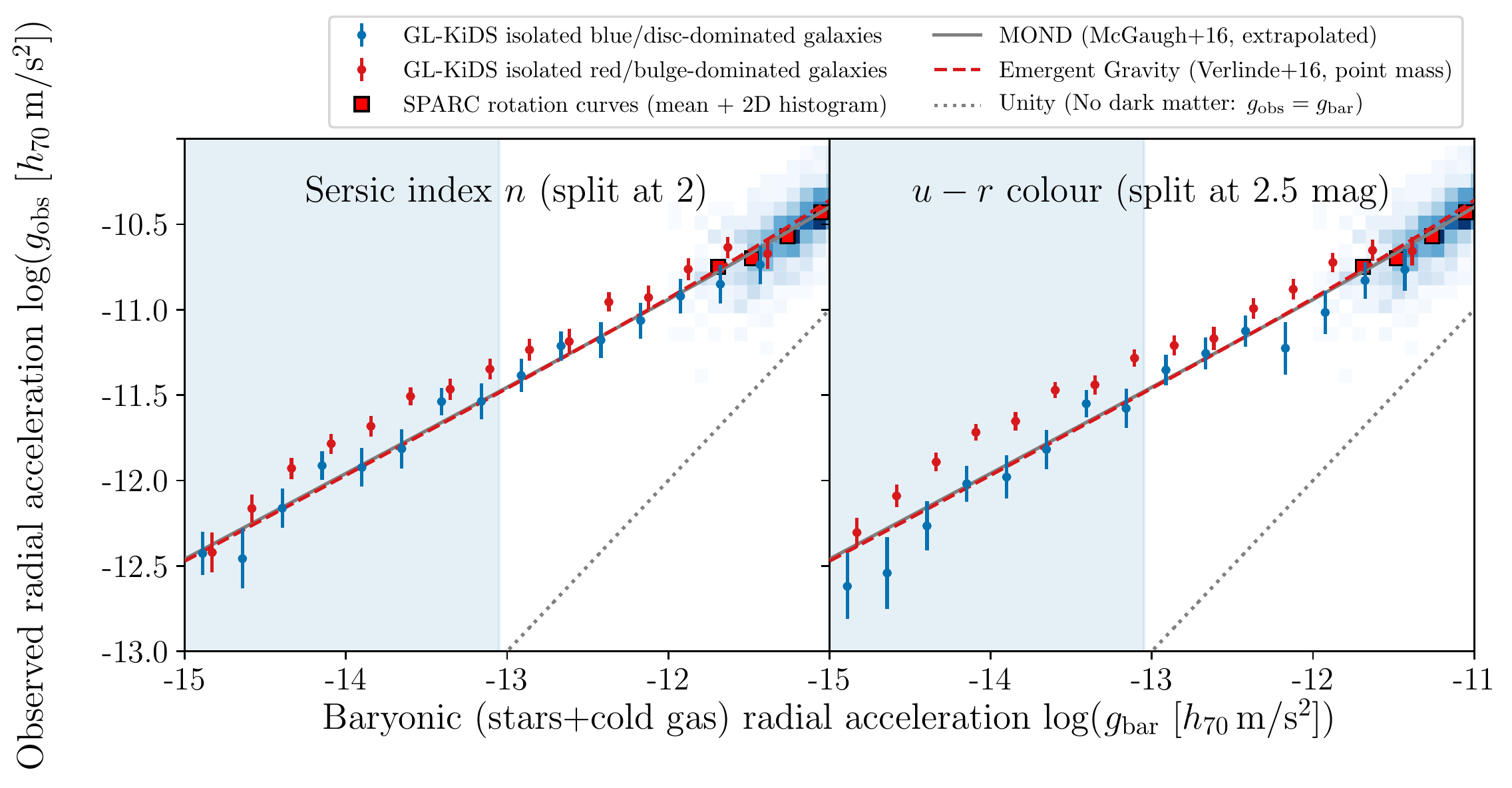}
	\caption{Measured RAR of the KiDS-bright  isolated lenses (points with $1\sigma$ error bars) divided into canonically early- and late-type galaxies. In the left panel, the lenses are split by S\'ersic index ($n\gtrless2$) into bulge-dominated (red points) and disc-dominated (blue points) galaxies. In the right panel they are split by $u-r$ colour ($m\un{u} - m\un{r} \gtrless 2.5$) into red and blue galaxies (with correspondingly coloured points). In both panels we find a significant difference between the RAR measurements of early and late galaxy types. The extrapolated MOND and EG predictions (grey solid and red dashed lines) and the SPARC data (red squares with 2D histogram) are shown as a reference.}
	\label{fig:RAR_kids_galtypebins}
\end{figure*}

The large size of the KiDS-bright lens sample gives us the opportunity to divide our lenses into different samples based on observed galaxy parameters. We determined the RAR for isolated galaxies split into two types based on either parameter: bulge-dominated and disc-dominated based on their S\'ersic index, and red and blue based on their $u-r$ colour. Although these selections are far from perfect representations of true morphological types, the red and bulge-dominated samples can roughly be identified with canonically early-type (pressure supported) galaxies and the blue and disc-dominated samples with late-type (rotationally supported) galaxies \cite[]{driver2006}\footnote{In general, the S\'ersic index $n$ does not separate early- and late-type galaxies because dwarf early- and late-type galaxies have similar values of $n$ \cite[]{graham2019}. However, dwarf early-type galaxies are not abundant in isolation \cite[]{janz2017}, which means that our isolated low-$n$ galaxy sample likely consists of late-type galaxies.}.

The $r$-band S\'ersic indices $n$ of all KiDS galaxies with $S/N > 50$ \citep[following][]{roy2018} were measured using the \textsc{2DPHOT} multi-purpose environment for 2D wide-field image analysis \mbox{\cite[]{barbera2008}}. For the colour split, we used the $u$ and $r$ magnitudes measured using the \textsc{GAaP} pipeline (see Section~\ref{sec:KiDS}). In Fig.~\ref{fig:galtypes_scatterplot} the $u-r$ colour versus stellar mass distribution of isolated galaxies shows the split based on S\'ersic index, which defines early-type galaxies as those with $n > 2$ and late-type disc-dominated galaxies as those with $n<2$. Based on the $u-r$ magnitude distribution of these two populations, we defined our split by galaxy colour as follows: galaxies with $m\un{u} - m\un{r} > 2.5 \magn$ are defined as red, and those with $m\un{u} - m\un{r} < 2.5 \magn$ as blue.

In both cases, we aimed to select two samples with the same stellar mass distribution, in order to isolate any possible effect of galaxy type on the RAR from that of $M_\star$. In Fig.~\ref{fig:galtypes_masshist} we show the $M_\star$ histogram of the two types (in this case based on galaxy colour). From both samples, we removed galaxies until only the overlapping section of both mass distributions remained. Ideally this should give us two samples (red and blue galaxies) with equal stellar mass distributions, shown by the light shaded blue region.

Fig.~\ref{fig:RAR_kids_galtypebins} shows the lensing RAR of equal-mass KiDS-bright galaxies split by S\'ersic index (left panel) and $u-r$ colour (right panel). For this result, we focus on establishing whether there exists a significant difference between the RAR of the two types. Contrary to previous plots, the effect of a $0.2 \dex$ global systematic bias in $M_\star$ (normally shown by a grey band) is omitted because this affects both measurements in the same way such that their relative difference does not change (the possibility of a colour- or S\'ersic index-dependent $M_\star$ bias is discussed below).

We indeed observe a significant difference between the RAR measurements of early and late galaxy types. To quantify this difference, we measured the reduced $\chi^2$ between the RAR measurements by replacing $g\un{obs}$ and $g\un{mod}$ in Eq. \ref{eq:chi2} with $g\un{obs,E}$ and $g\un{obs,L}$ from the early-type (red or bulge-dominated) and late-type (blue or disc-dominated) galaxy samples. The $\chi\un{red}^2$ equals $67.8 / 15 = 4.5$ for the lenses split by S\'ersic index, and $134.2 / 15 = 8.9$ for those split by $u-r$ colour. Taking the full covariance matrix into account we find that even the S\'ersic index split, which displays the smallest offset, results in RAR difference with a $5.7 \sigma$ significance. The mean ratio between the RAR measurements of the two types, $\log_{10}(\delta g\un{obs}^{\rm E/L}) = \log_{10}\left( \langle g\un{obs,E} / g\un{obs,L}\rangle \right)$, is $0.17 \dex$ and $0.27 \dex$ for the S\'ersic and colour splits respectively.

We address the question whether the observed difference of the RAR between early and late types could be caused by any bias in the stellar mass. To this end, we estimated the systematic stellar mass bias between the two types, defined as $\log_{10}(\delta M_\star^{\rm E/L}) = \log_{10}\left( \langle M_\star\rangle\un{E} / \langle M_\star\rangle\un{L} \right)$, that would be required to resolve the difference between their two RAR measurements. When trying to estimate the effect of this bias on the RAR, we had to take into account that $\delta M^{\rm E/L}_\star$ affects both the estimated acceleration from baryonic mass $g\un{bar}$ (directly) and the observed acceleration $g\un{obs}$ (indirectly, through the equal-mass selection). The bias in baryonic acceleration scales linearly with the bias in $M_\star$, such that: $\log_{10}(\delta g^{\rm E/L}\un{bar}) = \log_{10}(\delta M^{\rm E/L}_\star)$. Throughout this work, the observed relation between $g\un{bar}$ and $g\un{obs}$ at the scales measured by lensing has approximately followed $g\un{obs} \propto \sqrt{g\un{bar}}$. This means that we can roughly estimate the effect on $g\un{obs}$ as: $\log_{10}(\delta g^{\rm E/L}\un{obs}) \approx {\log_{10}(\delta M^{\rm E/L}_\star)} \, / \, {2}$. Since our measured difference $\delta g\un{obs} \gtrsim 0.2 \dex$, this means ${\log_{10}(\delta M^{\rm E/L}_\star)}$ should be $\gtrsim 2 \log_{10}(\delta g^{\rm E/L}\un{obs}) = 0.4 \dex$. That is, the observed difference could be resolved by a systematic stellar mass bias between the two types $\gtrsim 0.4 \dex$. We will now discuss different sources of a possible systematic bias, and estimate whether they could be the cause of the observed difference.

First, the statistical uncertainty in the $M_\star$ measurements could cause a systematic shift in the two $M_\star$ distributions resulting from Eddington bias \cite[]{eddington1913}. We estimated the size of this bias by adding a random offset to the true $\log_{10}(M_\star)$ measurements of KiDS-bright before selecting the two `equal' stellar mass distributions for red and blue galaxies. Based on our estimate of the statistical uncertainty in the KiDS-bright $M_\star$ (see Section~\ref{sec:KiDS-bright}), we drew the random offsets from a lognormal distribution with $\sigma = 0.12 \dex$. When looking at the underlying true stellar mass distributions we found that they are indeed not equal, but that the mean stellar masses $\lan M\un{\star,E} \ran$ and $\lan M\un{\star,L} \ran$ of the red and blue samples differ by only $0.025 \dex$. Of course, this method overlooks the fact that the measured $M_\star$ distribution already contains scatter, and is therefore not the true $M_\star$ distribution. Indeed when we apply the random offset multiple times, we see the Eddington bias decrease by $\sim5\%$ after every iteration. Therefore, the true Eddington bias is likely to be slightly larger, around $0.027 \dex$. This is still very small compared to the $\gtrsim 0.4 \dex$ bias needed, thus it is very unlikely that the difference we observe is caused exclusively by Eddington bias.

Second, there could be systematic errors in the KiDS-bright $M_\star$ measurements that differ between red and blue galaxies (due to e.g. systematic variation of the IMF, SPS model inaccuracies, or systematic errors in the measured redshifts or magnitudes). In order to estimate the size of any systematic biases in the stellar mass, we compared KiDS-bright's $M\un{\star,ANN}$ with GAMA's $M\un{\star,G}$ of exactly the same galaxies. Here $M\un{\star,ANN}$ is based on the nine-band KiDS+VIKING photometry and photometric redshifts $z\un{ANN}$ derived by training the \textsc{ANNz2} (Artificial Neural Network) machine learning method on the spectroscopic GAMA redshifts (see Section~\ref{sec:KiDS-bright}), while $M\un{\star,G}$ is based on the $ugrizZY$ SDSS+VIKING photometry combined with the spectroscopic GAMA redshifts (see Section~\ref{sec:GAMA}). After selecting our samples of blue and red galaxies with the same $M\un{\star,ANN}$ distribution as described above, we indeed found that the $M\un{\star,G}$ distributions are not exactly equal: $\langle M_\star \rangle\un{E} / \langle M_\star\rangle\un{L} = 1.4$, corresponding to $0.14 \dex$. This indicates that using different sets of observations and models to measure $M_\star$ can cause a systematic bias between red and blue galaxies, but that this effect is too small to reach the $\gtrsim 0.4 \dex$ difference in $M_\star$ needed to explain the $\gtrsim 0.2 \dex$ difference in the measured RAR.

In conclusion, even when combined the Eddington plus overall systematic measurement bias is at most $0.17 \dex$, not even half of what is needed. We note that this bias estimation has been carried out using the types split by $u-r$ colour; when split by S\'ersic index, the Eddington and other systematic biases between bulge- and disc-dominated galaxies are even smaller ($0.021$ and $0.12 \dex$ respectively).

\cite{sanchez2019} reported evidence of a varying IMF in massive early-type galaxies. As seen in fig.~19 of their work, this could cause the global mass-to-light-ratio of these galaxies to increase by as much as $0.09 \dex$ compared to a fixed Chabrier IMF. They find this effect only for their high-mass galaxy sample with a stellar mass of at least $M_\star > 2\times10^{11} \msun$, and not for their lower-mass sample. Since we limit all our galaxies to $M_\star < 10^{11} \hmsun$ (see Section~\ref{sec:KiDS-bright}), the varying IMF is not likely to apply to our early-type galaxy sample. However, even if this had been the case, this $0.09 \dex$ difference in $M_\star$ is small compared to the $\gtrsim 0.4 \dex$ needed to explain the difference in the RAR of early- and late-type galaxies.

The higher values of $g\un{obs}$ for red and bulge-dominated galaxies that we find in Fig.~\ref{fig:RAR_kids_galtypebins} are in qualitative agreement with earlier GGL studies. A recent KiDS-1000 lensing study by \cite{taylor2020} found that, within a narrow stellar mass range near the knee of the SHMR ($M_\star \sim 2-5\E{10} \hmsun$), galaxy halo mass varied with galaxy colour, specific star formation rate (SSFR), effective radius $R\un{e}$ and Sérsic index $n$. Although not explicitly mentioned, their figures~1 and 6 reveal that their early-type (red, low-SSFR) galaxies have larger halo masses than their late-type (blue, low-$n$, high-SSFR) galaxies of the same stellar mass. S\'ersic parameter coupling between $n$ and $R_{\rm e}$, for a fixed galaxy luminosity, may also contribute towards the trends seen among the early-type galaxies in their $M_{\rm halo}$--$n$ and $M_{\rm halo}$--$R_{\rm e}$ diagrams\footnote{The smaller average size for the early-type galaxies, compared to the late-type galaxies, is because of the different 3D-bulge-to-2D-disc ratios: a fixed stellar mass will fit into a smaller volume if distributed in a bulge rather than a disc.}. Much earlier \cite{hoekstra2005} measured the GGL signal of a sample of `isolated' Red-sequence Cluster Survey galaxies as a function of their rest-frame $B$-, $V$-, and $R$-band luminosity, and found that early-type galaxies have lower stellar mass fractions. In contrast, \mbox{\cite{mandelbaum2006}} found no dependence of the halo mass on morphology for a given stellar mass below $M_\star<10^{11} \msun$, although they did find a factor of two difference in halo mass between ellipticals and spirals at fixed luminosity.

Finding a significantly different RAR at equal $M_\star$ would have interesting implications for galaxy formation models in the $\lcdm$ framework. In the $\lcdm$ framework it is expected that the galaxy-to-halo-mass relation, and therefore the RAR, can be different for different galaxy types through their galaxy formation history \cite[]{dutton2010,matthee2017,posti2019,marasco2020}. Two parameters that correlate heavily with galaxy formation history are S\'ersic index and colour. 

Current MG theories do not predict any effect of galaxy morphological type on the RAR, at least on large scales. The MOND paradigm gives a fixed prediction for the relation between $g\un{bar}$ and $g\un{obs}$ given by Eq. \ref{eq:rar_mcgaugh}. Since the RAR is the observation of exactly this relation, in principle MOND gives a fixed prediction, independent of any galaxy characteristic. As discussed in Section~\ref{sec:MOND}, the main exception is the EFE that could be caused by neighbouring mass distributions. However, Fig.~\ref{fig:RAR_kids_gama_verlinde} shows that an increase in the EFE only predicts an increase in steepness of the downward RAR slope at low accelerations ($g\un{bar} < 10^{-12} \mpss$), while the observed RAR of both early- and late-type galaxies follow approximately the same slope across all measured accelerations. It is therefore unlikely that their amplitude difference can be explained through the EFE.

We will next discuss whether the observed difference in RAR between early and late types is at odds with EG, but first emphasise three caveats of this discussion.

First, the derivation of the EG formalism assumes a spherical mass distribution. Solutions for non-spherical systems do not exist yet. It is not excluded that solutions for large-scale triaxial ellipticals will differ from rotationally supported spiral galaxies. This requires further theoretical study.

Second, the current EG theory predicts ADM fields based exclusively on the static baryonic mass distribution, although very large-scale dynamics can potentially influence the excess gravitational force predicted by EG. It is unknown whether large-scale pressure supported (virialised) systems create an ADM distribution similar to that of rotationally supported galaxies.

Third, we assume here that, to first order, the uncertainty in the KiDS photometric redshifts affects the isolated galaxy selection of both galaxy types in the same way, allowing us to include the full acceleration range into our comparison. However, the well established morphology-density relation predicts a higher density of satellite and dwarf galaxies around early-type galaxies compared to the late types \cite[]{dressler1980,goto2003}, although we have minimised this effect by selecting isolated galaxies (see Appendix~\ref{app:Isolation}). It is not yet known whether and, if so, how these external gravitational fields affect the EG prediction.

To address this last caveat, the light blue shaded region in Fig.~\ref{fig:RAR_kids_galtypebins} shows the acceleration scales beyond the KiDS isolation criterion limit ($g\un{bar}<10^{-13}\mpss$), where the presence of satellites might play a role (see Appendix~\ref{app:Isolation}). But even when we remove all data points inside this region, we obtain a difference $\log_{10}(\delta g\un{obs}^{\rm E/L})$ of $0.14 \dex$ and $0.19 \dex$ for the S\'ersic and colour split respectively, where the latter has a significance of $3.2 \sigma$. Therefore, even at the scales where isolation is certain (corresponding to $R<0.3 \hMpc$), the difference remains significant.

To evaluate the possible effect of circumgalactic hot gas, we computed the RAR of early and late-type isolated galaxies (of the same stellar mass) while including a rough estimate of the hot gas contribution to $g\un{bar}$. We used the same model of the nominal hot gas distribution around our galaxies as discussed in Sect. \ref{sec:Results-MG_theories}: an isothermal halo within $100 \hkpc$, with a mass $M\un{gas} = M_\star$. When applying the same hot gas model to both early- and late-type galaxies, we find that there remains a $>6 \sigma$ difference between their RARs, both for the split by S\'ersic index and $u-r$ colour. However, for this particular gas model, we find that $g\un{bar}$ increases in such a way that the RAR of early-type galaxies moves to the right, close to the MG predictions where the RAR of late-type galaxies without circumgalactic gas resides. This means that, in the specific case where early-type galaxies have gaseous haloes with $M\un{gas} = M_\star$ while late-type galaxies (of the same stellar mass) have negligible hot circumstellar gas, this would reduce the difference in their RARs to $\sim4\sigma$. Fine-tuning the $M\un{gas}/M_\star$ ratio of early-type galaxies to a slightly higher value, while keeping $M\un{gas}/M_\star\approx0$ for late types, might remove the difference between their RARs. However, as discussed in Sect. \ref{sec:Results-MG_theories}, unbiased X-ray surveys of circumgalactic gas around isolated galaxies are still lacking, which makes it difficult to obtain representative observational data.

In conclusion, unless early-type galaxies have significant circumgalactic gaseous haloes while late types (of the same stellar mass) do not, the difference we find in the RARs of different galaxy types might prove difficult to explain within MG frameworks. In MOND, $g\un{bar}$ and $g\un{obs}$ should be directly linked through Eq.~\ref{eq:rar_mcgaugh} without any dependence on galaxy type. In EG the effect might be a consequence of yet unexplored aspects of the theory, such as a non-symmetric mass distribution or the effect of large-scale dynamics. To explore whether this is the case, however, more theoretical work is needed. Through the derivative in Eq.~\ref{eq:eg_mdm}, EG does include a dependence on the slope of the baryonic density distribution. A shallower slope of $M\un{bar}(r)$ increases $M\un{ADM}$ and thus $g\un{obs}$, which might solve the current tension if early-type  galaxies have significantly shallower baryonic mass distributions that extend far beyond $30 \hkpc$, such as gaseous haloes (although \citealp{brouwer2017} did not find evidence for a significant effect of the baryonic mass distribution on the EG prediction; see their section~4.3). In addition, EG is currently only formulated for spherically symmetric systems. It would be interesting to investigate whether discs and spheroidal galaxies yield different predictions, and whether these differences would extend beyond $30 \hkpc$.

In a $\lcdm$ context, our findings would point to a difference in the SHMR for different galaxy types. Recently \cite{correa2020} used SDSS data with morphological classifications from Galaxy Zoo to find that, at fixed halo mass (in the range $10^{11.7} - 10^{12.9} \msun$), the median stellar mass of SDSS disc galaxies was a factor of $1.4$ higher than that of ellipticals. They found this to be in agreement with the EAGLE simulations, where haloes hosting disc galaxies are assembled earlier than those hosting ellipticals, therefore having more time for gas accretion and star formation.

\subsection{The RAR as a function of stellar mass}
\label{sec:Results-Stellar_mass}

\begin{figure*}
	\centering
	\includegraphics[width=17cm]{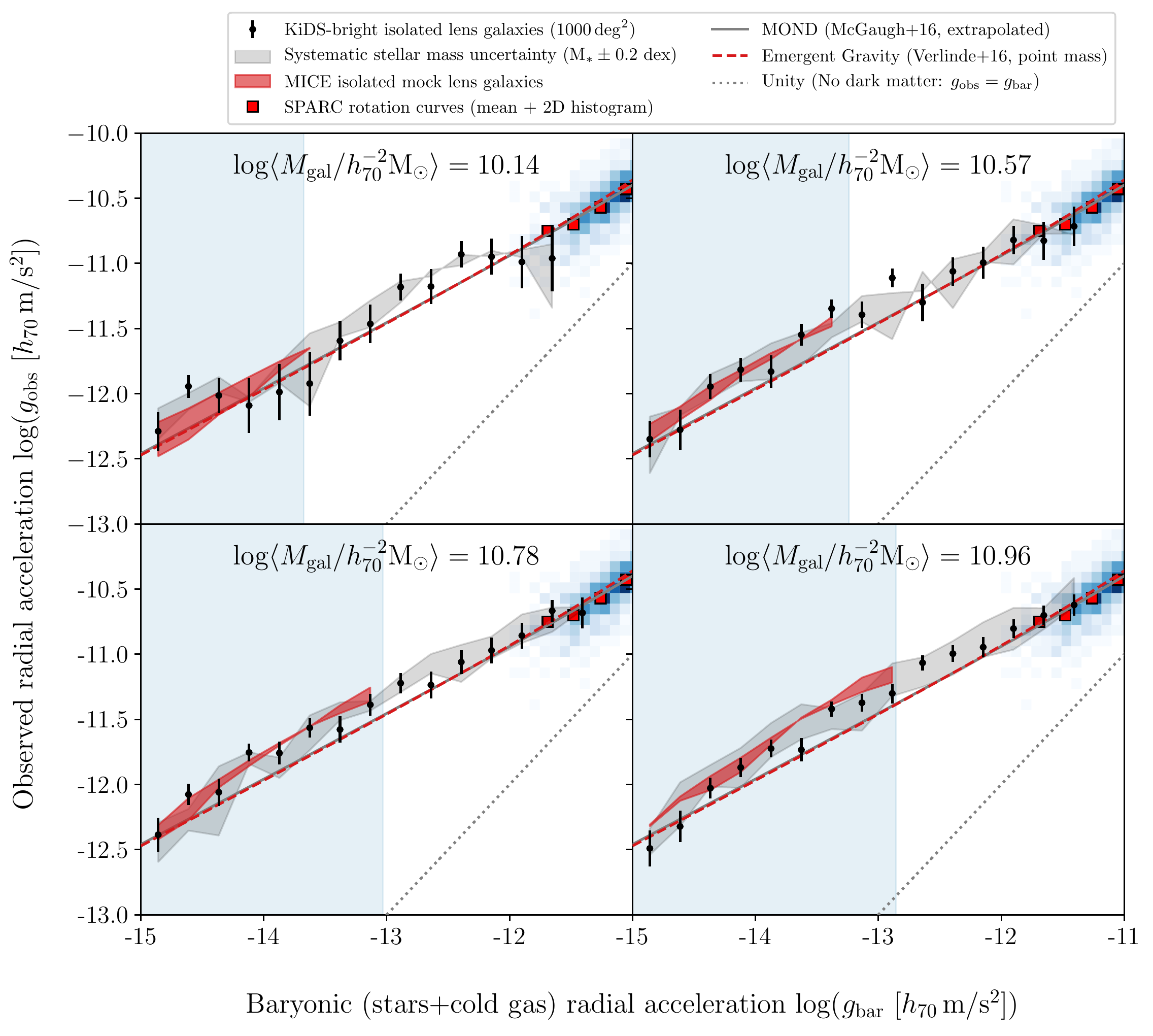}
	\caption{Measured RAR of isolated KiDS-bright lenses (black points with $1\sigma$ error bars) divided into four stellar mass bins. The mean galaxy mass (stars+cold gas) of the lenses is shown at the top of each panel. At increasing stellar mass, the measurements seem to rise above the predictions from MOND (grey solid line) and EG (red dashed line). However, at scales larger than $R > 0.3 \hMpc$ (light blue shaded region) this could be caused by false positives in the isolated galaxy sample due to the KiDS-bright redshift uncertainty.}
	\label{fig:RAR_kids_mice_mstarbins}
\end{figure*}

In addition to splitting by galaxy type, it is interesting to create the RAR for galaxy samples with different stellar mass $M_\star$ (including very low-mass galaxies, `dwarfs', in Section~\ref{sec:Results-Dwarfs}). In the $\lcdm$ paradigm, where baryonic and dark matter are described as separate substances, there can in theory be a difference in the SHMR depending on galaxy observables such as stellar mass, which could cause a shift in the measured RAR. This is in contrast with most MG models, which predict a fixed RAR (as is the case for MOND, and for EG at scales beyond the galaxy disc). In this section, we separated our isolated KiDS-bright lenses into four samples based on $M_\star$. We selected our $M_\star$-bins to obtain a similar $S/N$ ratio of the lensing signal in each bin, resulting in the following limits: $\log_{10}(M_\star/\hmsun) = [8.5,10.3,10.6,10.8,11.0]$. 

Fig.~\ref{fig:RAR_kids_mice_mstarbins} shows the lensing measurements and predictions for isolated galaxies split in four stellar mass bins. For each bin the mean galaxy mass (stars+cold gas) of the lenses, $\log_{10}\lan M\un{gal}/\hmsun \ran = [10.14, 10.57, 10.78, 10.96]$, is shown at the top of the panel. Quantifying the difference between MOND (the extended M16 fitting function) and our measurement at all scales results in: $\chi\un{red}^2 = 117.0 / 60 = 1.9$, which (noting that the prediction for EG is very similar) excludes both models at the $\sim4.5 \sigma$ level. This result should be taken with caution, however, as at accelerations $g\un{bar}$ that correspond to scales larger than $R > 0.3 \hMpc$ (light blue shaded region) an increasing signal is to be expected since at these distances satellite galaxies missed by our isolation criterion might affect the measurement. Galaxies with higher stellar masses reside in denser neighbourhoods, and therefore tend to have more satellites \cite[see e.g.][]{baldry2006, bolzonella2010, brouwer2016}.

The reduced $\chi^2$ values using only the data within $R < 0.3 \hMpc$ are $\chi\un{red}^2 = 49.9 / 31 = 1.6$ for MOND and $51.7 / 31 = 1.7$ for EG respectively (corresponding to a standard deviation of $2.4$ and $2.5\sigma$). Considering the stellar mass uncertainty ($\Updelta M_\star=\pm0.2 \dex$), which, if it acts to reduce the observed RAR, results in $\chi\un{red}^2 = 0.97$ for the extended M16 fitting function (with similar results for EG): a good fit. If the stellar mass uncertainty increases the observed RAR, we find $\chi\un{red}^2=4.6$: a poor fit. This again highlights the grave importance of accurate baryonic mass measurements in determining the RAR, in addition to deep lensing surveys that can detect satellites down to very faint magnitudes. This could be achieved by future cosmology telescopes such as \emph{Euclid} \cite[]{laureijs2011} and \emph{The Vera C. Rubin Observatory}, previously called Large Synoptic Survey Telescope \cite[LSST;][]{lsst2012}. As for the MICE simulation, it matches our measurements reasonably well in every $M_\star$ bin. For the result that includes the photometric redshift uncertainty $\sigma\un{z}$ in the isolated galaxy selection, we find $\chi\un{red}^2 = 49.7 / 30 = 1.7$ ($2.5 \sigma$).

\subsection{The RAR of low-mass (dwarf) late-type galaxies}
\label{sec:Results-Dwarfs}

\begin{figure*}
	\centering
	\includegraphics[width=17cm]{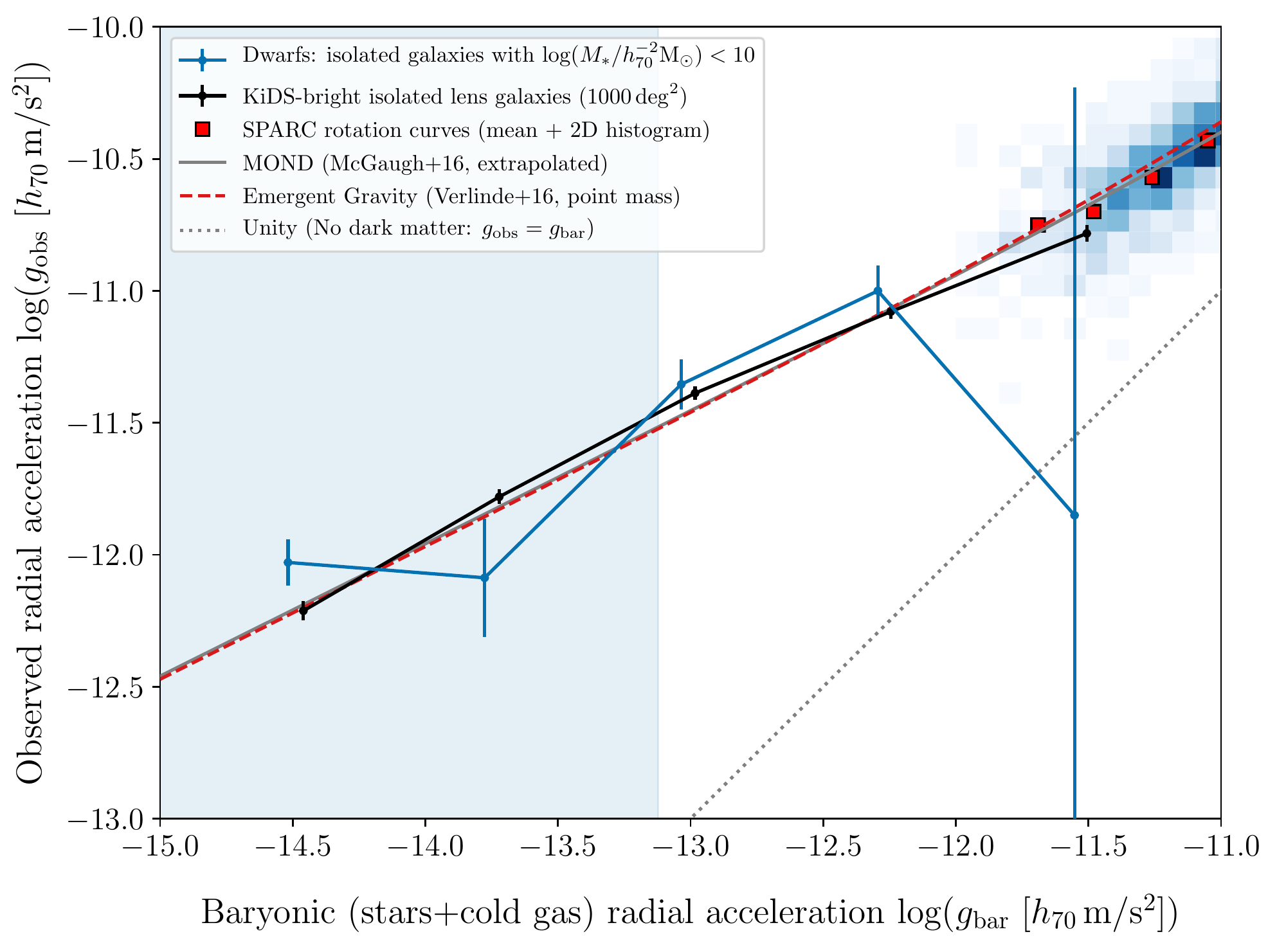}
	\caption{Measured RAR of KiDS-bright lenses (points with $1\sigma$ error bars), respectively for isolated dwarfs ($\log(M_\star/\hmsun)<10$, blue) and the full isolated galaxy sample ($\log(M_\star/\hmsun) < 11$, black). Due to the low $S/N$ ratio of the dwarf lensing signal, the number of $g\un{bar}$-bins is reduced from $15$ to $5$. We find that the RAR of dwarfs is consistent with that of our regular sample, and with the extrapolated MOND and EG predictions (grey solid and red dashed lines), which are shown as a reference.}
	\label{fig:RAR_kids_dwarfs}
\end{figure*}

As a final exploration of different galaxy masses, we attempt to measure the RAR for the lightest lenses in KiDS-bright. Low-mass galaxies are of particular interest to DM and MG researchers as extreme examples that might show eccentric behaviour \citep[e.g.][]{oman2016,dokkum2018,guo2020}, as well as those who attempt to extend the RAR to lower accelerations using galaxy rotation curves \citep{lelli2017b,paolo2019}. We therefore select a sample of dwarfs: isolated galaxies with a stellar mass $M_\star<10^{10} \hmsun$ (whereas the full sample of isolated galaxies has $M_\star < 10^{11} \hmsun$, see Section~\ref{sec:KiDS-bright}). As can be seen in Fig.~\ref{fig:galtypes_scatterplot}, this sample is dominated by blue, disc-dominated galaxies based on their colours and S\'ersic indices ($m\un{u} - m\un{r} > 2.5 \magn$ and $n<2$), which means they are likely to be late-type. Since these galaxies are few, and have an even smaller effect on the path of light rays than more massive ones, we needed to reduce the number of bins in $g\un{bar}$ from $15$ to $5$ to obtain sufficient $S/N$ radio in each bin. Fig.~\ref{fig:RAR_kids_dwarfs} shows the resulting RAR measurement of dwarfs compared to the full isolated sample. We do not show the effect of the $\Updelta M_\star=\pm0.2\dex$ systematic uncertainty because this would affect both results in the same way. We find that, within its large error bars, the RAR of the dwarfs is consistent with that of the full isolated sample; they both approximately follow the $g\un{obs}\propto\sqrt{g\un{bar}}$ relation expected by the extended MOND and EG predictions, which are shown as a reference. Hence, we do not find a significant difference in the RAR of dwarf galaxies.

\section{Discussion and conclusions}
\label{sec:Discussion_Conclusion}

Galaxy-galaxy lensing observations from the fourth data release of the Kilo Degree Survey (KiDS-1000) have extended the RAR of isolated galaxies by nearly $2$ orders of magnitude in gravitational acceleration $g\un{obs}$, compared to previous measurements based on rotation curves \cite[most notably][M16]{mcgaugh2016}. To compute the lensing RAR, we converted our ESD profiles $\Updelta\Upsigma(R)$ into the observed gravitational acceleration $g\un{obs}$, and our galaxy masses (measured using nine-band KiDS+VIKING photometry) into $g\un{bar}$. These measurements allowed us to perform unprecedented tests of two MG models: MOND and EG, as well as tests of DM using the MICE (N-body + semi-analytic) and BAHAMAS (hydrodynamical) simulations. Our conclusions from these observational tests are as follows:

\begin{itemize}
	
	\item Fig.~\ref{fig:Vrot_kids}: We find that lensing rotation curves of isolated galaxies, as inferred from GGL measurements, remain approximately flat at scales far beyond the visible disc ($0.03 < R < 3 \hMpc$). At the accelerations corresponding to the outskirts of observable galaxies ($R \approx 30\hkpc$), our lensing results are in excellent agreement with the SPARC rotation curves \mbox{\cite[]{lelli2016b}}. These two measurements are obtained by two very different methods, providing independent corroboration of each result.
	
	\item Fig.~\ref{fig:RAR_kids_gama_verlinde}: At the low accelerations corresponding to GGL scales, the lensing RAR of isolated galaxies approximately follows a $g\un{obs} \propto \sqrt{g\un{bar}}$ relation. This is in agreement with the expectations from EG (Eq. \ref{eq:rar_verlinde}) and MOND (which we take to be the M16 fitting function, Eq. \ref{eq:rar_mcgaugh}, extrapolated to larger scales). At low accelerations both these models predict a direct relation between observed and baryonic acceleration of this form, with a very similar proportionality constant\footnote{The proportionality constant $c H_0 / 6$ in EG is almost equal to the value of $g\un{\dagger}$ found by M16, which is again equal to the $a_0=1.2\E{-10} \mpss$ canonical in MOND.} of $\sim1.2\E{-10} \, \mpss$. This reinforces the results of \cite{brouwer2017}, who found that EG provides a good description of ESD profiles measured using $180 \deg^2$ of KiDS-GAMA data, but with a five times larger survey area. However, this result only remains valid if no massive ($M\un{gas} \gtrsim M_\star$) extended baryon distributions, such as as-yet undetected gaseous haloes, are common around our isolated lens galaxies.
	
	\item Fig.~\ref{fig:RAR_kids_mice_bahamas}: We find that the BAHAMAS simulation underestimates our KiDS-bright lensing RAR. The discrepancy relative to MICE is caused by a bias in the stellar-to-halo-mass-relation (SHMR) of isolated galaxies in BAHAMAS, which is absent in MICE: BAHAMAS galaxies have stellar masses typically three times higher at fixed halo mass than their non-isolated counterparts. Determining which of the two models more accurately captures the true SHMR is clearly crucial to the interpretation of our measurements in the $\lcdm$ context. Interestingly, the BAHAMAS RAR still has approximately the correct low-acceleration slope, rather than a steeper slope as would naively be predicted based on the $\rho\propto r^{-3}$ outer slopes of the simulated DM haloes. The prediction from MICE (only feasible at low accelerations due to the limited resolution of the simulated lensing measurements) matches our RAR measurements very well.
	
	\item The additional lensing power at large radii with respect to the prediction from \citet[][see Appendix~\ref{app:Results_N17}]{navarro2017} might be caused by large-scale structure along the line-of-sight to the source, in spite of our efforts to select isolated galaxies. This highlights the crucial importance of simulating the entire measurement process (where possible) when making theoretical predictions, both in $\lcdm$ and MG, before they can be ruled out. In addition, the need for accurate isolated galaxy selection highlights the importance of large spectroscopic surveys, such as the upcoming 4MOST  \cite[]{Rdejong2019} and Dark Energy Spectroscopic Instrument \cite[DESI;][]{ruizmacias2020} surveys.
	
	\item Fig.~\ref{fig:RAR_kids_galtypebins}: When we split galaxies into two types based on S\'ersic index or $u-r$ colour, we find at least a factor of $1.5$ ($\simeq0.2 \dex$) difference between the respective lensing RAR measurements with a significance of at least $5.7 \sigma$. This observed difference could be resolved by a $\gtrsim 0.4 \dex$ systematic bias between the stellar masses of the two types. However, we calculated that the expected $M_\star$ bias (due to Eddington bias or systematic biases in the $M_\star$ measurement) is at most $0.17 \dex$. This variation in the RAR based on galaxy type, which is in agreement with \cite{taylor2020} and \cite{correa2020}, could be difficult to explain for MG models that predict a fixed relation between baryonic mass and the total gravitational potential.
	
	\item Fig.~\ref{fig:RAR_kids_mice_mstarbins}: The lensing RAR for galaxy samples split by stellar mass $M_\star$ demonstrated a slight upward trend, away from the fixed predictions of MOND and EG, with increasing $M_\star$. This could be caused by satellite or companion galaxies missed by the isolated galaxy selection due to the KiDS-bright redshift uncertainty, however. With the inclusion of the KiDS isolation criterion limit and accounting for uncertainty in the stellar mass, we find a reasonable agreement between the MG models and observations. This highlights the crucial importance of accurate baryonic mass measurements in determining the RAR, in addition to deep lensing surveys that can detect satellites to down to very faint magnitudes (such as the future \emph{Euclid} space telescope and \emph{Vera C. Rubin Observatory}). The MICE prediction, which is corrected for the KiDS-bright redshift uncertainty, again matches well to our data.
	
	\item Fig.~\ref{fig:RAR_kids_dwarfs}: We find no significantly different RAR, relative to the entire isolated lens sample, for a subsample of the lightest KiDS-bright lenses: isolated dwarf ($M_\star < 10^{10} \hmsun$) galaxies.

	\item Throughout this work, we find that the field of GGL has reached a level of accuracy in the measurement of $g\un{obs}$ greater than that of the baryonic acceleration $g\un{bar}$. The fact that we have no accurate measurements of the additional hot gas at large radii, and the ambiguity around the cosmological missing baryons, forces us to limit $g\un{bar}$ to the contributions of stars and cold gas. In addition, the current $0.2 \dex$ systematic uncertainty in $M_\star$ prevents us from definitively excluding any of the models we test. This shows that, if we want to have any hope of testing DM and MG models using the next generation of cosmological lensing surveys (such as \emph{Euclid} and LSST), we also need to focus on the models and observations needed to accurately measure the baryonic mass distribution in and around galaxies.
\end{itemize}

We find that galaxy lensing rotation curves continue approximately flat out to $R=3\hMpc$ (where observations are bound to encounter lensing due to surrounding galaxies), which is difficult to explain in a $\lcdm$ framework that predicts simple NFW-like haloes because of their $r^{-3}$ outer slope (see the N17 model in Appendix~\ref{app:Results_N17}). However, our analysis of the MICE and BAHAMAS simulations shows that the combination of the lenses and the additional structure along the line-of-sight can yield an ESD profile consistent with an $\sim r^{-2}$ density profile for isolated galaxies, even though the lenses have an intrinsic $\sim r^{-3}$ outer profile.

Throughout our analysis we find that the extrapolated M16 fitting function (Eq. \ref{eq:rar_mcgaugh}), which approximately corresponds to the prediction of both MG models (EG and MOND), holds to scales of $3 \hMpc$ for isolated galaxies. A fundamental limitation of this measurement is that the additional diffuse gas surrounding galaxies remains difficult to measure, and has therefore not been included in most of this study. By implementing a rough order of magnitude estimate of the hot gas contribution to $g\un{bar}$, an isothermal distribution with $M\un{gas} = M_\star$ within $100 \hkpc$, we found that this causes an overall downward shift of the RAR and a steeper downward slope at very low accelerations (see Fig.~\ref{fig:RAR_kids_gama_verlinde}, and also Fig.~\ref{fig:missing-baryons} for a broader discussion of missing baryons). Although the MOND external field effect (EFE) causes a similar steepening of the RAR, we find that the idealised EFE fitting function of \cite{chae2020} is not steep enough the explain the effect of gaseous haloes. Therefore, a convincing detection of additional gaseous components with a nominal mass of $M\un{gas} \gtrsim M_\star$ would move the observed RAR away from the MG predictions ($g\un{bar} \propto \sqrt{g\un{obs}}$) at very low accelerations ($g\un{bar}<10^{-13} \mpss$) and towards the DM predictions (where $g\un{bar}$ and $g\un{obs}$ are independent). A robust non-detection of such massive gaseous haloes in general would likely strengthen the position of MG models. Finding them for early-type galaxies only would reduce the difference between the RAR of early- and late-type galaxies, which otherwise remains unexplained in MG frameworks.

In conclusion, we find that the lensing RAR is a promising method to be used by future cosmological surveys to distinguish between MG and DM models. This can be done by measuring the RAR including large-scale baryonic mass observations; by simply performing the same comparison with even more accurate lensing and stellar mass measurements; or by further exploring the offset that we have found between the RARs of different galaxy types. All these options require that systematic biases in the stellar and other baryonic mass measurements be reduced.

\section*{Acknowledgements}

Based on observations made with ESO Telescopes at the La Silla Paranal Observatory under programme IDs 177.A-3016, 177.A-3017, 177.A-3018 and 179.A-2004, and on data products produced by the KiDS consortium. The KiDS production team acknowledges support from: Deutsche Forschungsgemeinschaft, ERC, NOVA and NWO-M grants; Target; the University of Padova, and the University Federico II (Naples).

GAMA is a joint European-Australasian project based around a spectroscopic campaign using the Anglo-Australian Telescope. The GAMA input catalogue is based on data taken from the Sloan Digital Sky Survey and the UKIRT Infrared Deep Sky Survey. Complementary imaging of the GAMA regions is being obtained by a number of independent survey programs including GALEX MIS, VST KiDS, VISTA VIKING, WISE, Herschel-ATLAS, GMRT and ASKAP providing UV to radio coverage. GAMA is funded by the STFC (UK), the ARC (Australia), the AAO, and the participating institutions. The GAMA website is \url{www.gama-survey.org}.

We are indebted to Ian McCarthy, who provided the BAHAMAS data products used in our analysis. Bob Sanders provided us useful comments about the relation between MOND and the M16 fitting function, and about the deflection of photons in MOND. We would also like to thank Federico Lelli, who provided the idea for the lensing rotation curves shown in Fig.~\ref{fig:Vrot_kids}. Finally, we would like to thank the anonymous referee for insightful questions and useful comments, which helped to increase the value of this work.

This work has made use of CosmoHub \cite[]{carretero2017,tallada2020}. CosmoHub has been developed by the Port d'Informaci{\'o}n Cient{\'i}fica (PIC), maintained through a collaboration of the Institut de F{\'i}sica d'Altes Energies (IFAE) and the Centro de Investigaciones Energ{\'e}ticas, Medioambientales y Tecnol{\'o}gicas (CIEMAT), and was partially funded by the ``Plan Estatal de Investigaci{\'o}n Cient{\'i}fica y T{\'e}cnica y de Innovaci{\'o}n'' program of the Spanish government.

KAO acknowledges support by the Netherlands Foundation for Scientific Research (NWO) through VICI grant 016.130.338 to M. Verheijen, and support from the European Research Council (ERC) through Advanced Investigator grant to C.S. Frenk, DMIDAS (GA 7786910). MBi is supported by the Polish National Science Center through grants no. 2020/38/E/ST9/00395, 2018/30/E/ST9/00698 and 2018/31/G/ST9/03388, and by the Polish Ministry of Science and Higher Education through grant DIR/WK/2018/12. CH acknowledges support from the European Research Council under grant number 647112, and support from the Max Planck Society and the Alexander von Humboldt Foundation in the framework of the Max Planck-Humboldt Research Award endowed by the Federal Ministry of Education and Research. HHo acknowledges support from Vici grant 639.043.512, financed by the Netherlands Organisation for Scientific Research (NWO). AHW is supported by an European Research Council Consolidator Grant (No. 770935). MA acknowledges support from the European Research Council under grant number 647112. AD acknowledges the ERC Consolidator Grant (No. 770935). BG acknowledges support from the European Research Council under grant number 647112 and from the Royal Society through an Enhancement Award (RGF/EA/181006). HHi is supported by a Heisenberg grant of the Deutsche Forschungsgemeinschaft (Hi 1495/5-1) as well as an ERC Consolidator Grant (No. 770935). KK acknowledges support from the Royal Society and Imperial College. HYS acknowledges the support from NSFC of China under grant 11973070, the Shanghai Committee of Science and Technology grant (No. 19ZR1466600) and Key Research Program of Frontier Sciences, CAS (No. ZDBS-LY-7013). TT acknowledges support from the European Research Council under grant number 647112, as well as funding from the European Union's Horizon 2020 research and innovation programme under the Marie Sk\l{}odowska-Curie grant agreement (No. 797794). JLvdB is supported by an ERC Consolidator Grant (No. 770935). The work of MV is funded by the canton of Geneva and the Swiss National Science Foundation, through Project Grants 200020 182513 and NCCR 51NF40-141869 (SwissMAP).

This work is partly based on tools and data products produced by GAZPAR operated by CeSAM-LAM and IAP.

Computations for the $N$-body simulations were performed in part on the Orcinus supercomputer at the WestGrid HPC consortium (\url{www.westgrid.ca}), in part on the GPC supercomputer at the SciNet HPC Consortium. SciNet is funded by: the Canada Foundation for Innovation under the auspices of Compute Canada; the Government of Ontario; Ontario Research Fund - Research Excellence; and the University of Toronto.

We are grateful to \url{https://math.stackexchange.com} user Paul Enta for providing an expression for one of the integrals needed in Appendix~\ref{app:PPL_method}.

This work has made use of {\scshape python} (\url{www.python.org}), including the packages {\scshape numpy} (\url{www.numpy.org}) and {\scshape scipy} (\url{www.scipy.org}). Plots have been produced with {\scshape matplotlib} \cite[]{hunter2007}.

\emph{Author contributions:} All authors contributed to the development and writing of this paper. The authorship list is given in three groups: the lead authors (M. Brouwer, K. Oman, E. Valentijn), followed by two alphabetical groups. The first alphabetical group includes those who are key contributors to both the scientific analysis and the data products. The second group covers those who have either made a significant contribution to the data products, or to the scientific analysis.




\bibliographystyle{aa}
\bibliography{biblio}



\appendix

\section{Isolated galaxy selection and validation}
\label{app:Isolation}

After performing the measurement of the RAR using GGL, our final goal was to compare the results to the different analytical models (Section~\ref{sec:Theory}) and N-body simulations (Section~\ref{sec:Simulations}) that make specific predictions on the galaxy-halo connection. While the simulations were designed to describe galaxies in their cosmological environment, the analytical models mainly focus on the description of individual galaxies. This means that, in order to test these models, we need to select galaxies that are relatively isolated. We defined our isolated lenses such that they do not have any satellites with more than a fraction $f\un{M_\star} \equiv M\un{\star,sat}/M\un{\star,lens} = 0.1$ of their stellar mass within a spherical radius $r\un{sat}=3\hMpc$ (see Section~\ref{sec:KiDS-bright}).

Here we validate our isolation criterion using the KiDS-bright and MICE datasets. We find that increasing the value of $r\un{sat}$ does not yield any decrease in the `two-halo term': the GGL signal at larger scales ($>0.3\hMpc$) corresponding to the contribution of satellites. This is true both when all lens masses are considered, and when they are restricted to a specific stellar mass: $\log(M_\star/\hmsun)=10.5\pm0.1$. Using both the KiDS-bright and MICE galaxies, we reduce the satellite mass fraction to $f\un{M_\star}=0$ (corresponding to no visible satellites). This also yields no decrease in the two-halo term of the ESD profile since galaxies with $f\un{M_\star}\ll0.1$ are not likely to be observed in a flux-limited survey. When we restrict the total stellar mass $M\un{\star,tot}$ of all satellites within $r\un{sat}$ to $f\un{M\un{\star,tot}}<0.1$ this does not significantly affect the isolated lens sample (i.e. the samples selected with KiDS-bright are $>99\%$ overlapping). Using the MICE and KiDS data we also experimented with selecting lenses that are isolated within a conical frustum, defined by a projected radius $R$ and line-of-sight distance range $\Updelta D$ around the lens. However, significantly increasing $\Updelta D$ beyond $3 \hMpc$ has no effect on the ESD profile, until it reduces our number of selected lenses to the point where the $S/N$ does not allow for a significant measurement. Finally, we apply our isolation criterion to the GAMA survey, to compare our current isolated sample with the `isolated centrals' that we used in \cite{brouwer2017}. These were selected using a more elaborate isolation criterion, which was driven by the Friends-of-Friends (FoF) group finding algorithm of \cite{robotham2011}. We find that the two isolated galaxy samples are more than $80\%$ overlapping.

\begin{figure}
	\resizebox{\hsize}{!}{\includegraphics{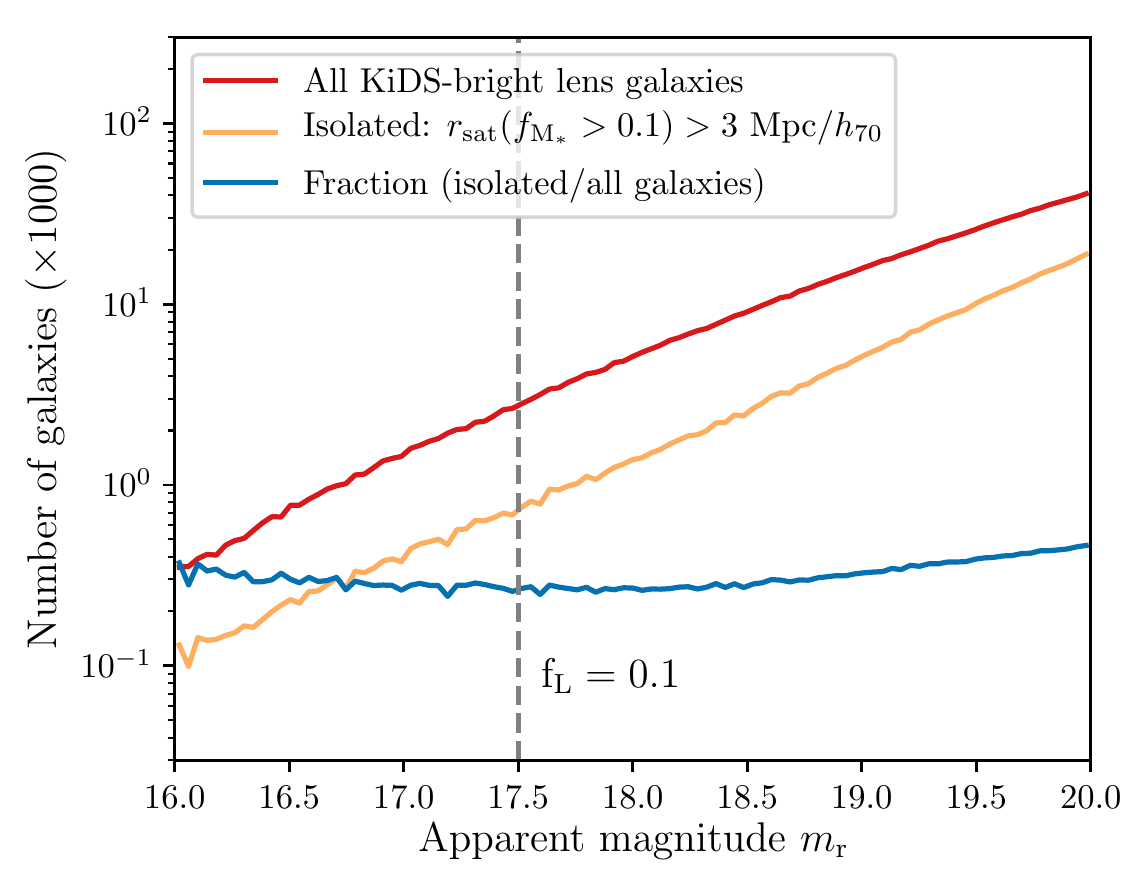}}
	\caption{Histogram of the number of isolated galaxies (orange line) compared to the total number of galaxies (red line), as a function of apparent $r$-band magnitude $m\un{r}$. The dashed vertical line represents the magnitude $m\un{bright}$, below which all satellites with a luminosity fraction larger than $f\un{L} \equiv L\un{sat}/L\un{lens}=0.1$ compared to the lens are still detected. Beyond this limit, the fraction of isolated galaxies (blue line) slightly increases because satellites fainter than the flux limit are not detected, which can cause lenses close to the magnitude limit ($m\un{lim}=20 \magn$) to be falsely identified as isolated.}
	\label{fig:isolation_test_fraction}
\end{figure}

\begin{figure}
	\resizebox{\hsize}{!}{\includegraphics{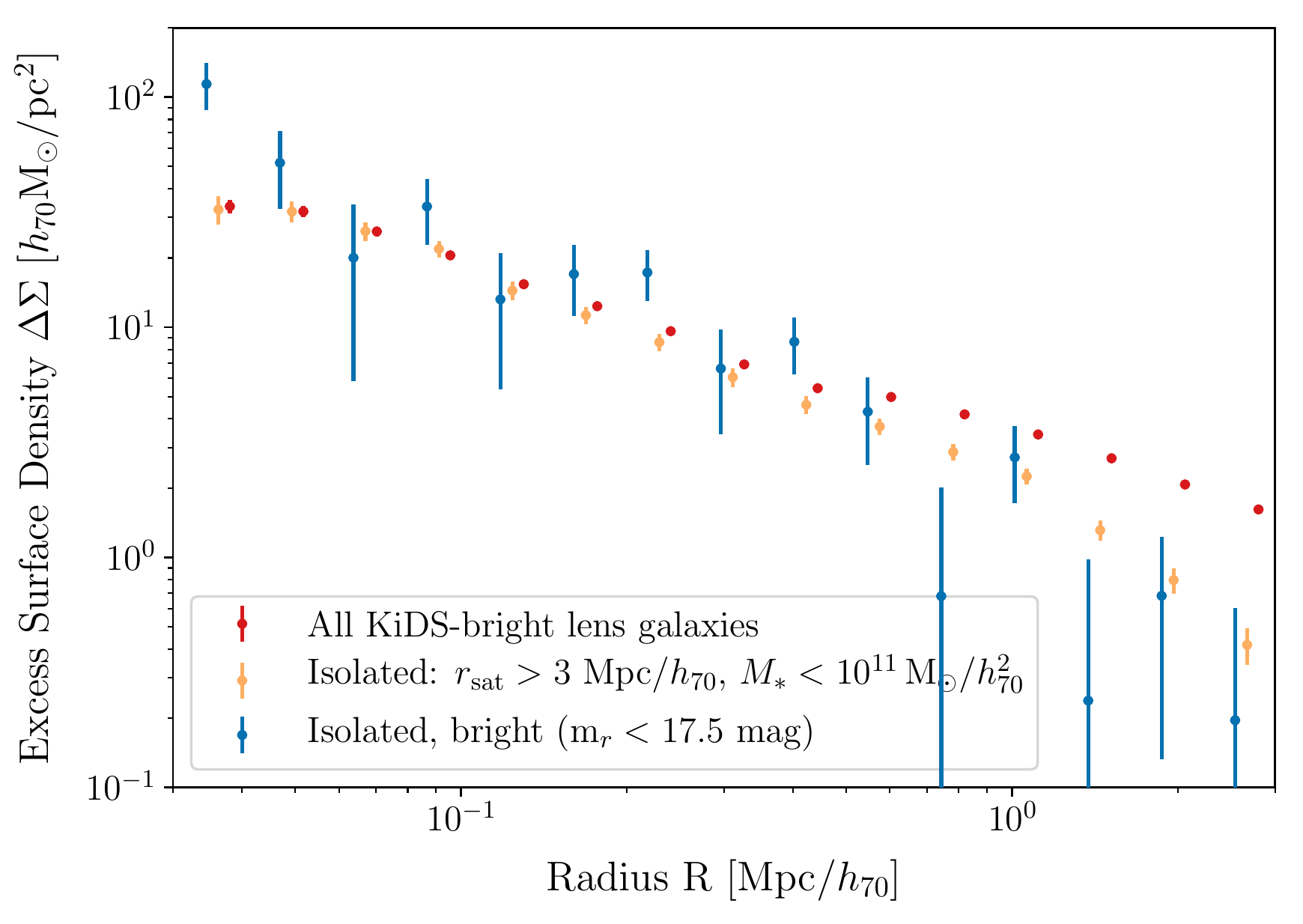}}
	\caption{Measured ESD profiles of our full sample of isolated KiDS-bright galaxies (red points with $1\sigma$ error bars), compared to that of a more reliable `bright' sample (blue, $m\un{r}<17.5 \magn$), which allows us to see all satellites down to luminosity fraction $f\un{L} \equiv L\un{sat}/L\un{lens}=0.1$. This is done to assess the effect of the KiDS-bright magnitude limit ($m\un{r}<20 \magn$) on the isolation criterion. Due to the smaller number of lenses, the ESD profile of the bright isolated sample is noisier. Nevertheless, its behaviour on both small and large scales is consistent with the ESD profile of the full isolated sample (orange), indicating that the effect of the magnitude limit is limited.}
	\label{fig:isolation_test_ESD}
\end{figure}

However, because both the GAMA survey and the samples designed to mimic it (KiDS-bright and MICE) are flux-limited, satellites that are fainter than the flux limit are not detected. This can cause lenses that are close to the magnitude limit ($m\un{lim}=20 \magn$) to be falsely identified as isolated. This problem is illustrated in Fig.~\ref{fig:isolation_test_fraction}, which shows that the fraction of galaxies assigned to the isolated lens sample increases for higher values of the apparent $r$-band magnitude $m\un{r}$. The dashed vertical line represents the magnitude $m\un{bright}$, below which all satellites with a luminosity fraction larger than $f\un{L} \equiv L\un{sat}/L\un{lens}=0.1$ compared to the lens are still detected. In the case of KiDS-bright:
\begin{equation}
m\un{bright} = m\un{lim} - 2.5 \log_{10}(f\un{F}=0.1) = 17.5 \magn \, .
\end{equation}
Applying $m_r < m\un{bright}$ provided us with an isolated lens sample that should be free of false positives, allowing us to estimate their effect on the ESD profiles. In Fig.~\ref{fig:isolation_test_ESD} we compare the ESD profiles of all galaxies and isolated galaxies with the more reliable `bright' sample. The mean stellar masses of the lens samples are very similar for the bright and the full sample: $\log\meanb{M_\star/\hmsun} = 10.78$ and $10.77$ respectively, and slightly lower for the full isolated sample: $\log\meanb{M_\star/\hmsun} = 10.61$. Due to the smaller number of lenses (only $3800$), the ESD errors and scatter of the bright isolated sample are much larger than those of the full isolated sample. Nevertheless, it is clear that their ESD profiles show consistent behaviour at both small and large scales. Compared to the total (non-isolated) galaxy sample, both isolated samples show significantly lower lensing signals at large scales (the two-halo term, corresponding to the contribution of satellites). The high level of consistency between the ESD profiles of the full and bright isolated samples indicates that the effect of false positives due to the magnitude limit is limited. In addition, by comparing the expected percentage of true isolated galaxies ($32.0\%$, found in the bright sample) with the higher percentage found in the `faint' sample ($36.8\%$, for galaxies with $m_r > 17.5 \magn$), we estimated that the expected percentage of false positives is less than $20\%$ of the full sample of isolated galaxies.

\begin{figure}
	\resizebox{\hsize}{!}{\includegraphics{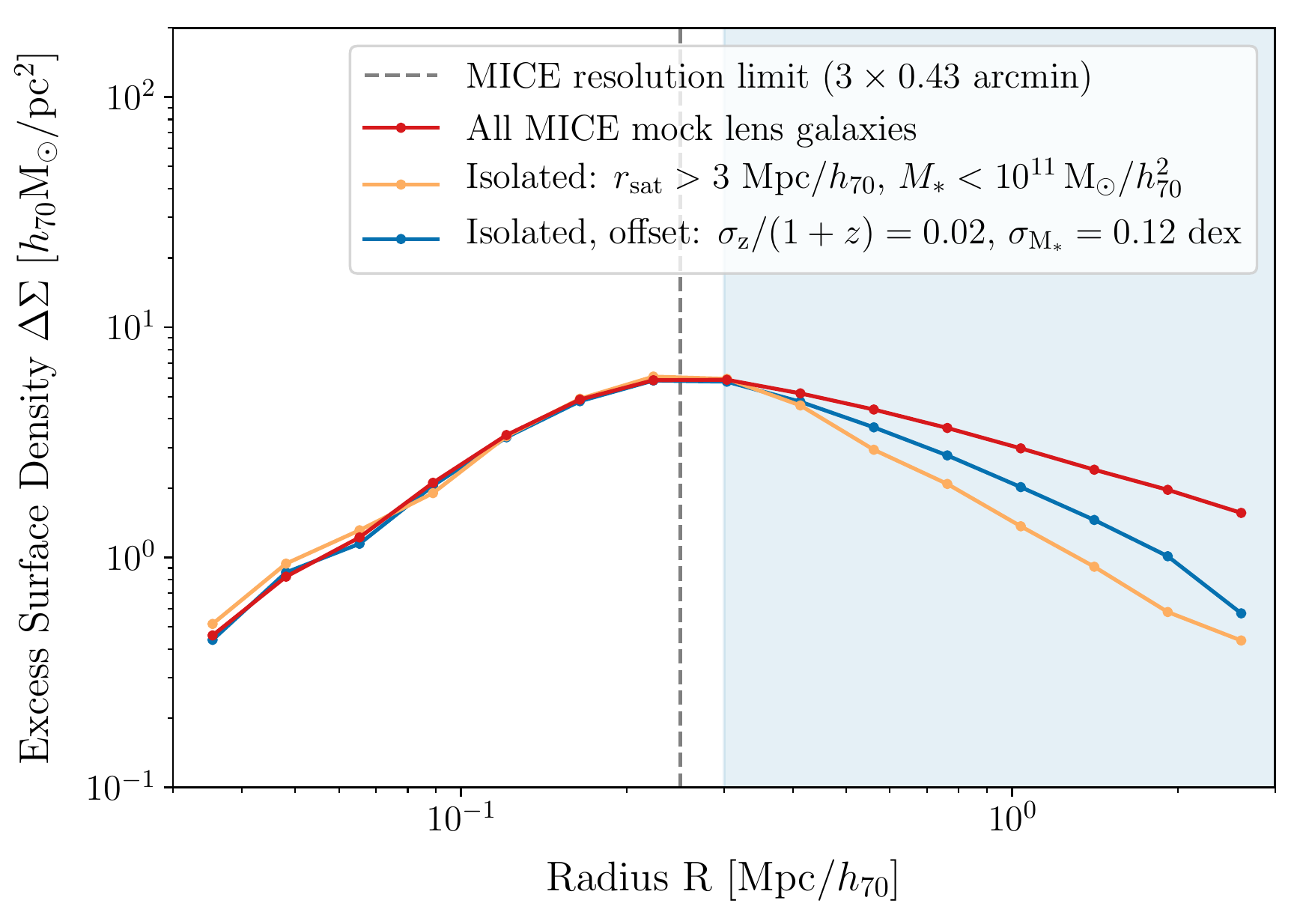}}
	\caption{Simulated ESD profile of the offset isolated MICE sample (blue line), created by using MICE galaxies with randomly offset redshifts ($\sigma\un{z}/[1+z] = 0.02$) and stellar masses ($\sigma\un{M_\star} = 0.12 \dex$) when selecting the isolated lenses. This is done in order to mimic the effect of the KiDS-bright measurement uncertainties on the isolation criterion. Compared to the ESD profile of the truly isolated MICE sample (orange line) the offset sample has a $\sim30\%$ higher signal at large scales due to the contribution of satellites. We therefore take extra care with KiDS-bright results at $R > 0.3 \hMpc$ (light blue shaded region). Nevertheless, the ESD of the offset isolated MICE sample is significantly lower than that of all MICE galaxies (red line), created without any isolation criterion. In addition, we show the radius corresponding to three times the resolution of the MICE simulation (dashed vertical line), which in the case of the isolated MICE sample is $R<0.25\hMpc$. Throughout this work, we only use the MICE results beyond this radius.}
	\label{fig:isolation_test_offset}
\end{figure}

\begin{figure*}
	\includegraphics[width=17cm]{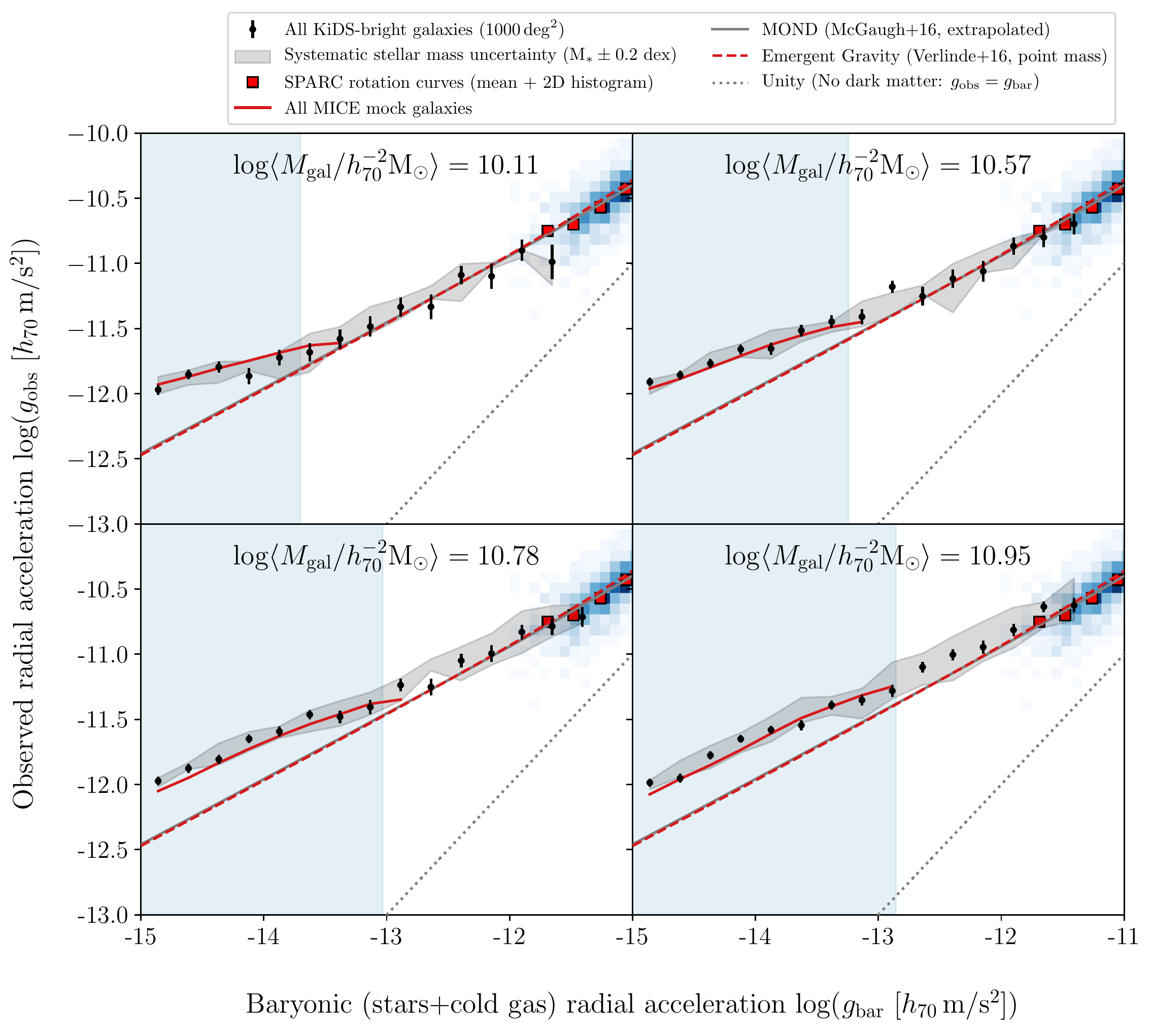}
	\caption{Measured RAR of all GL-KiDS lenses (black points with $1\sigma$ error bars) divided into four stellar mass bins, with the mean galaxy mass (stars+cold gas) of the lenses shown at the top of each panel. This figure primarily shows the effect on the RAR when the isolation criterion is not applied, which is quite significant and depends on the stellar mass of the galaxies (which correlates with galaxy clustering). The extrapolated M16 and EG predictions (grey solid and red dashed lines) function merely as a reference, showing the approximate location of the isolated galaxy RAR. They do not represent predictions in this case because $g\un{bar}$ is calculated using only the baryonic masses of the main lens galaxies (without including the baryonic masses of the satellites). The predictions from the MICE simulation (red line) match with our observations, which shows that the clustering simulated within MICE, driving the low-acceleration upturn due to the two-halo term, is indeed quite accurate.}
	\label{fig:RAR_kids_mice_mstarbins_all}
\end{figure*}

Nevertheless, we used the MICE simulations to perform one additional test. We selected the isolated sample of MICE lenses using satellite galaxies that extend to $m\un{r}<22.5 \magn$, such that all satellites with $f\un{L}>0.1$ can be observed. This paints a similar picture as the bright KiDS sample: although the much smaller sample of isolated galaxies selected using the faint satellites greatly increases the scatter, we find no consistent decrease in the lensing signal at $>0.3\hMpc$ scales compared to the original sample of isolated MICE galaxies. All these tests demonstrate the overall robustness of our isolation criterion. In addition, we note that this issue is only relevant when comparing our observations to the theoretical models (EG, MOND and N17). When comparing to the simulations (BAHAMAS and MICE), applying the same isolation criterion to both data and mocks ensured that any issues with the isolated galaxy selection are mimicked.

The major difference between the isolated galaxy selection of the GAMA and mock galaxies compared to KiDS-bright is that for GAMA and the mocks the true redshift values are known, whereas the \textsc{ANNz2} photometric redshifts of KiDS-bright are only known within a certain standard deviation $\sigma\un{z}$ (see Section~\ref{sec:KiDS-bright}). Since these photometric redshifts were used to calculate the galaxy distances $D(z)$ (using a flat $\lcdm$ cosmology, ignoring peculiar velocities) they directly affect the observed spherical distances $r$ between the galaxies, a key ingredient of the isolation criterion. The redshift uncertainty also affects the KiDS-bright stellar mass estimates, which influence both the isolation criterion (through $f\un{M_\star}$) and the application of the stellar mass limit: $\log(M_\star/\hmsun) < 11$. We assessed the effect of these uncertainties on the isolated galaxy selection by adding a normally distributed random offset with $\sigma\un{z}/(1+z) = 0.02$ to the MICE redshifts, and $\sigma\un{M_\star} = 0.12 \dex$ to its stellar masses. We find that the effect of the mass uncertainty is negligible, but that of redshift uncertainty is significant. Because the random redshift offset decreases the galaxy clustering, it increases the number of galaxies selected by the isolation criterion, adding galaxies that are not truly isolated to the lens sample (as well as excluding some truly isolated galaxies).

The ESD profile of the offset isolated MICE sample is shown in Fig.~\ref{fig:isolation_test_offset}, compared to the ESD profiles of all MICE galaxies (without any isolation criterion) and the truly isolated MICE sample. At scales $R > 0.3 \hMpc$, the ESD of the isolated sample selected using the offset MICE data is $\sim 30\%$ higher than that of the truly isolated MICE galaxies. When comparing our KiDS-bright lensing measurements to the MICE simulation, we always take this effect into account by mimicking the redshift offset in the simulation. However, for our comparison with the analytical models (MOND, EG and N17) this process is more difficult. When testing these models, we can only use the ESD profile of isolated KiDS-bright lenses within $R < 0.3 \hMpc$. For the mean galaxy mass of the KiDS-bright isolated sample ($\log[M\un{gal}/\hmsun] = 10.69$) this corresponds to a baryonic acceleration of $g\un{gal}>7.56\E{-14} \mpss$. For each RAR measurement resulting from isolated KiDS-bright lenses we will indicate the range in $g\un{bar}$ where the measurement is reliable, based on the mean $M\un{gal}$ of the appropriate lens sample.

Finally, to indicate the effect of selecting isolated galaxies on our lensing RAR measurements, Fig.~\ref{fig:RAR_kids_mice_mstarbins_all} shows the RAR of KiDS-bright and MICE galaxies for all lens galaxies without applying the isolation criterion. Because the clustering of galaxies (and hence the effect of the satellite galaxies) correlates with their stellar mass, we divided the lens galaxies into the same four stellar mass bins as used in Section~\ref{sec:Results-Stellar_mass}: $\log_{10}(M_\star/\hmsun) = [8.5,10.3,10.6,10.8,11.0]$. In that section, Fig.~\ref{fig:RAR_kids_mice_mstarbins} shows the RAR of isolated galaxies in the same stellar mass bins. In both cases, $g\un{bar}$ is calculated using only the baryonic masses of the main lens galaxies (i.e. the baryonic masses of the satellites are not included in the x-axis of Fig.~\ref{fig:RAR_kids_mice_mstarbins_all}). Comparing these two results shows that the effect of our isolated galaxy selection on $g\un{obs}$ is very striking: the isolated RAR measurements in Fig.~\ref{fig:RAR_kids_mice_mstarbins} approximately follow the extrapolated M16 and EG predictions, while the non-isolated RAR measurements in Fig.~\ref{fig:RAR_kids_mice_mstarbins_all} lie well above these lines at low accelerations ($g\un{bar}<10^{-13} \mpss$). As expected, the strength of this two-halo term (which shows the amount of clustering) increases with increasing galaxy stellar mass. Again the MICE simulation was able to predict our measurements: $\chi\un{red}^2 = 51.3 / 33 = 1.6$ ($2.3 \sigma$). This shows that the clustering simulated within MICE, which drives the low-acceleration upturn due to the two-halo term, is indeed quite accurate. This was to be expected since the clustering in MICE is constructed to reproduce the SDSS observations at $z<0.25$ \cite[]{zehavi2011}.

\section{Calculating $g\un{obs}$ from an ESD profile}
\label{app:PPL_method}

To calculate $g\un{obs}$ from the ESD profile throughout this work, we have used a simple analytical method that assumes that DM haloes can be roughly approximated with a singular isothermal sphere density model (see Section~\ref{sec:Conversion}). To make sure this conversion is robust, we compared it to a more elaborate numerical approach that fits a piece-wise power law (PPL) to the stacked ESD profile, without any assumption on the averaged halo shape except for spherical symmetry. We validate both methods using mock surface density maps from the BAHAMAS simulation in Section~\ref{sec:Conversion_test}.

The PPL method assumes a self-consistent form for the volume density profile $\rho(r)$ and parametrises it as a piece-wise power law constrained to be continuous. This comes at the cost of needing to invert the non-linear function $\Updelta\Upsigma(\rho)$, which we achieve via an iterative method. We chose to parametrize $\rho(r)$ in terms of $N$ pairs of values $(r_n,\rho_n)$ such that the slope $a_n$ and normalisation $b_n$ of the power law profile segments are:
\begin{align}
\ln\rho &= a_n \ln(r) + b_n \\
a_n &= \frac{\ln(\rho_{n+1})-\ln(\rho_n)}{\ln(r_{n+1})-\ln(r_n)}\\
b_n &= \ln(\rho_n) - a_n\ln(r_n)\\
(a_n, b_n) &=
\begin{cases}
(a_0, b_0) & {\rm if}\ r < r_0\\
(a_n, b_n) & {\rm if}\ r_n \leq r < r_{n+1}\\
(a_{N-1}, b_{N-1}) & {\rm if}\ r \geq r_N \, .
\end{cases}
\label{eq:rho_piecewise_powerlaw}\end{align}

\onecolumn

The ESD profile was measured in a series of discrete radial bins with edges $R_m$. The representative value at the centre of the bin\footnote{Here we define the bin centre as $\frac{1}{2}(R_m+R_{m+1})$, not the logarithmic centre $\sqrt{R_mR_{m+1}}$, which ensures accuracy in the calculation of the mean enclosed surface density.} is $\Updelta\Upsigma_m=\overline{\Upsigma}_m-\Upsigma_m$, where $\overline{\Upsigma}_m$ is the mean surface density within $\frac{1}{2}(R_m+R_{m+1})$ and $\Upsigma_m$ is the surface density averaged over the interval $[R_m,R_{m+1})$. We give an expression for this discrete ESD profile in terms of the parametric form for $\rho(r)$ given in Eq.~\ref{eq:rho_piecewise_powerlaw}.

The mean enclosed surface density is:
\begin{align}
\overline{\Upsigma}_m &= \frac{1}{\pi R_mR_{m+1}}\left[I_1(0, \sqrt{R_0R_1}, \tilde{a}_0, \tilde{b}_0) + 
\sum_{k=0}^m I_1(\sqrt{R_mR_{m+1}},\sqrt{R_{m+1}R_{m+2}}, \tilde{a}_m, \tilde{b}_m)\right]\\
\tilde{a}_m &= \frac{\ln(\Upsigma_{m+1})-\ln(\Upsigma_m)}{\frac{1}{2}\left(\ln(R_{m+2})-\ln(R_m)\right)}\\
\tilde{b}_m &= \ln(\Upsigma_m) - \frac{1}{2}\tilde{a}_m\ln(R_mR_{m+1})\\
I_1(R_i,R_j,\tilde{a},\tilde{b}) &= \frac{2\pi e^{\tilde{b}}}{\tilde{a}+2}\left(R_j^{a+2} - R_i^{a+2}\right) \, ,
\end{align}
\noindent and the local surface density is given by:
\begin{align}
\Upsigma_m &= \sum_{n=0}^{N-1}
\begin{cases}
0 & {\rm if}\ r_{n+1} < R_m\\
\frac{4e^{b_n}}{R_{m+1}^2-R_m^2} \left(-I_2(r_{n+1},R_m,a_n)\right) & {\rm if}\ r_n < R_m\ {\rm and}\ R_m \leq r_{n+1} < R_{m+1}\\
\frac{4e^{b_n}}{R_{m+1}^2-R_m^2} \left(I_2(r_{n+1},R_{m+1},a_n)-I_2(r_{n+1},R_m,a_n)\right) & {\rm if}\ r_n < R_m\ {\rm and}\ r_{n+1} \geq R_{m+1}\\
\frac{4e^{b_n}}{R_{m+1}^2-R_m^2} (I_2(r_{n+1},r_n,a_n)-I_2(r_{n+1},R_m,a_n)\\ \quad +I_2(r_n,R_m,a_n)+I_2(r_{n+1},R_{m+1},a_n)\\ \quad -I_2(r_{n+1},r_n,a_n)) & {\rm if}\ R_m \leq r_n < R_{m+1}\ {\rm and}\ r_n \geq R_{m+1}\\
\frac{4e^{b_n}}{R_{m+1}^2-R_m^2} (I_2(r_{n+1},R_{m+1},a_n)-I_2(r_{n+1},R_m,a_n)\\ \quad -I_2(r_n,R_{m+1},a_n)+I_2(r_n,R_m,a_n)) & {\rm if}\ r_n \geq R_{m+1}\\
\frac{4e^{b_n}}{R_{m+1}^2-R_m^2} (I_2(r_{n+1},r_n,a_n)-I_2(r_{n+1},R_m,a_n)\\ \quad +I_2(r_n,R_m,a_n)-I_2(r_{n+1},r_n,a_n)) & {\rm if}\ r_n \geq R_m\ {\rm and}\ r_{n+1} < R_m\\
\end{cases}\\
I_2(r,R,a) &=
\begin{cases}
-\frac{1}{3}R^{a+3}\left(\frac{r^2}{R^2}-1\right)^{\frac{3}{2}}{}_2{\rm F}_1\left(\frac{3}{2},-\frac{a}{2};\frac{5}{2};1-\frac{r^2}{R^2}\right) & {\rm if}\ r\ {\rm is}\ {\rm finite}\\
\frac{\sqrt{\pi}}{2}\frac{\Gamma\left(-\frac{a+1}{2}\right)}{\Gamma\left(-\frac{a}{2}\right)}\frac{R^{a+3}}{a+3} & {\rm if}\ r=\infty \, ,
\end{cases}
\end{align}
where ${}_2{\rm F}_1(\cdot,\cdot;\cdot;\cdot)$ is the Gaussian hypergeometric function and $\Gamma(\cdot)$ is the Gamma function. We assumed that the power law slope in the innermost bin continues to $r=0$, and that the slope in the outermost bin continues to $r\rightarrow\infty$. When inverting $\Updelta\Upsigma(\rho)$, we imposed uninformative priors on the power law slopes except those constraints required to guarantee that the total mass is finite and that the calculation is numerically stable.

In order to invert $\Updelta\Upsigma_m(\rho_n)$, we took as constant the values $R_m$, $\Updelta\Upsigma_{{\rm obs},m}$ and $r_{n}$. We then proposed an initial guess $\rho_n$, which we perturbed iteratively, calculating the corresponding $\Updelta\Upsigma_m$ at each iteration and comparing with $\Updelta\Upsigma_{{\rm obs},m}$ via the likelihood function:
\begin{align}
\ln\mathcal{L} &\propto -\frac{1}{2}(\Updelta\Upsigma\un{obs}-\Updelta\Upsigma)^{\rm T}C^{-1}(\Updelta\Upsigma\un{obs}-\Updelta\Upsigma) \, ,
\end{align}
where $C$ is the covariance matrix for the $\Updelta\Upsigma\un{obs}$. We used the package {\sc emcee} \citep{foreman-mackey13} to estimate the posterior probability distribution of $\rho_n$, and subsequently of the corresponding $g_{{\rm obs},n}$ via integration of the volume density profile.

\twocolumn

\section{The RAR of the N17 model}
\label{app:Results_N17}

\begin{figure*}
	\centering
	\includegraphics[width=17cm]{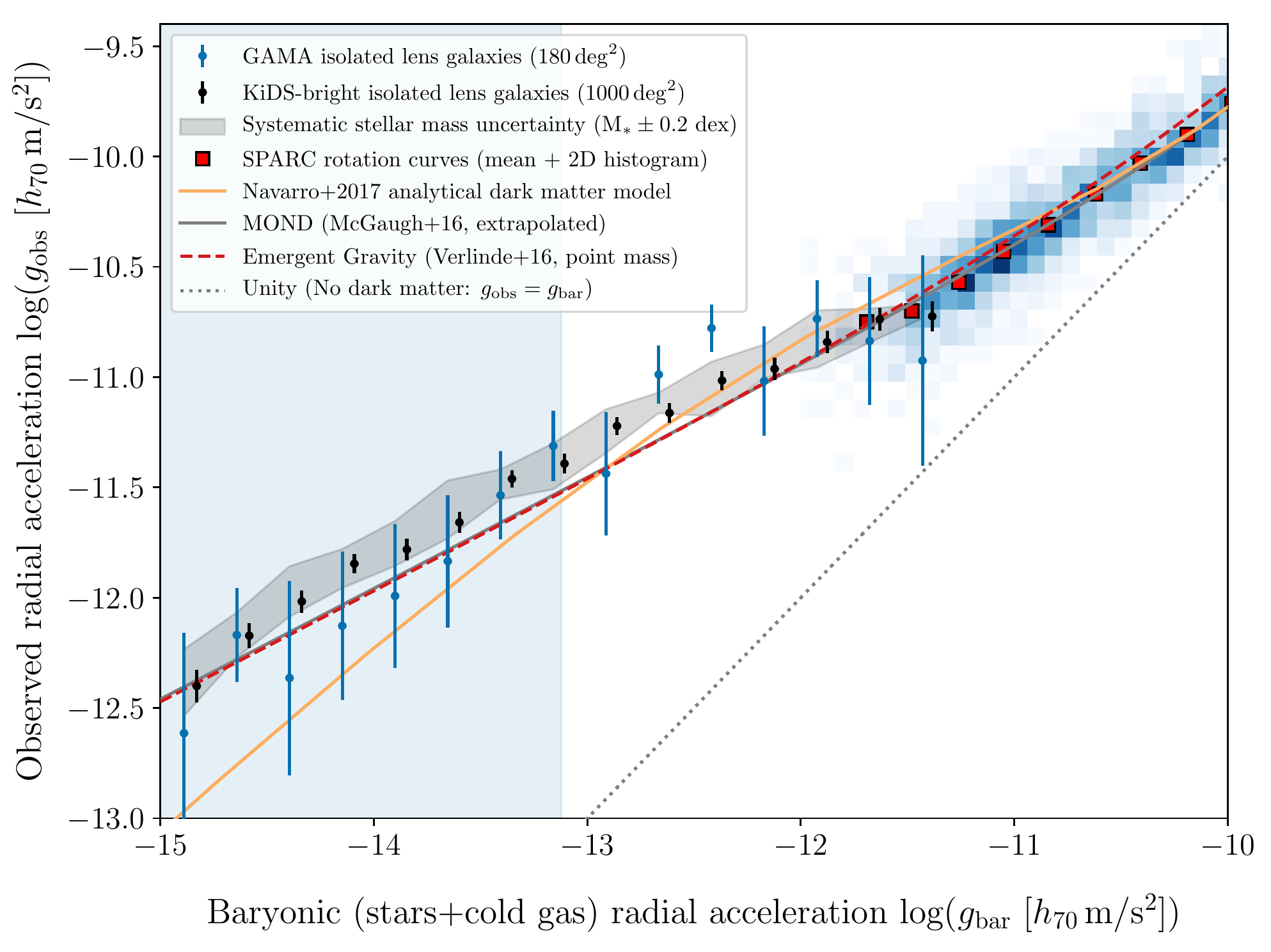}
	\caption{Measured RAR of the spectroscopic GAMA and photometric KiDS-bright isolated lens samples (blue and black points with $1\sigma$ error bars). We compare our results to the analytical $\lcdm$-based model created by N17. At higher accelerations there is a good match between the N17 model and the M16 observations, as expected. However, at lower accelerations the model bends down with respect to the lensing measurements, due to the steep outer slope of the NFW density profile ($\rho \propto r^{-3}$). Large-scale contributions to the total mass distribution, such as the average cosmic DM density, could slightly mitigate this discrepancy. However, these are not implemented into the simple analytical N17 model, which was created to reproduce the RAR at the small scales measured by rotation curves.}
	\label{fig:RAR_kids_gama_Navarro}
\end{figure*}

We used the lensing RAR to test the analytical prediction from the $\lcdm$-based model created by N17. In Fig.~\ref{fig:RAR_kids_gama_Navarro} we show the RAR predicted by this model for a galaxy with a baryonic mass equal to the average stellar + cold gas mass of the lens sample ($\log_{10}\meanb{M\un{gal}} = 10.69$). At higher accelerations there is a good match between the model and the M16 RAR measurements from galaxy rotation curves, which is expected since the N17 model is designed and confirmed to reproduce these results. However, at the lower accelerations unique to our lensing measurements the N17 model underpredicts the $g\un{obs}$ amplitude in comparison to our measurements. Due to their large error bars, the GAMA data can still accommodate the analytical prediction: $\chi\un{red}^2 = 0.90$. The KiDS-bright result, however, excludes the N17 prediction with $\chi\un{red}^2 = 4.8$, corresponding to $4.3 \sigma$. Here we have removed all data points beyond the KiDS isolation limit ($R>3 \hMpc$); therefore, the strong disagreement between N17 and the data is unlikely to be caused by contamination from satellites.

Because of the significant difference in the slope of the model and the data, even taking the $\Updelta M_\star=\pm0.2\dex$ uncertainty into account does not result in a better fit. This strong downward slope results from the $r^{-3}$ radial dependence of the Navarro-Frenk-White (NFW) density profile at large scales (where an $r^{-2}$ density profile would instead follow the same slope as the MG predictions in Fig.~\ref{fig:RAR_kids_gama_verlinde}). This effect could be slightly mitigated by taking into account the average DM density of the Universe, which would result in an upward turn towards an $r^{-1}$ slope at $g\un{bar}<10^{14}\mpss$ (as shown by the BAHAMAS prediction of the RAR in the lower panel of Fig.~\ref{fig:missing-baryons}). However, components contributing to the large-scale DM profile are not included in the N17 model, which was created to reproduce the RAR at the small scales measured by rotation curves. It is clear from this exercise that, while succeeding to describe the RAR at small scales, this simple model is not sufficient to reproduce the results at the larger scales probed by weak lensing. This requires more elaborate modelling within the $\lcdm$ paradigm, represented by large cosmological simulations such as BAHAMAS and MICE (see Section~\ref{sec:Simulations}). In Section~\ref{sec:Results-Simulations} we made a fairer comparison using these two simulations, which can mimic the measurement more faithfully.

\end{document}